\documentclass[a4paper,11pt]{article}
\pdfoutput=1 % if your are submitting a pdflatex (i.e. if you have
\usepackage{jcappub} % for details on the use of the package, please
\usepackage[T1]{fontenc} % if needed
\usepackage{orcidlink}

\newcommand{\namaster}{{\tt NaMaster}}
\newcommand{\bbpower}{{\tt BBPower}}
\newcommand\planck{\textit{Planck }}

\usepackage{siunitx}
\DeclareSIUnit{\belm}{Bm}
\DeclareSIUnit{\beli}{Bi}

\usepackage[super]{nth}
\usepackage{xcolor}

\usepackage{lineno}
\usepackage{amsmath, bm}

\title{The Simons Observatory: Quantifying the impact of beam chromaticity on large-scale $B$-mode science}

\author[a]{Nadia Dachlythra\orcidlink{0009-0006-7382-1434},}
\author[b]{Kevin Wolz\orcidlink{0000-0003-3155-6151},}
\author[c]{Susanna Azzoni\orcidlink{0000-0002-8132-4896},}
\author[b]{David Alonso\orcidlink{0000-0002-4598-9719},}
\author[d]{Adriaan J. Duivenvoorden\orcidlink{0000-0003-2856-2382},}
\author[e,f]{Alexandre E. Adler\orcidlink{0000-0002-5736-5524},}
\author[g,h]{Jon E. Gudmundsson\orcidlink{0000-0003-1760-0355},}
\author[i,j,k]{Carlo Baccigalupi\orcidlink{0000-0002-8211-1630},}
\author[i]{Alessandro Carones\orcidlink{0009-0004-1436-841X},}
\author[a,l]{Gabriele Coppi\orcidlink{0000-0002-6362-6524},}
\author[c]{Samuel Day-Weiss,\orcidlink{0009-0003-5814-2087}}
\author[m]{Josquin Errard\orcidlink{0000-0002-1419-0031},}
\author[n,o]{Nicholas Galitzki\orcidlink{0000-0001-7225-6679},}
\author[p]{Martina Gerbino\orcidlink{0000-0002-3538-1283},}
\author[q]{Remington G. Gerras\orcidlink{0009-0009-0876-9168},}
\author[r]{Carlos Hervias-Caimapo\orcidlink{0000-0002-4765-3426},}
\author[s]{Selim C. Hotinli\orcidlink{0000-0003-0061-8188},}
\author[a,l]{Federico Nati\orcidlink{0000-0002-8307-5088},}
\author[t]{Bruce Partridge\orcidlink{0000-0001-6541-9265},}
\author[c]{Yoshinori Sueno\orcidlink{0000-0002-3644-2009},}
\author[u]{Edward J. Wollack\orcidlink{0000-0002-7567-4451}.}

\affiliation[a]{Department of Physics, University of Milano-Bicocca, Piazza della Scienza 3, 20126, Milano, Italy}
\affiliation[b]{Department of Physics, University of Oxford, Denys Wilkinson Building, Keble Road, Oxford, OX1 3RH, UK}
\affiliation[c]{Joseph Henry Laboratories of Physics, Jadwin Hall, Princeton University, Princeton, NJ, USA 08544}
\affiliation[d]{Max-Planck-Institut fur Astrophysik, Karl-Schwarzschild Str. 1, 85741 Garching, Germany}
\affiliation[e]{Department of Physics, University of California, Berkeley, 366 LeConte Hall Berkeley, CA 94720, USA}
\affiliation[f]{Physics Division, Lawrence Berkeley National Laboratory, 
1 Cyclotron Road, Berkeley, CA 94720, USA}
\affiliation[g]{Science Institute, University of Iceland, 107 Reykjavik, Iceland}
\affiliation[h]{The Oskar Klein Centre, Department of Physics, Stockholm University, AlbaNova, SE-10691 Stockholm, Sweden}
\affiliation[i]{The International School for Advanced Studies (SISSA), via Bonomea 265, I-34136 Trieste, Italy}
\affiliation[j]{The National Institute for Nuclear Physics (INFN), via Valerio 2, I-34127, Trieste, Italy}
\affiliation[k]{The Institute for Fundamental Physics of the Universe (IFPU), Via Beirut 2, I-34151, Trieste, Italy}
\affiliation[l]{Istituto Nazionale di Fisica Nucleare, INFN, Sezione Milano-Bicocca, Piazza della Scienza 3, 20126, Milano, Italy}
\affiliation[m]{Universit\'e Paris Cit\'e, CNRS, Astroparticule et Cosmologie, F-75013 Paris, France}
\affiliation[n]{Department of Physics, University of Texas at Austin, Austin, TX, 78712, USA}
\affiliation[o]{Weinberg Institute for Theoretical Physics, Texas Center for Cosmology and Astroparticle Physics, Austin, TX 78712, USA}
\affiliation[p]{INFN Sezione di Ferrara, Via Giuseppe Saragat, 1, 44121, Ferrara}
\affiliation[q]{Department of Physics, University of Southern California, Los Angeles, CA 90089, USA}
\affiliation[r]{Instituto de Astrof\'isica and Centro de Astro-Ingenier\'ia, Facultad de F\'isica, Pontificia Universidad Cat\'olica de Chile, Av. Vicu\~na Mackenna 4860, 7820436 Macul, Santiago, Chile}
\affiliation[s]{Perimeter Institute for Theoretical Physics, 31 Caroline St N, Waterloo, ON N2L 2Y5, Canada}
\affiliation[t]{Department of Physics and Astronomy, Haverford College, 370 Lancaster Ave, Haverford PA 19041 USA}
\affiliation[u]{NASA Goddard Space Flight Center, 8800 Greenbelt Road, Greenbelt, MD 20771, USA}

\emailAdd{konstantina.dachlythra@unimib.it}\emailAdd{kevin.wolz@physics.ox.ac.uk}\emailAdd{sa5705@princeton.edu}\emailAdd{David.Alonso@physics.ox.ac.uk}\emailAdd{adriaand@mpa-garching.mpg.de}\emailAdd{aadler@lbl.gov}\emailAdd{jegudmunds@hi.is}\emailAdd{bacci@sissa.it}\emailAdd{acarones@sissa.it}\emailAdd{gabriele.coppi@unimib.it}\emailAdd{dayweiss@princeton.edu}\emailAdd{josquin@apc.in2p3.fr}\emailAdd{nicholas.galitzki@austin.utexas.edu}\emailAdd{marti.gerbino@gmail.com}\emailAdd{gerras@usc.edu}\emailAdd{carlos.hervias@uc.cl}\emailAdd{selimcanhotinli@gmail.com}\emailAdd{federico.nati@unimib.it}\emailAdd{bpartrid@haverford.edu}\emailAdd{ys5857@princeton.edu}\emailAdd{edward.j.wollack@nasa.gov}

\abstract{
The Simons Observatory (SO) Small Aperture Telescopes (SATs) will observe the Cosmic Microwave Background (CMB) temperature and polarization at six frequency bands. Within these bands, the angular response of the telescope (beam) is convolved with the instrument’s spectral response (commonly called bandpass) and the signal from the sky, which leads to the band-averaged telescope beam response, which is sampled and digitized. The spectral properties of the band-averaged beam depend on the natural variation of the beam within the band, referred to as beam chromaticity. 
In this paper, we quantify the impact of the interplay of beam chromaticity and intrinsic frequency scaling from the various components that dominate the polarized sky emission on the tensor-to-scalar ratio, $r$, and foreground parameters.
We do so by employing a parametric power-spectrum-based foreground component separation algorithm, namely \bbpower{}, to which we provide beam-convolved time domain simulations performed with the \texttt{beamconv} software while assuming an idealized version of the SO SAT optics. We find a small, $0.02\sigma$, bias on $r$, due to beam chromaticity, which seems to mostly impact the dust spatial parameters, causing a maximum $0.77 \sigma$ bias on the dust $B$-mode spectra amplitude, $A_{d}$, when employing Gaussian foreground simulations. However, we find all parameter biases to be smaller than $1\sigma$ at all times, independently of the foreground model. This includes the case where we introduce additional uncertainty on the bandpass shape, which accounts for approximately half of the total allowed gain uncertainty, as estimated in previous work for the SO SATs.}
\begin{document}
\maketitle
\flushbottom

\section{Introduction}
\label{sec:intro}

Measurements of the Cosmic Microwave Background (CMB) polarization are at the forefront of cosmological research, as they offer insights into the origins and evolution of the Universe. Of particular interest is the measurement of the CMB $B$-mode polarization anisotropies (the parity-odd component of the CMB polarization) as a probe of the very early Universe and a strong test of the inflationary paradigm. Cosmic inflation is the current leading theory explaining the initial density perturbations, and the Universe's large-scale homogeneity and flatness. It postulates a brief period of exponential expansion of an initially small patch of the infant Universe into its current observable size. The primordial quantum fluctuations, stretched by this expansion to cosmological scales, leave traces in the spacetime metric in the form of both scalar and tensor perturbations. The latter, also referred to as gravitational waves, are the only primordial source of CMB $B$-mode fluctuations, with scalar modes only sourcing the parity-even $E$-modes. By measuring both components of the CMB polarization, we can quantify the amplitude of primordial gravitational waves generated by inflation, and thus probe the high-energy physical processes that caused it. This amplitude is commonly parametrized in terms of the ratio between the amplitudes of the power spectrum of primordial tensor and scalar perturbations, $r$ \citep{Crittenden_1993, Hu_1997, Zaldarriaga_1997}.

Cosmological $B$-modes are significantly smaller than the $E$-mode perturbations. In addition, on the large, degree-sized scales where the sensitivity to the primordial signal is highest, the polarized sky is dominated by emission from Galactic foregrounds \citep{WMAP_3_fore, Planck_dustfore_2015, planck_diff_sep_comp_2020}. Thus, the accuracy with which we can measure $r$ is tightly linked to our ability to properly characterize these foregrounds and, crucially, the impact of all instrumental effects on the measured signal. To cleanly separate the cosmological signal from any other contaminants, the CMB community has developed a large variety of component separation methods (e.g. \citep{Aumont_smica, Commander3, Delabrouille_2009, Stompor_2009, planck_diff_sep_comp_2020}) that rely mostly on the frequency-dependent Spectral Energy Distributions (SEDs) of the foregrounds to distinguish them from the CMB's nearly perfect blackbody spectrum \citep{Fixsen_2009}. However, the performance of any of these methodologies depends strongly on the impact of instrumental systematics on the data and on our ability to mitigate or model them. For $B$-mode experiments, great care must be taken when modeling the systematic effects related to the telescope’s optics. In particular, improper modeling of the telescope's spatial response, namely the telescope's Point-Spread-Function (PSF), commonly referred to as the telescope's beam, can cause leakage of the stronger CMB temperature signal into the fainter polarization data, as well as mixing between the $E$- and $B$-modes \citep{beamconv_2018}.

An important aspect of beam modeling is linked to the beam's intrinsic frequency dependence. As CMB telescopes typically observe radiation over wide frequency bands instead of monochromatic frequency channels, it is natural to expect some variation in the beam properties within each band. The impact of this beam ‘chromaticity’ is further complicated when the beam is convolved with frequency-dependent sky components with differing spectral signatures. It is worth noting that, due to these differing spectral signatures, the chromaticity effects will vary slightly for each component. Further complications arise when additional beam non-idealities are present. Representative examples include pronounced beam power at large angles away from the beam center (sidelobes), non-negligible polarization in the direction orthogonal to the detector's polarization direction (cross-polarization), as well as beam asymmetry \citep{bicep2_systematics, beamconv_2018, Lungu_2022}. For sufficiently complex beams and foregrounds, if these effects are not taken into account in the analysis, they may give rise to significant biases in the inferred model parameters. Therefore, these effects must be quantified to determine the most reliable estimate of the instrumental beam to employ in the component separation analysis \citep{Miller_beam_asymmetry, Shimon_2008}.

Although these beam-related biases are typically sub-percent at the power spectrum level, they can still threaten the high-accuracy requirements of current and next-generation CMB missions \citep{Leloup_2024} such as the Simons Observatory (SO) \citep{SO_overview_2019}. By observing the polarized microwave sky from the Chilean Atacama desert in six frequency bands centered between \SI{27}{} and \SI{280}{\giga\hertz}, the SO Small Aperture Telescopes (SATs) \citep{Galitzki_2024} aim to constrain $r$ with a statistical uncertainty of $\sigma(r)\simeq0.003$ or better. %The SO Large Aperture Telescope (LAT) will instead target the higher-$\ell$ multipoles focusing on measuring the lensed $E$-mode signal for delensing the $B$-modes and constraining late-time cosmological parameters, as well as performing Galactic science. 
Given the broad observing bands of the SATs, with $\sim 25\%$ relative fractional bandwidths, it is crucial to model the impact of the anticipated beam frequency dependence on the large-scale $B$-mode spectra. We do so in this paper, starting from time-domain simulations convolved with chromatic beams that we then analyse using a power-spectrum-based component separation pipeline. An estimation of the corresponding impact on scientific objectives relevant to the SO Large Aperture Telescope (LAT) \citep{SO_LATr_2021, Zhu_2021} like the effective number of relativistic species, $N_{\mathrm{eff}}$, and amplitude of unresolved radio sources in temperature, $a_{s}$, is presented in \cite{Giardello_2024}. That work assumes a LAT-like experiment and Gaussian beams with perturbed Full-Width-Half-Maximum (FWHM) values, which are applied to the simulated data at the power spectrum level.

The paper is structured as follows: Section \ref{beam_fdependence} introduces the individual components that form the telescope's band-averaged beam and describes the specifics of the beam simulations. Section \ref{methods} elaborates on the sky model assumptions and methods employed for the beam-convolved simulations, the power spectra estimation, and the foreground component separation algorithm. The results are presented in Section \ref{results}, including the impact of beam systematics of non-frequency-dependent beams, the inclusion of the beam chromaticity effect as a function of foreground complexity, and any further biases caused by the integration of bandpass uncertainty in the model. Finally, Section \ref{conclusion} summarizes the paper and outlines potential directions for future studies.

%Errors in the estimate of beam response impact cosmological analysis for CMB experiments. This is discussed widely in the literature (see e.g., \cite{}). Beam response depends on frequency and as part of cosmological analysis, it is standard to implement a band-averaged beam model that accounts for the finite bandwidth of a CMB experiment. The millimetre sky is composed of multiple astrophysical sources with different spectral energy distributions (SEDs) which effectively weigh the frequency-varying beam response differently. This varying frequency weighting of them beam response might lead to errors in CMB data analysis if not. Although uncertainty in the determination of the spectral response function of CMB experiments at the detector level has been explored in the literature (see \cite{}), we are unaware of analysis that probes the dependence on  beam chromaticity. $

%\begin{itemize}
%    \item We like CMB
%    \item $B$-modes especially
%    \item threats: foreground, noise and systematics (what each one does bad)
 %   \item frequency coverage and expert foreground component separation analysis - can happen in power spectra, maps etc.
 %   \item fgc analysis should include noise estimates and systematics  
 %   \item the spatial response of the instrument, the beam... 
 %   \item beam systematics + literature causes both T-to-P and E-to-B.. 
     
%\end{itemize}

\section{Spectral dependence of beam response}
\label{beam_fdependence}
%\todo{Suggestion: Introduce beam, bandpass and bandpass-integrated sky signal. Then talk about intrinsic chromaticity, how it couples with the sky component SEDs. Then introduce all higher-order effects we study here (beam asymmetry, cross-polar beams, sidelobes, bandpass errors).} 

The SO SATs observe the microwave sky at frequency bands of approximately $25\%$ width around the center frequencies. Assuming a detector on the focal plane characterized by an instrumental bandpass, $\tau(\nu)$, and beam response, $B(\theta, \phi, \nu)$, at frequency $\nu$, we can derive the band-averaged beam of the detector as follows:

\begin{equation}
\label{chrom_beam}
B_{S}(\theta, \phi) = \frac{\int \tau(\nu)B(\theta, \phi, \nu)S(\nu)d\nu}{\int \tau(\nu) S(\nu)d\nu}.   
\end{equation}

\noindent This formula assumes integration only over frequency and incorporates the SED of the observed sky component denoted by $S(\nu)$ expressed in antenna temperature units. The beam is described in terms of both the polar angle, $\theta$, and the azimuthal angle, $\phi$, allowing for beam asymmetry by introducing a dependency on the position angle in the time-ordered data. Note that, while the source emission will not be spatially constant, we avoid adding its orientation dependence in this equation, which describes the synthesis of the band-averaged beam. The spatial dependence of $S(\nu)$ is taken into account properly in the sky simulations.

To generate the SAT Physical Optics (PO) beam simulations, we assume the same optical configuration (three-lens refracting telescope) as for the work described in \cite{Dachlythra_2024} and employ the  \texttt{TICRA TOOLS}\footnote{TICRA, Landemærket 29, Copenhagen, Denmark (\url{https://www.ticra.com)}.} software. Specifically, we produce monochromatic far-field beam maps for four frequency bands roughly centered at \SI{93}{}, \SI{145}{}, \SI{225}{}, and \SI{280}{\giga\hertz}, respectively. Each of these Mid- and Ultra-High-Frequency (MF and UHF) bands has $25\%$ relative width around its center frequency, and to each band correspond five monochromatic beam maps evenly spread across the full frequency range of the band. Note that we have two MF and two UHF bands as the SATs employ dichroic detectors. For the nominal case, we assume a detector located at the center of the focal plane. We then integrate across the full band in each case, employing the original or perturbed versions of the simulated bandpasses presented in \cite{Abitbol_2021}. The beams of the Low-Frequency (LF) bands, which are roughly centered at \SI{27}{} and \SI{39}{\giga\hertz}, are approximated using Gaussian curves of FWHM equal to \SI{91}{\arcminute} and \SI{63}{\arcminute}, respectively. This approach is sufficient at this stage since the SO SATs will dedicate only a small part of their full scanning time to LF observations. Furthermore, we ensure in Appendix \ref{cl_based} that the partial treatment of only the MF and UHF bands as chromatic PO beams does not impact the analysis results. We do, however, plan to generate PO simulations for the LF beams and integrate them into the pipeline in future work. 

The resulting beam models include both asymmetric and cross-polar components, as well as faint far sidelobes. Note, however, that these beam models are still the product of ideal simulations of the SO SAT optics\footnote{More realistic optics simulations could, for example, include non-ideal baffling elements or reflective Half-Wave-Plates (HWPs), like the effects discussed in \citep{Didier_2019}. }. In reality, we may expect more pronounced beam non-idealities arising from the optical setup which could impact the cosmological analysis. Appendices \ref{eb_corr} and \ref{det_loc} discuss how the results could vary if we scale up the cross-polarization and ellipticity of the simulated beams. In Appendix \ref{eb_corr}, we present the $EB$ spectra of a single sky realization convolved with only co-polar beams, the nominal beams including their simulated cross-polar components, and a perturbed version of the latter, where we dramatically increased the cross-polarization amplitude by one order of magnitude. We find the resulting $EB$ correlation increasing non-negligibly only in the last case, and for multipoles $\ell \gtrsim 150$, while the degree scales are not largely impacted. To increase the beam ellipticity, we simulate one pixel at the edge of the focal plane, probing the asymmetrical illumination of the aperture. We compare the beam profiles for a center and edge pixel in Appendix \ref{det_loc}, and find a minor bias from the increased ellipticity beam on the fitted parameters.  

\begin{figure}
    \centering
    \includegraphics[width=\textwidth]{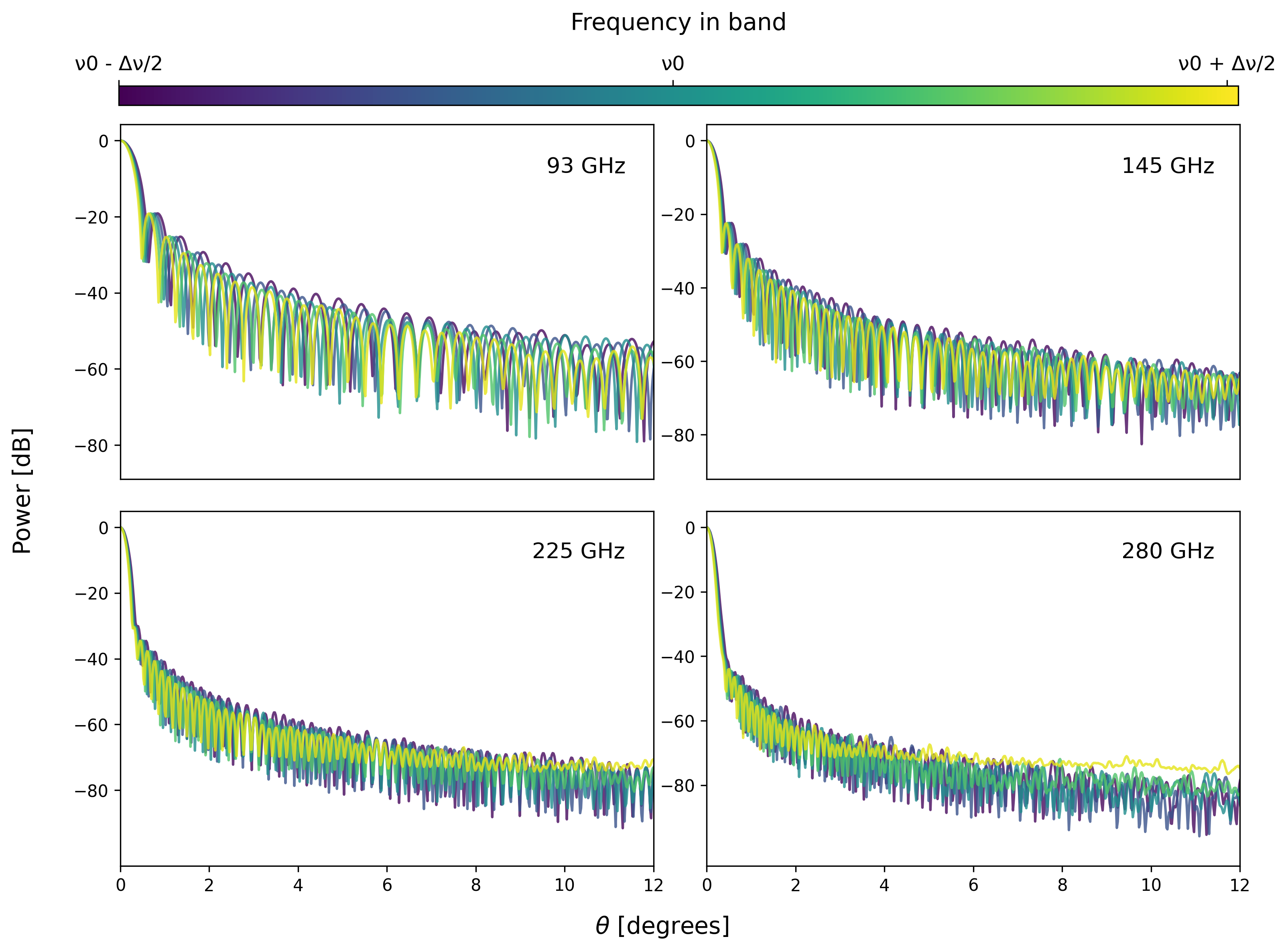}    \caption{\label{fig:chromatic_input_beams}Logarithmic profiles of five monochromatic co-polar beam maps generated for each of the 93 (top left), 145 (top right), 225 (bottom left), and \SI{280}{\giga\hertz} bands (bottom right), using \texttt{TICRA TOOLS} software. Each set of (five) maps is produced at frequencies of uniform spacing across the full frequency range of the corresponding band. All bands are assumed to have 25$\%$ fractional width, and all beam maps are generated for a detector on the center of the focal plane of a simulated three-lens refracting telescope.}
\end{figure}

The logarithmic profiles of the monochromatic co-polar beams for the MF and UHF bands are shown in Figure \ref{fig:chromatic_input_beams}. From the figure, we observe the profiles `contracting' with increasing frequency within each band, as expected for a diffraction limited beam. The asymptotic yellow curve in the bottom right panel, representing the highest frequency of the \SI{280}{\giga\hertz} band, highlights the challenge posed by the numerical precision of the PO simulations. Since the computational costs of the simulations scale as frequency squared, the accuracy with which the simulation converges on a solution is also frequency-dependent. At the limits of the machine's precision, very low beam power will, therefore, be approximated to be at the convergence threshold, creating a constant-looking sidelobe at large angles. 

\begin{figure}
    \hspace*{-1cm}
    \includegraphics[width=1.1\textwidth]{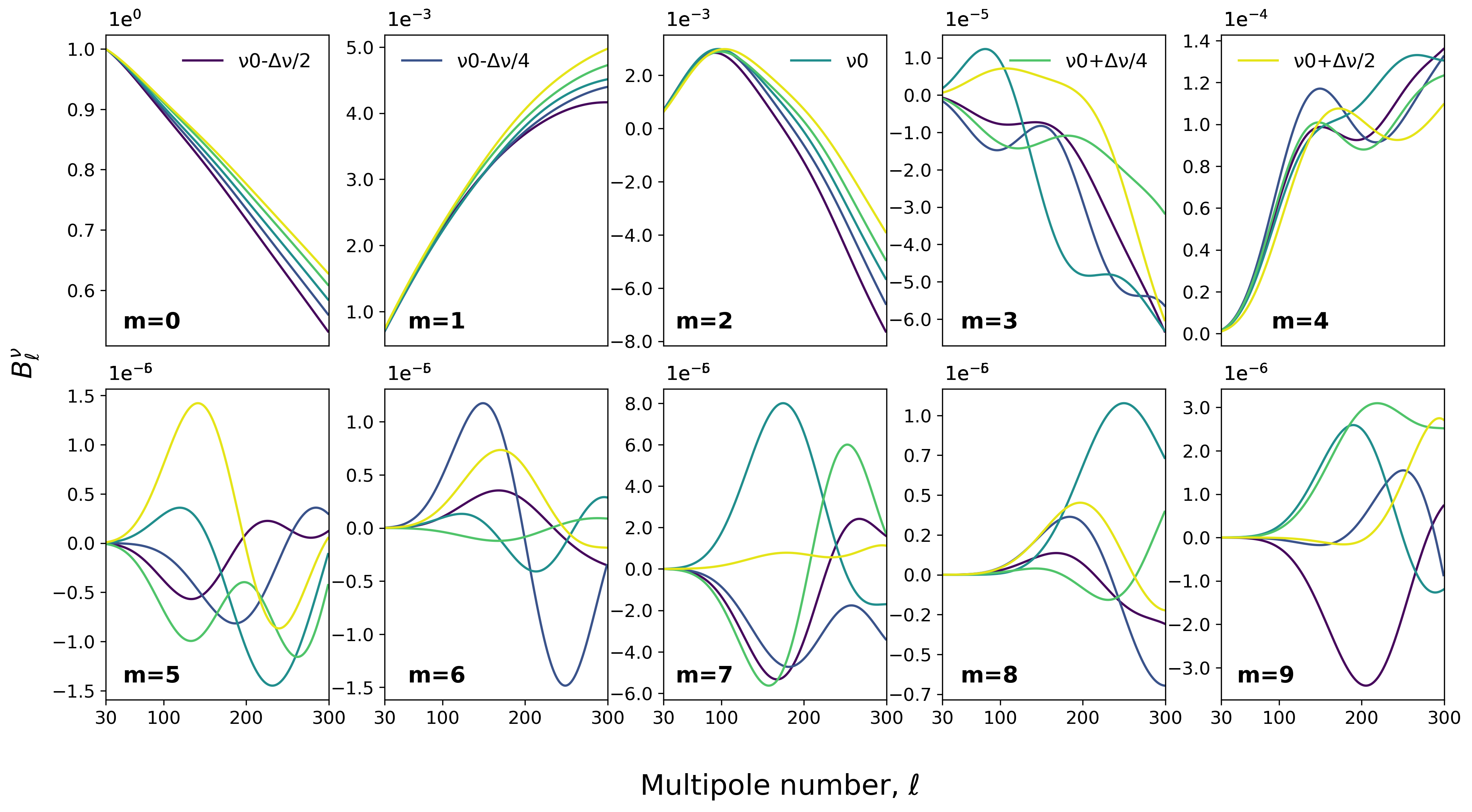}
    \caption{\label{fig:first_10_az_modes} The harmonic transforms of the first ten azimuthal modes of five monochromatic beam simulations sampled uniformly across a $25\%$-relative-width frequency band centered on \SI{93}{\giga\hertz}. The beam transfer functions have been truncated to a multipole range spanning $\ell$=30-300 and normalized with respect to the peak amplitude of the symmetric mode.}
\end{figure}

The smooth in-band variation of the beam reduces the number of sub-frequencies needed to adequately capture its chromaticity. The relative variation of the main beam between any pair of adjacent sub-frequencies in a band is roughly $\sim 6\%$. This holds true for all four frequency bands, independently of how far apart the adjacent simulations are in frequency space, enabling us to probe the chromaticity effect in all bands with the same number of sub-frequency simulations. In this way, we ensure we do not favor some frequency bands with respect to others. As a parametric component separation method like $\bbpower$ is expected to absorb any beam chromaticity bias in the parameters of the most challenging sky component to model, changing the number of sub-frequency samples consistently across all bands is unlikely to significantly affect the results. This has been confirmed by performing the chromaticity analysis only using the center and edge frequency beam simulations for each band (three sub-frequencies instead of five). Whether to employ a different number of sub-frequencies between bands to sufficiently characterize the CMB and foreground spectra separate from beam effects, is a valid question but one that extends beyond the scope of this paper. We therefore choose five sub-frequency simulations for all bands, as illustrated in Figure \ref{fig:chromatic_input_beams}. Note that, while these will be the beam models used throughout the paper, we also estimate the best-fit forecast parameters in the case where we have underestimated the chromaticity of the beam by a smaller and larger value of $10\%$ and $25\%$ change, respectively, within each band (see Appendix \ref{increased_chrommaticity}).

\begin{figure}
    \hspace*{-1cm}
    \includegraphics[width=1.1\textwidth]{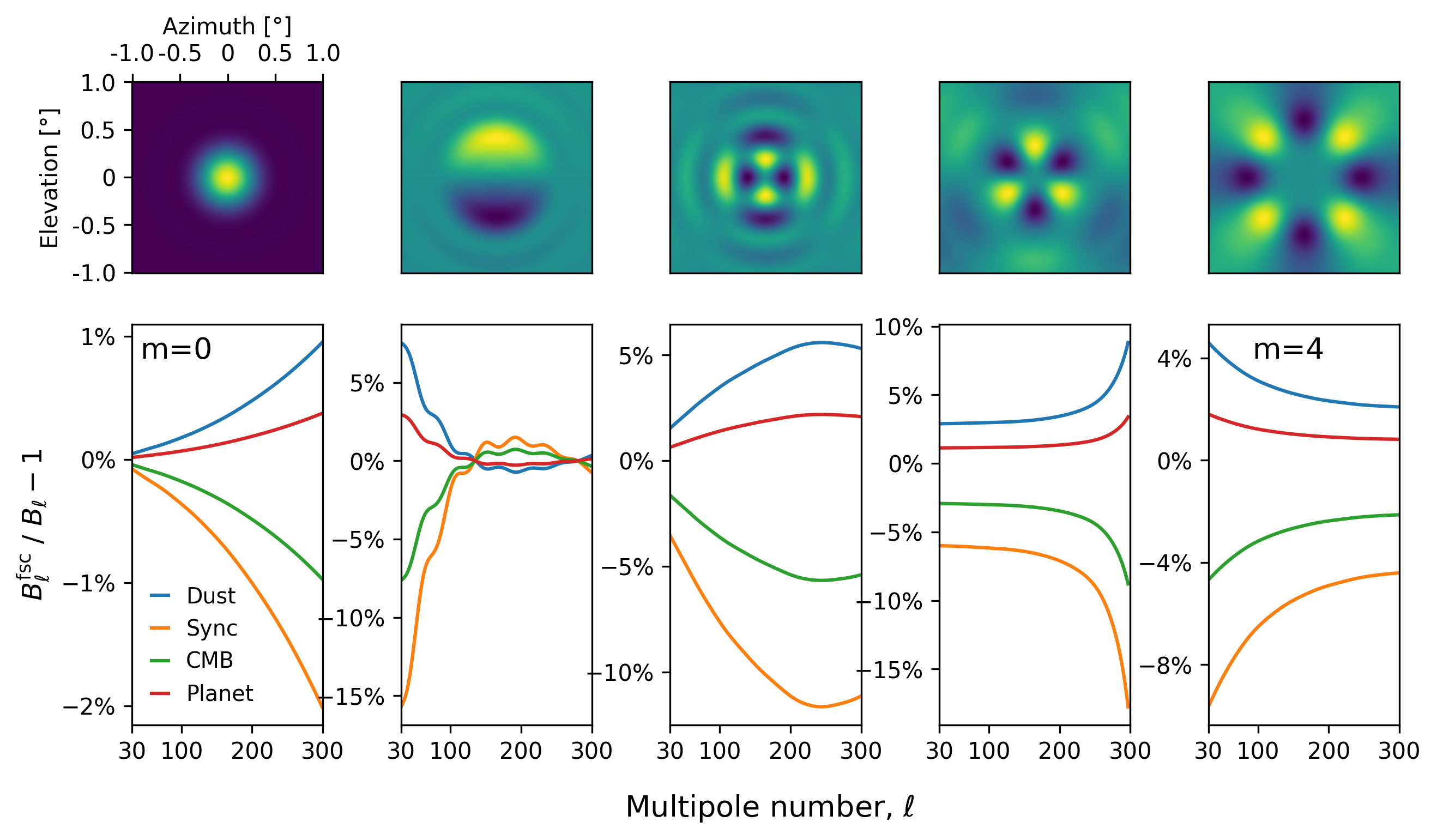}
    \caption{\label{fig:f_sed_blms} Top: The first five azimuthal modes of the band-averaged \SI{93}{\giga\hertz} beam shown in map space. Bottom: The percent relative error between the frequency-scaled, $B_{\ell}^{\mathrm{fsc}}$, and constant-SED beam harmonic transforms, $B_{\ell}$, per mode. The different curves represent the cases where the frequency scaling matches the dust (blue), synchrotron (orange), CMB (green) and planet (red) SEDs. The band-averaged beams with and without frequency scaling are estimated from the monochromatic beam modes of $m=0..4$ shown in the first row of Figure \ref{fig:first_10_az_modes}.}
\end{figure}

Figure \ref{fig:first_10_az_modes} illustrates the frequency-dependent beam variation in the harmonic domain utilizing the monochromatic beams of the \SI{93}{\giga\hertz} band (MF1), shown in the top left panel of the previous figure, as a representative example. The plots show the harmonic transforms for the first ten azimuthal beam modes, $m\leq 9$, spanning multipoles in the range $\ell=30-300$. This is the $\ell$ range we will be employing throughout this paper for consistency with \cite{wolz2023simons}. For the symmetric part of the beam (m=0), increasing the frequency widens the beam transfer function. The impact on the rest of the modes, nevertheless, is not as straightforward to capture. To decide on the number of beam modes we employ in the analysis, we refer to the expected calibration accuracy for the MF and UHF bands of the SO SATs. The beam of the \SI{93}{\giga\hertz} band that is shown in this example is expected to be calibrated within a $\sim 1.2\%$ accuracy for multipoles spanning $\ell=30-700$ and better than $0.6\%$ for $\ell=50-200$ (see Table 3 of \cite{Dachlythra_2024}). In principle, this means we can ignore any modes that are smaller than three orders of magnitude with respect to the main beam. However, we choose to allow for a margin of error and include all azimuthal modes that contribute at least $10^{-4}$ relative to the 
m=0 beam. These correspond to the first five azimuthal modes for the MF1 beam. We ensure this approach is consistent across all four frequency bands.

The combined frequency scaling of the beam and the sky components results in a net frequency-dependent scaling of the band-averaged beams, as shown in Equation \ref{chrom_beam}. That depends on the observation and is generally different from simply averaging the beam over the bandpass. To illustrate the interplay between the intrinsic beam chromaticity and the frequency dependence of different sky components, we compute the band-averaged beam transfer function by assuming a source SED that corresponds either to pure Galactic dust, synchrotron radiation, CMB or planets (commonly used as beam calibrators). The bottom panel of Figure \ref{fig:f_sed_blms} presents the percentage ratio of the frequency-scaled transfer function, $B_{\ell}^{\mathrm{fsc}}$, relative to the nominal band-averaged transfer function, $B_{\ell}$, which assumes a constant SED. Both cases are produced employing the first five azimuthal beam modes, as displayed in Figure \ref{fig:first_10_az_modes}.  These modes are also shown in map space for the constant SED beam in the top panel of the same figure. In each case, the sky component's SED is the one also used in the employed component separation pipeline (see Section \ref{skymodel}). Note that the \texttt{TICRA TOOLS} simulations employed in this work only include the beam intrinsic frequency dependence, corresponding to the case where the calibrated beam which will be used in the analysis has already been corrected for the SED of the calibration source (see Section 4.5 of \citep{hfi_beams_2015}). From the figure, we observe the symmetric beam component varying as much $\sim 2\%$ at the highest multipoles, whereas the frequency-scaled asymmetric modes show deviations from the corresponding azimuthal mode's constant-SED beam of up to $\sim 10\%-15\%$ (the non-frequency scaled m=1-4 modes remain, however, three to five orders of magnitude smaller than the corresponding m=0 beam). 

\section{Methods}\label{methods}

To assess the impact that beam chromaticity has on the cosmological analysis, we produce beam-convolved simulations in the time domain, bin them into maps, and estimate their pseudo-$C_{\ell}$, $B$-mode power spectra, which we then utilize to derive the best-fit values for the tensor-to-scalar ratio, $r$, and foreground parameters. We do so for both chromatic and achromatic beams and quote the bias between the two cases on the forecast parameters in terms of each parameter's expected uncertainty in the achromatic beams scenario. 

\subsection{Sky model}\label{skymodel}
The assumed sky model is comprised of CMB, Galactic dust, and synchrotron emission. To simulate CMB sky maps, we draw Gaussian random realizations of the CMB power spectra assuming the \planck best-fit $\Lambda$CDM parameters \citep{Planck_2018-6} and no tensor fluctuations ($r=0$). In our fiducial analysis, we also simulate dust and synchrotron as (uncorrelated) Gaussian realizations of their corresponding $EE$ and $BB$ spectra as measured by the \planck and WMAP experiments \citep{Planck_2018_11, planck_diff_sep_comp_2020, WMAP_3_fore} (see Section 3.3 of \cite{wolz2023simons}). We do, however, repeat part of our analysis using realistic, non-Gaussian foreground realizations (see Section \ref{non_Gauss_sims}), as this is particularly important for beams carrying pronounced far sidelobes that may pick up the signal from strongly emitting Galactic sources.

The sky SEDs are given by a blackbody law for the CMB, a modified blackbody law (at fixed temperature $T_d=20$ K) for thermal dust, and a power law for synchrotron. The free parameters in the sky model, which will be later fitted for, are the tensor-to-scalar ratio, $r$, the amplitude of lensing $B$-modes, $A_{\mathrm{lens}}$, the spectral indices for dust and synchrotron, $\beta_{d}$ and $\beta_{s}$, the amplitude and spectral tilt of the dust and synchrotron $B$-mode power spectra, denoted as $A_{d}$, $A_{s}$, $\alpha_{d}$ and $\alpha_{s}$, and the dust-synchrotron correlation parameter, $\varepsilon_{ds}$. The CMB power spectrum employed for the Gaussian realizations is computed as:
\begin{equation}
    C_\ell^{\rm c} = A_{\rm lens}C_\ell^{\rm lens} + r C_\ell^{\rm tens} .
\end{equation}
The templates employed for $C_\ell^{\rm lens}$ and $C_\ell^{\rm tens}$ correspond to lensing-only $B$-modes and tensor-only $B$-modes of amplitude $r$ = 1, respectively, and are precomputed by \texttt{CAMB} \citep{camb}. The amplitudes and scaling factors for Galactic dust and synchrotron are linked through a power law that describes the input spectra for these two components:
\begin{equation} \label{eq:meth:clfg}
C_{\ell}^{d/s} = \frac{A_{d/s}}{\ell(\ell+1)/(2\pi)}\left(\frac{\ell}{\ell_{0}}\right)^{\alpha_{d/s}},
\end{equation}
with $\ell_{0}$ = 80. 

It is worth noting that the foreground models used here are relatively simple, particularly since they assume foreground sources that are perfectly correlated across frequencies. Frequency decorrelation, which could be sourced by spatially varying foreground spectra, is a major source of contamination for primordial B-mode searchers if not accounted for. We ignore this effect in this work by design, as it allows us to isolate the effective frequency decorrelation induced by beam chromaticity. As we shall see, the effects of beam chromaticity are relatively small and therefore would not modify the resulting constraints on $r$ in more complex foreground scenarios. The impact of foreground frequency decorrelation and techniques to account for it are further discussed in \cite{Azzoni_2021, Azzoni_2023, wolz2023simons}.

\subsection{Beam-convolved simulations}
\label{beamconv_sims}
  Random realizations of the different sky components are generated from a known model of their power spectra, as described above\footnote{The code used for the sky simulations is available at \url{https://github.com/susannaaz/BBSims}}. These are produced by employing the \texttt{synfast} function of the \texttt{HEALPix} \footnote{\url{http://healpix.sourceforge.net}} package \citep{Gorski_2005, Zonca2019}, which is used throughout the paper and convolution code. We generate maps of NSIDE=256 at five single frequencies uniformly sampled across a 25$\%$ bandwidth around the center frequency of each MF and UHF band. These single-frequency sky models are provided as input to \texttt{beamconv}\footnote{\url{https://github.com/AdriJD/beamconv}} \citep{beamconv_2018} along with the corresponding simulated SAT beams for the same frequencies. The software is equipped with a ground-based scanning function, which we employ to approximate the scan strategy of the SATs by simulating a telescope located in the Atacama Desert in Chile, performing observations informed by the nominal SAT scan strategy (see Section 2.3 of \cite{Ade2019}). We use a wide Field-Of-View (FOV) of \SI{35}{\degree} matching the corresponding size of the actual telescope, populate the focal plane with two hundred detector pairs that are spread uniformly across a square grid, and simulate year-long observations sampling the sky at \SI{50}{\hertz}. While there is dedicated software for SO time-domain simulations, namely the Time-Ordered Astrophysics Scalable Tools (\texttt{TOAST}\footnote{\url{https://github.com/hpc4cmb/toast}}) and \texttt{sotodlib}\footnote{\url{https://github.com/simonsobs/sotodlib}} packages, we choose to employ \texttt{beamconv} for this study due to the latter's reduced computational cost and the fact that no noise or atmospheric simulations are needed. At this stage, we choose not to simulate the SAT Half-Wave-Plate (HWP) since our main aim is to assess the impact of beam chromaticity independently of the systematics arising from the beam and HWP coupling. For HWPs comprised of three birefringent layers, as is the case for one of the SO SATs, the coupling between the beam chromaticity and the frequency-dependent angle offset associated with this type of HWP could introduce additional challenges \citep{beamconv_hwp_2021}.

\begin{figure}
    \centering
    \includegraphics[width=\textwidth]{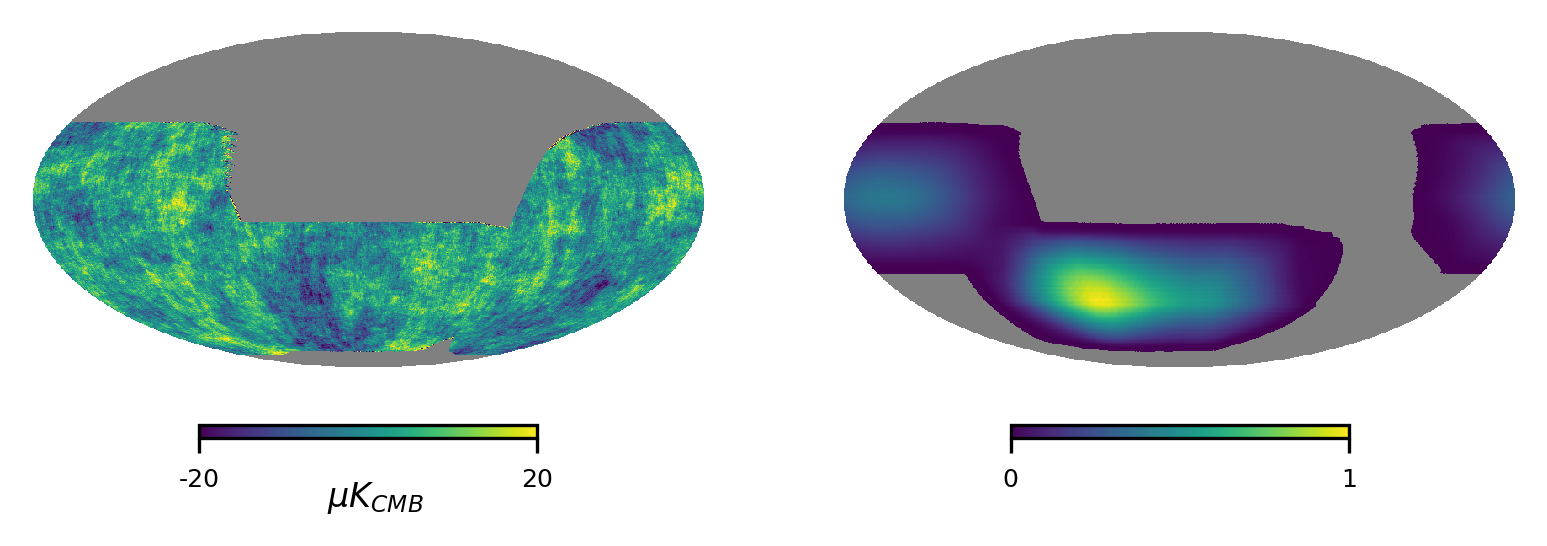}
    \caption{\label{fig:beamconv_output_maps} Left: Beam-convolved Stokes $Q$ map at \SI{280}{\giga\hertz} produced by simulating the SAT scan strategy with \texttt{beamconv}, assuming an input sky model including CMB and Gaussian foregrounds. Right: The apodized custom mask that is generated from the detector hits for the simulated SAT-like scan strategy. Both maps are shown in Equatorial coordinates.}
\end{figure}

  The sky templates are produced only for the Stokes Q and U components while a zero temperature map is provided to \texttt{beamconv}. This is done to exclude any contributions from $T$-to-$P$ leakage in the analysis and for consistency with the work of \cite{wolz2023simons}. After the input polarization maps are beam-convolved in the time domain, they are transformed back into maps using a `naive' mapmaker that simply bins the time-ordered data into pixels given the pointing. An example of a monochromatic map at \SI{280}{\giga\hertz} produced this way is shown in Equatorial coordinates in the left panel of Figure \ref{fig:beamconv_output_maps}. In total, for each sky realization, we obtain six maps corresponding to the six SO frequency bands; four of them constructed using PO chromatic beams (MF, UHF) and two using achromatic Gaussian beams (LF). The band-averaged map of each of the MF and UHF bands is the result of applying the (nominal or perturbed) bandpass to the corresponding (five) monochromatic maps. %\dam{Would it be illustrative to show the map before beaming it? Or are they indistinguishable? Would it be worth plotting the mask separately -- rather than the masked map?}. 
  The right panel of the same figure shows the apodized mask that is applied to the beam-convolved maps and is created from the detector hits of the simulated scan strategy. This mask is fairly similar to the SO SAT mask defined in \cite{Ade2019} and used in \cite{wolz2023simons}. However, the latter was an early estimate that did not employ the scan strategy, and, thus, we choose to use the mask that is informed from the hits map produced in \texttt{beamconv}. % marginally wider than the simulated sky coverage we achieve in the \texttt{beamconv} maps. The presence of pixels with zero values close to the edges of the mask can pose challenges when correcting for cut-sky effects, and thus, the custom mask is preferred. 

  Masking the sky results in mode-mixing, which we need to estimate and correct for when reconstructing the angular spectra. Assuming no noise contributions to the data, the pseudo-$C_{\ell}$ power spectra, $\tilde{C_{\ell}}$, can be connected with true spectra, $C_{\ell}$, as follows \citep{AnneXFaster,Alonso_2019}:

  \begin{equation}
    \langle {\tilde{C_{\ell}}}\rangle = \sum_{\ell'}M_{\ell \ell^{'}}B_{\ell'}^{2}\langle{C_{\ell'}} \rangle.
  \end{equation}
 
  \noindent The matrix $M_{\ell \ell^{'}}$ is the mode-mixing kernel, capturing the statistical coupling between different harmonic-space modes caused by the sky mask while $B_{\ell'}$ represents the beam transfer function. We make use of the pseudo-$C_\ell$ power spectrum estimator as implemented in \namaster{}\footnote{\url{https://github.com/LSSTDESC/NaMaster}} \citep{Alonso_2019}. In particular, we make use of $B$-mode purification \citep{2006PhRvD..74h3002S}, which projects out the contribution from true (i.e. full-sky) $E$-modes that leak into the observed $B$-modes due to mode mixing caused by the sky mask. $B$-mode purification reduces the power spectrum variance on the large scales by order 50\% or more, depending on the noise level, and is therefore crucial for achieving SO's projected precision of $\sigma(r)=0.002-0.003$ \citep{Ade2019}. Beam convolution, on the other hand, reduces the observed signal power spectrum on the small scales, effectively up-weighting the instrumental noise contribution, and increasing the uncertainty on the parameters sensitive to $\ell$>150, such as $A_{\rm lens}$. The beam deconvolution in \namaster{} is performed by including only the symmetric component of the beam in the mode-mixing matrix, $M_{\ell\ell'}$. Hence, we can expect a slight bias when estimating the power spectra of maps convolved with asymmetric beams using this method (see Section \ref{beam_systematics}), or when using the incorrect beam transfer function due to the chromaticity effects discussed above.%Note that the beam used to deconvolve the power spectra in Sections \ref{Gauss_sims},\ref{non_Gauss_sims} is the harmonic transform of the symmetric component of the beam models shown in Section \ref{beam_fdependence} and includes both the co- and cross-polar response. The only exception is in Section \ref{beam_systematics}, where maps convolved with beams, both with and without cross-polar components, are analyzed separately.

  Finally, some degree of bias is to be expected when employing time-domain simulations instead of an analytic approach using the (true) component spectra. Simple binning mapmakers like the one \texttt{beamconv} uses are limited by the degree of cross-linking they can get in each pixel. They are also subject to sub-pixel errors \citep{Naess_2023}.  The bias on the resulting $BB$ spectra between the two approaches is shown in Appendix \ref{cl_based} and is further quantified in terms of the corresponding impact on the tensor-to-scalar ratio and foreground parameters.

\subsection{$B$-mode analysis pipeline}

%For the foreground component separation, we use a harmonic-space-based algorithm, namely \texttt{BBPower}, which is comprehensively described in \cite{wolz2023simons} (referred to as `Pipeline A'). The code estimates the best-fit values and uncertainty of the nine parameters mentioned earlier by applying a Gaussian likelihood across the $BB$ spectra of simulated sky maps. The posterior distributions are sampled using the \texttt{emcee} code \citep{emcee_code, wolz2023simons}. In the context of the SO SATs, \texttt{BBPower} assumes six frequency bands represented only by their center frequencies. 

For the foreground component separation, we use a power-spectrum-based algorithm as implemented in \bbpower
\footnote{\url{https://github.com/simonsobs/BBPower}}, which is introduced in \citep{Abitbol_2021, Azzoni_2021, Azzoni_2023} and in \citep{wolz2023simons}, where it is referred to as `Pipeline A'. The code consists of two main stages, one to compute $BB$ power spectra from maps, and another one to perform power-spectrum-based component separation and parameter inference. We use both stages when running on the binned time-domain simulations, and only the second stage when cross-checking with (true) analytical power spectra. 

Stage 1 uses \namaster{} to compute $B$-mode-purified cross-frequency power spectra from a set of input maps of the partial sky. The purification algorithm requires a smoothed input sky mask, which we obtain by apodizing the hits map corresponding to our simulated SAT scanning strategy described above. As in \cite{wolz2023simons}, we choose an analysis range of $\ell \in [30,\, 300]$, and adopt multipole bins of width 10 to increase the numerical stability of the mode coupling matrix. One aspect to note is the fact that if the asymmetric modes of the beam are significant, then the standard purification would not be optimal since it would not clean up the ambiguous $B$-modes caused by the asymmetric beam. This could lead to additional variance in the estimator and potentially bias the $B$-mode power spectra.

Stage 2 carries out component separation and parameter inference at the same time. Specifically, given the full set of $B$-mode auto- and cross-power spectra between the six different SO SAT frequencies, \bbpower{} builds a forward model for these measurements incorporating the contribution from both CMB and foregrounds. Using a likelihood for the measured spectra, it then derives the posterior distribution of all foreground and CMB parameters at the same time. The methodology is similar to that used by BICEP-{\sl Keck} \citep{2015PhRvL.114j1301B}. Here we use Gaussian likelihood to describe the distribution of the cross-frequency bandpower amplitudes\footnote{Although the power spectrum of a Gaussian field generally follows a Wishart distribution, the binned bandpowers contain a sum over sufficiently many independent modes so that the central limit theorem guarantees approximate Gaussianity. As verified in \cite{wolz2023simons}, choosing a Gaussian likelihood in the present setup ($\ell>30$, $\Delta\ell=10$, SAT sky patch) is sufficiently accurate.}. The forward model that constitutes the mean of this distribution is built by combining the contributions of all sky components ($c$: CMB, $s$: synchrotron, $d$: dust):
\begin{equation}
    C_\ell^{\nu_i\nu_j} = \sum_{x\in\{c,\, s,\, d\}} S_x(\nu_i) S_x(\nu_j) C_\ell^x + \varepsilon_{\rm ds} \sqrt{C_\ell^{\rm d}C_\ell^{\rm s}}\,[S_{\rm s}(\nu_i)S_{\rm d}(\nu_j) + S_{\rm d}(\nu_i) S_{\rm s
}(\nu_j)] \,.
\end{equation}
The first term above contains the autocorrelations of all sky components, while the second term describes the correlation between dust and synchrotron emission, quantified by the parameter $\varepsilon_{\rm ds}$. The auto-spectra of the different components are described by the same models outlined in Section \ref{skymodel}. Note that we expect the combination of potential model mismatch with beam chromaticity to be worrisome only in cases of fairly complex foreground models like the ones including frequency de-correlation. The model mismatch in such cases can be compensated using the method of moments extension to the model \citep{Azzoni_2021} and dust marginalization techniques \citep{Alonso_2019}. The small mismatch that can be seen between the Gaussian and non-Gaussian cases we study in this work can be absorbed by employing wide priors for the parameters of the (fairly generic) fitting model. The full set of nine parameters entering the likelihood is:
\begin{equation}
\{ A_{\rm lens},\, r,\, \, \beta_{\rm d},\, \varepsilon_{\rm ds}, \, \alpha_{\rm d},\, A_{\rm d}, \, \beta_{\rm s}, \, \alpha_{\rm s}, \, A_{\rm s}\} \, .
\end{equation}
The theoretical prediction is evaluated at all integer multipoles and then convolved with the bandpower window functions provided by \namaster. The bandpower covariance is the same as the one described in \cite{wolz2023simons}, estimated from 500 simulations of coadded Gaussian CMB, foregrounds, and SAT-like noise in the `baseline-optimistic' scenario \citep{Ade2019}. The latter assumes isotropic noise, incorporating contributions from both white and 1/f noise components, with observations covering $10\%$ of the sky over a total duration of five years for the MF and UHF bands and one year for the LF bands, respectively. The posterior distribution is sampled using the \texttt{emcee} Markov Chain Monte-Carlo code \citep{emcee_code}, assuming prior distributions given in Table \ref{tab:prior_table} \footnote{The (unphysical)  negative values for $r$ are employed to track volume effects from biases in other parameters (as explained in \cite{wolz2023simons}). Volume effects in the posteriors can bias the cosmological parameters due to noisy (foreground) data and will be investigated in future work.}.

\begin{table*}[h!]
\centering
\begin{tabular}{|c c c|}
\hline
%\noalign{\smallskip}
Parameter & Prior type & Bounds \\
\hline
$A_{\rm lens}$ & TH & [0.0, 2.0] \\
$r$ & TH & [-0.1, 0.1] \\
$\beta_d$ & G & 1.54 $\pm$ 0.11 \\
$\epsilon_{ds}$ & TH & [-1.0, 1.0] \\
$\alpha_d$ & TH & [-1.,0.] \\
$A_d$ & TH & [0,$\infty$) \\
$\beta_s$ & G & -3 $\pm$ 0.3 \\
$\alpha_s$ & TH & [-2.,0.] \\
$A_s$ & TH & [0,$\infty$) \\
%\noalign{\smallskip}
\hline
%\noalign{\smallskip}
\noalign{\smallskip}

\end{tabular}
\caption{Parameter priors employed by \bbpower{} for all foreground models assumed throughout the paper. Top-hat and Gaussian priors are labeled `TH' and `G', respectively.}% In the TH case, the prior is expressed as [minimum, maximum] while in the G case as mean value $\pm$ standard deviation.} 
 \label{tab:prior_table}
\end{table*}

\newpage

\section{Results}
\label{results}
%\todo{We can always restructure this section, but the idea would be to first present the validation of symmetric beam results against the analytical $C_\ell$-based results. Only then move to more sophisticated beamconv sims.} 
The results presented in this section are produced by employing simulations that are a mix of only CMB, Galactic dust, and synchrotron in all cases. To avoid mixing potential beam-related bias with the noise bias, we deliberately exclude any noise contributions from the simulations. The covariance estimate used by the component separation pipeline is the one described in \cite{wolz2023simons} and includes the noise variance, cosmic variance, and foreground uncertainty estimates. After assessing the significance of non-ideal beam properties for this analysis, we demonstrate the impact of beam chromaticity on the forecast parameters for both Gaussian and non-Gaussian foreground models, considering cases both with the nominal and perturbed bandpasses.

\subsection{Bias from beam systematics}
\label{beam_systematics}

\begin{figure}
    \includegraphics[width=\textwidth]{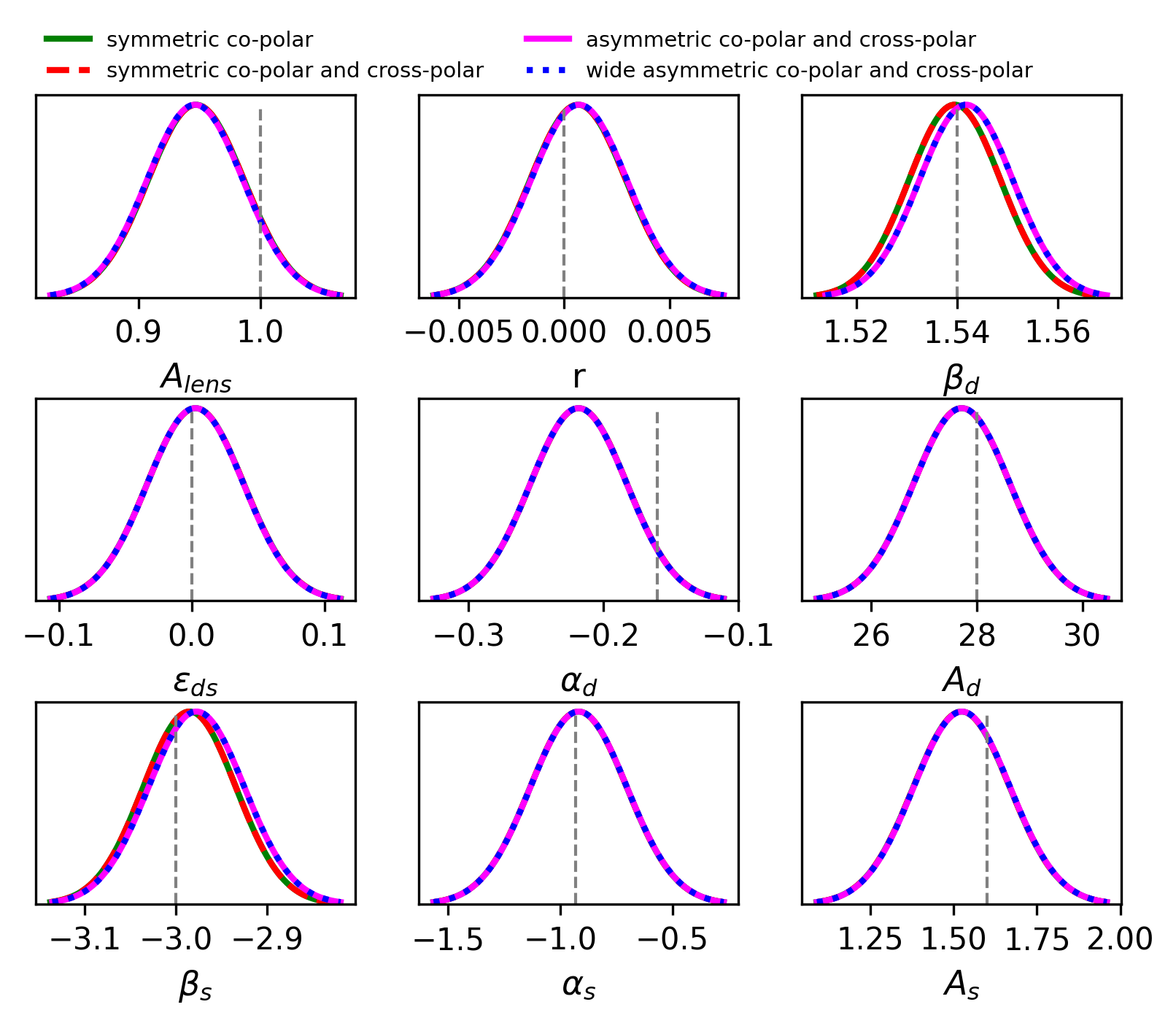}
    \caption{ \label{fig:beam_syst_posteriors} The posterior distributions of the nine model parameters as estimated from ten sets of six \texttt{beamconv} maps where the MF, UHF frequency maps were convolved with symmetric co-polar (green), symmetric co- and cross-polar (red), asymmetric co- and cross-polar (magenta) and wide asymmetric co- and cross-polar beams (blue). The true values for these parameters are also shown for reference (grey vertical lines). The component separation method is applied only to the center frequencies of each band for a net evaluation of the beam systematics impact that is decoupled from the beam chromaticity effect.} %\dam{Any way to make the vertical lines go all the way to the top? E.g. plt.axvline?} \nadiacomment{I feel like it is common to normalize to the maximum of the distribution instead of spanning the full y-axis. Or alternative to only be from y=0 to the nearest point on the posterior.}} 
\end{figure}

In order to quantify the impact of different beam non-idealities in the inferred model parameters, we employ sky realizations that include CMB and Gaussian foregrounds and carry out four distinct types of \texttt{beamconv} simulations only for the center frequencies of the MF, UHF bands using (non-chromatic) beams of increasing complexity. The LF beams are also approximated as Gaussian beams in this case. The different types of simulations include (i) symmetric co-polar beams, (ii) symmetric co- and cross-polar beams, (iii) asymmetric co- and cross-polar beams, and (iv) asymmetric co- and cross-polar beams, including (low-amplitude) far sidelobes. For cases (i)-(iii), we truncate the beams to a maximum angle of \SI{4}{\degree} while for case (iv), we use the full angular range of the beam simulations, extending to \SI{12}{\degree}. The beam transfer function employed for the deconvolution is the one of case (i) in all cases. We chose to use a small number of sky realizations for computational efficiency, and we find them to be enough to capture any significant parameter biases caused by beam non-idealities. This is also aided by the fact that the simulations used are noiseless, although they still include cosmic variance. The estimated best-fit parameters for the four beam cases, assuming only the center frequencies of the bands and averaged over ten different sky realizations \footnote{We observed throughout the paper that while a larger number of simulations could improve the accuracy of the forecast parameters in the achromatic beam scenario, the bias on the parameters due to the inclusion of chromatic beams converges rapidly, typically within the range of five to ten simulations.}, are depicted in Figure \ref{fig:beam_syst_posteriors} through their one-dimensional posteriors. These are obtained by averaging the posterior distributions of each parameter across all sky realizations.

The inclusion of the simulated symmetric cross-polar beam (red) does not appear to change the estimated values of the forecast parameters as compared to the symmetric co-polar beam simulations (green). This is likely due to the small relative amplitude of the cross-polar component (see Appendix \ref{eb_corr}). However, the results for the asymmetric beam simulations do display small, but still visible, deviations, both for the narrow (magenta) and wide beams (blue). In particular, we can see two distinct sets of distributions forming for the spectral indices of dust and synchrotron ($\beta_d$ and $\beta_s$), depending on whether or not the employed beams include asymmetric components. This is because an asymmetric beam induces a frequency-dependent asymmetry in the maps, and the most direct way for the likelihood to interpret this is through a shift in the spectral parameters of the foreground components. The bias induced by beam asymmetry is approximately $0.2\sigma$ for both $\beta_d$ and $\beta_s$, where $\sigma$ represents the average uncertainty across all sky realizations in the symmetric co-polar beam scenario. This type of bias significantly outweighs any additional bias from the beam's far sidelobes, which are much fainter than the main beam (by six to eight orders of magnitude compared to the amplitude at the beam center). All other parameters are biased by less than $0.05\sigma$, across the various cases studied. Note, however, that these statements solely refer to the beam systematics derived from our ideal optics simulations and may not accurately reflect the challenges present in real-world conditions. The in-band beam chromaticity, though, is a consequence of the diffraction-limited optics of the SATs and is not expected to change, even if, for example, we observe larger amplitudes in the far sidelobes when modeling the beams from planet observations. The reason for assessing the beam systematics present in the simulations was, thus, only to showcase they are not strong enough to mask the chromaticity effect we want to capture. Another relevant remark to make here is that beam asymmetry can introduce $T$-to-$P$ leakage, which the HWP is expected to prevent. The contribution of the HWP in mitigating this type of leakage is thoroughly discussed in \cite{Takakura_2017}. Moreover, achieving uniform coverage across multiple HWP angles is expected to reduce beam asymmetry, which in turn helps mitigate its impact on the spectral indices of the galactic foregrounds. Consequently, the previous results could be further verified by also simulating temperature scans and including the SATs HWP in the assumed optical setup.

Any residual bias between the symmetric co-polar beam simulations and the input parameter values is primarily due to the integration of \texttt{beamconv} into the pipeline and, to a lesser extent, to deviations of these beams from perfect Gaussian distributions. The former is further examined in Appendix \ref{cl_based}, where we observe small biases in all parameters, similar to what is seen in Figure \ref{beam_systematics}. Notably, when employing \texttt{beamconv} in the simulations, the posteriors of $A_{\mathrm{lens}}$ and $\alpha_{d}$ show a visible difference. While we expect a small difference in all parameters due to the use of a mask motivated by the scan strategy (as opposed to the simple estimate for which these parameters were recovered in \cite{wolz2023simons}), $A_{\mathrm{lens}}$ and $\alpha_{d}$ are particularly affected by the loss of power at small scales which occurs due to the sparse distribution of detectors on the focal plane (we use two hundred detector pairs as opposed to the few thousand detectors present on the actual SAT focal plane). This choice was made with respect to the computational efficiency we wanted to achieve \footnote{A full set of chromatic beam simulations for a given foreground model can be performed within a week.}. However, any bias between the analytic and \texttt{beamconv} cases is included both in the chromatic and non-chromatic beam simulations and does not, thus, impact the results.

\subsection{Chromatic beams}
\label{sec:symmetric_beams}

\subsubsection{Gaussian foregrounds}
\label{Gauss_sims}

We now repeat our analysis, applying our component separation pipeline to ten sets of six \texttt{beamconv} maps (one per observing band) that were simulated employing the approximated SAT scanning strategy described in the previous section.
The simulations are performed for the center frequencies of the LF bands assuming Gaussian beams and for five single frequencies within each of the MF and UHF bands, using monochromatic PO beams that include asymmetry, cross-polarization, and far sidelobes. The best-fit tensor-to-scalar ratio and foreground parameters are then derived from band-averaged maps that were created by applying the reference bandpasses from \cite{Abitbol_2021} to the MF, UHF monochromatic maps and compared against the case where only the center-frequency maps were used for all bands. Note that the beam used to deconvolve the power spectra in the current and following sections is the harmonic transform of the symmetric center-frequency or band-averaged beam (for the achromatic and chromatic cases, respectively) and includes both the co- and cross-polar response. 

The average values of the best-fit parameters from these ten simulation sets
%of six band-integrated maps convolved with frequency-dependent beams
are shown in Table \ref{tab:forecast_params_chrom} and are denoted as `Chromatic' or `C'. The corresponding parameters, where frequency bands are represented solely by their center frequencies, are also provided for reference. These are labeled as `Achromatic' or `A', and are presented alongside the input values assumed in the simulated sky configuration. The last column presents the bias between the chromatic and achromatic beams scenario in terms of the anticipated uncertainty for each parameter in the achromatic case.

\begin{table}[h!]
\begin{center}
\begin{tabular}{ |c|c|c|c|c| } 
  \hline
 Parameter &  Input & Achromatic (A) & Chromatic (C) & $(\mathrm{C} - \mathrm{A})/\sigma(\mathrm{A})$  \\
  \hline
$A_{\mathrm{lens}}$ & 1 & 0.94 $\pm$ 0.04 & 0.93 $\pm$ 0.04 & 0.18 \\
$r$ $(10^{-3})$ & 0 & 1.320 $\pm$ 2.282 & 1.369 $\pm$ 2.305 & 0.02 \\
$\beta_{d}$ & 1.54 & 1.540 $\pm$ 0.009 & 1.538 $\pm$ 0.008 & 0.17 \\
$\varepsilon_{ds}$ $(10^{-3})$ & 0 & -3.375 $\pm$ 34.65 & -0.35 $\pm$ 32.33 & 0.09 \\
$\alpha_{d}$ & -0.16 & -0.229 $\pm$ 0.035 & -0.239 $\pm$ 0.031 & 0.27 \\
$A_{d}$ & 28 & 27.8 $\pm$ 0.9 & 27.1 $\pm$ 0.7 & 0.77 \\
$\beta_{s}$ & -3.0 & -2.992 $\pm$ 0.051 & -2.979 $\pm$ 0.051 & 0.24 \\
$\alpha_{s}$ & -0.93 & -0.918 $\pm$ 0.197 & -0.920 $\pm$ 0.194 & 0.01 \\
$A_{s}$ & 1.6 & 1.57 $\pm$ 0.14 & 1.56 $\pm$ 0.14 & 0.06 \\
\hline
\end{tabular}
\end{center}
\caption{\label{tab:forecast_params_chrom}The average best-fit parameters as estimated from ten simulation sets of six \texttt{beamconv} maps presented for the `Chromatic' and `Achromatic' MF/UHF beams scenarios. The input values of the fitted parameters are also presented in the second column for reference. The last column presents the beam chromaticity bias on the best-fit values in terms of their expected uncertainty in the achromatic beams case.}
\end{table}

Upon examining the table, we conclude that the most noticeable effects of beam chromaticity apply to the Galactic dust parameters, with the amplitude of the dust $BB$ spectra inheriting a $0.77 \sigma$ bias when chromatic beams are employed in the simulations. On the contrary, the tensor-to-scalar ratio, $r$, appears to be largely insensitive to the beam's frequency dependence. This is due to the fact that the constraining power on $r$ is at degree scales (around $\ell \sim 80$), thus the beam systematics that are more pronounced at the higher multipoles, like chromaticity (see Figure \ref{fig:f_sed_blms}), have very little effect. This is different from the impact on other parameters like the lensing amplitude, which does depend on what happens at $\ell \sim 300$. Furthermore, the small values associated with the synchrotron spatial parameters should be interpreted with caution throughout the analysis results in this and the following sections. This is because they are primarily informed by the LF bands, where the simulations did not account for the beam chromaticity effect. All parameters, however, show deviations smaller than $1\sigma$. 

Finally, it is worth noting that the beam chromaticity bias on $r$ is not expected to increase even if a non-zero input value of $r$ is chosen in the fiducial model. We would, however, find a larger value for $\sigma(r)$ due to the added cosmic variance. Moreover, the most significant biases are observed on the dust spatial parameters ($A_{d}$ and $\alpha_{d}$), which are known to be largely uncorrelated with $r$, unlike the spectral index $\beta_{d}$, which remains largely unaffected. Nevertheless, a follow-up study could ensure that this is the case. The interplay between the input $r$ (0 versus 0.01) and foreground anisotropy was studied using the same pipeline in \cite{wolz2023simons}, see Table 5.

\subsubsection{Non-Gaussian foregrounds}
\label{non_Gauss_sims}

\begin{figure}
    \centering
    \includegraphics[width=\linewidth]{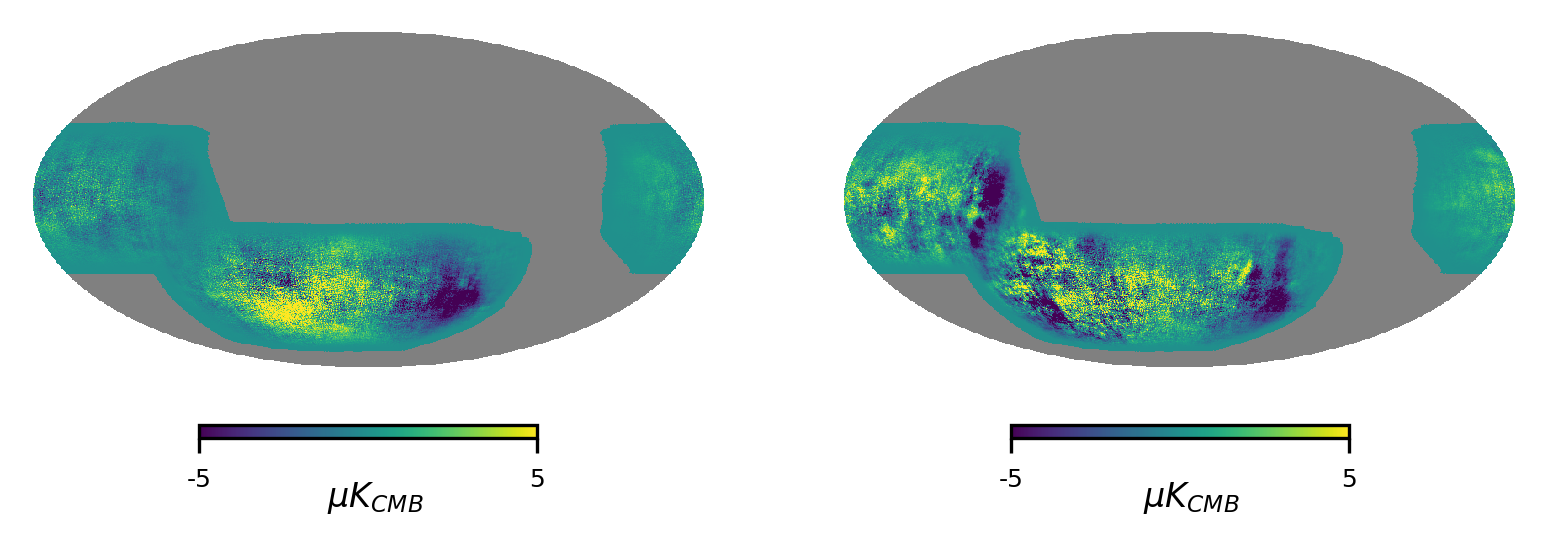}
    \caption{\label{fig:diff_maps} Stokes Q maps of sky simulations employing non-Gaussian foreground models at \SI{27}{\giga\hertz} (left) and \SI{280}{\giga\hertz} (right), after applying the custom mask.}
    
\end{figure}

We proceed by assuming non-Gaussian foregrounds and repeating the steps described in the previous section. In this case, we use the \texttt{PySM} module to produce template maps of dust emission and synchrotron radiation that describe, once again, a modified black-body distribution and power-law, respectively. These foreground models correspond to models `d0' and `s0' of the \texttt{PySM} package and are commonly used as representative non-Gaussian test models \citep{Hensley_2022, wolz2023simons}. The spectral indices of synchrotron and dust are constant across the sky, but the template maps now include the large-scale anisotropies caused by the Milky Way. Much of the emission from the Galactic plane is removed from the simulated maps by applying the mask described in Section \ref{beamconv_sims}. Figure \ref{fig:diff_maps} shows the Stokes Q maps generated for the lowest (left) and the highest (right) band center frequencies, namely \SI{27}{} and \SI{280}{\giga\hertz}, using the `d0s0' model. An important remark at this stage is that these non-Gaussian simulations display a larger dynamic range that could be picked up by the beam sidelobes. The best-fit parameters obtained by combining chromatic beams with non-Gaussian foreground models are presented in the same manner as for the Gaussian case. Table \ref{tab:forecast_params_chrom_nong} summarizes the average values from ten simulation sets, utilizing both achromatic and chromatic beams. It also presents the biases between these two cases, estimated as a function of the expected uncertainty per parameter.

\begin{table}[h!]
\begin{center}
\begin{tabular}{ |c|c|c|c|c| } 
  \hline
 Parameter &  Input & Achromatic (A) & Chromatic (C) & $(\mathrm{C} - \mathrm{A})/\sigma(\mathrm{A})$  \\
  \hline
$A_{\mathrm{lens}}$ & 1 & 0.94 $\pm$ 0.04 & 0.93 $\pm$ 0.04 & 0.14 \\
$r$ $(10^{-3})$ & 0 & 0.916 $\pm$ 2.273 & 0.930 $\pm$ 2.283 & 0.01 \\
$\beta_{d}$ & 1.54 & 1.539 $\pm$ 0.007 & 1.539 $\pm$ 0.007 & 0.01 \\
$\varepsilon_{ds}$ $(10^{-3})$ & - & 35.7 $\pm$ 49.5 & 46.7 $\pm$ 46.2 & 0.22 \\
$\alpha_{d}$ & - & -0.318 $\pm$ 0.036 & -0.335 $\pm$ 0.033 & 0.47 \\
$A_{d}$ & - & 34.1 $\pm$ 1.1 & 33.4 $\pm$ 1.0 & 0.53 \\
$\beta_{s}$ & -3.0 & -3.05 $\pm$ 0.09 & -3.05 $\pm$ 0.09 & 0.15 \\
$\alpha_{s}$ & - & -1.579 $\pm$ 0.303 & -1.597 $\pm$ 0.293 & 0.06 \\
$A_{s}$ & - & 0.63 $\pm$ 0.14 & 0.63 $\pm$ 0.14 & 0.02 \\
\hline
\end{tabular}
\end{center}
\caption{\label{tab:forecast_params_chrom_nong}The average best-fit parameters as estimated from ten simulation sets of six \texttt{beamconv} maps produced using sky templates of CMB and non-Gaussian foreground models convolved with both `Achromatic' (A) and `Chromatic' beams (C) for the MF/UHF bands. Selected input parameters are presented in the second column while the last column describes the beam chromaticity bias on the best-fit values with respect to their expected uncertainty for achromatic beams.}
\end{table}

Note that we do not provide input values for $\varepsilon_{ds}$, $\alpha_{d}$, $A_{d}$, $\alpha_{s}$, and $A_{s}$, in the second column as, in this case, the spatial properties of the dust and synchrotron maps are determined by the data on which {\tt PySM} is based, rather than any input power spectrum. The best-fit values for the power law parameters of the dust and synchrotron power spectra instead depend on the particular sky region targeted. Furthermore, one should consider that, even though the Galactic plane is largely removed when masking the data, strong emission from that region may still leak into the unmasked footprint via the beam sidelobes (especially at the mask edges), potentially impacting the fitted parameters. However, it is important to note that the values of $\alpha_{d/s}$ and $A_{d/s}$ used to generate the Gaussian simulations discussed earlier are determined as the best-fit values derived from a large sample of `d0s0' maps (without any additional beam convolution) and the analysis mask described in \cite{wolz2023simons}. The choice to use a Gaussian likelihood also for non-Gaussian foregrounds is motivated by the same work, where the authors tested that the chi-squared statistic computed for non-Gaussian foregrounds and using the simulation-based covariance discussed ealier agreed with the expectations for Gaussian data (see Appendix A of the same publication). 

As in the case of Gaussian foregrounds, the largest bias corresponds to the dust spatial parameters, $A_{d}$ and $\alpha_d$, at approximately $0.5\sigma$ for both parameters. It is, however, interesting to notice the increased bias of the dust-synchrotron correlation parameter, $\varepsilon_{ds}$ ($0.22\sigma$ as compared to $0.09\sigma$ in the Gaussian foreground scenario). Independently of the foreground model assumed, the two components are convolved with the same beam and the fitting code can interpret the additional frequency scaling introduced by the beam as additional correlation between the foregrounds. This result is more obvious for the non-Gaussian simulations (that already include some level of dust-synchrotron correlation) described in Table \ref{tab:forecast_params_chrom_nong} but note that it also applies to the quoted values of Table \ref{tab:forecast_params_chrom}. The beam frequency scaling increases also the correlation between the Gaussian dust and synchrotron components. In fact, the slight anti-correlation observed in the achromatic beam scenario is nearly eliminated when beam chromaticity is included in the simulations.
%David:It is also interesting to note that, in this case, the dust-synchrotron correlation parameter $\varepsilon_{ds}$  deviates further from the best-fit value found for achromatic beams than was found for Gaussian simulation. This is not entirely suprising, as these Gaussian simulations do include the level of correlation between these two components present in the {\tt PySM} templates, and thus differences in the response of this parameter to beam non-idealities can be expected.
%\dam{Note that I changed the interpretation here, but I've left the old text commented out, in case my interpretation is wrong if I'm missing something.}
%An interesting remark to make is that $\varepsilon_{ds}$ deviates further from its true, zero value when we simulate scanning with chromatic beams. This contrasts with the Gaussian foreground case. We conclude that the simplicity of the Gaussian models, along with the significant increase in the number of frequency maps provided to the code in the chromatic beams scenario, allowed the fitting algorithm to more effectively differentiate between the synchrotron and dust components.
Overall, the chromaticity bias on all forecast parameters remains approximately below the $\sim0.5\sigma$ level, when we assume non-Gaussian foreground models. 

\subsection{Impact of bandpass uncertainty}
\label{bandpass_uncertainty}
Bandpass uncertainty can arise due to variations in the fabrication process of the filters or from inherent systematics present in their measurements conducted using a Fourier transform spectrometer \citep{Shitvov_2022, Planck_2013-9}. It can also be the result of frequency-dependent reflections between elements in the optical system. Spectrally averaging over a finite bandwidth can help in mitigation. However, care needs to be employed to ensure the methodology used to characterize the receiver elements does not differ from the as-used configuration at a level that influences the calibration data acquired \citep{Burleigh:16}. Time-varying atmospheric fluctuations further contribute to bandpass variations by inducing alterations in the frequency-dependent atmospheric transmission. These factors collectively influence fluctuations in both the effective gain and central frequency of the bandpass as explained in \cite{Ward_bpass_variations}. A detailed analysis of the bandpass calibration requirements for the SO SATs is provided in \cite{Abitbol_2021}, where the allowed bandpass variation is quantified in terms of effective frequency and gain. According to this work, both of these parameters must be known to percent level or better. Note that these constraints are designed with respect to avoid biases of the order of $\Delta r \sim 10^{-3}$ on the tensor-to-scalar ratio, $r$, measuring which is the primary scientific goal for the SATs.

The gain can be expressed as the integrated product of the telescope's effective area, $A_{\mathrm{eff}}$, the instrumental bandpass, and the instrument's beam response:

\begin{equation}
g = \int A_{\mathrm{eff}} \tau(\nu) B(\theta,\phi, \nu)\mathrm{d}\nu.    
\end{equation}

\noindent Since the authors of \cite{Abitbol_2021} employ achromatic beams in the analysis, we attribute the majority of gain variations to arise from bandpass variations. However, in this work, our interest is not in redefining the bandpass uncertainty requirements in terms of the $r$-constraint. Instead, we focus on quantifying potential biases across all fitted parameters arising from the interplay of bandpasses with systematic chromaticity and chromatic beams. We, thus, employ a simple model where we perturb the nominal bandpasses by adding a slope, parametrized by a spectral index, $\beta_{\tau}$. %We do not include any effective gain variation to prevent degenerate biases between the latter and perturbed versions of the bandpasses for small values of the index. 
In this way, we construct a set of ten perturbed bandpass curves, $\tau(\nu)^{\mathrm{pert}}$, for each of the MF and UHF bands as follows:

\begin{equation}
\label{delta_b}
\tau(\nu)^{\mathrm{pert}} = \tau(\nu) \left(\frac{\nu}{\nu_{0}} \right) ^{\beta_{\tau}}.
\end{equation}

\noindent The parameter $\nu_{0}$ refers to the same effective frequencies assumed throughout the analysis, while we assign $\beta_{\tau}$ ten values uniformly spanning the range [-1,1]. 

\begin{figure}[t!]
    \centering
    \includegraphics[width=\textwidth]{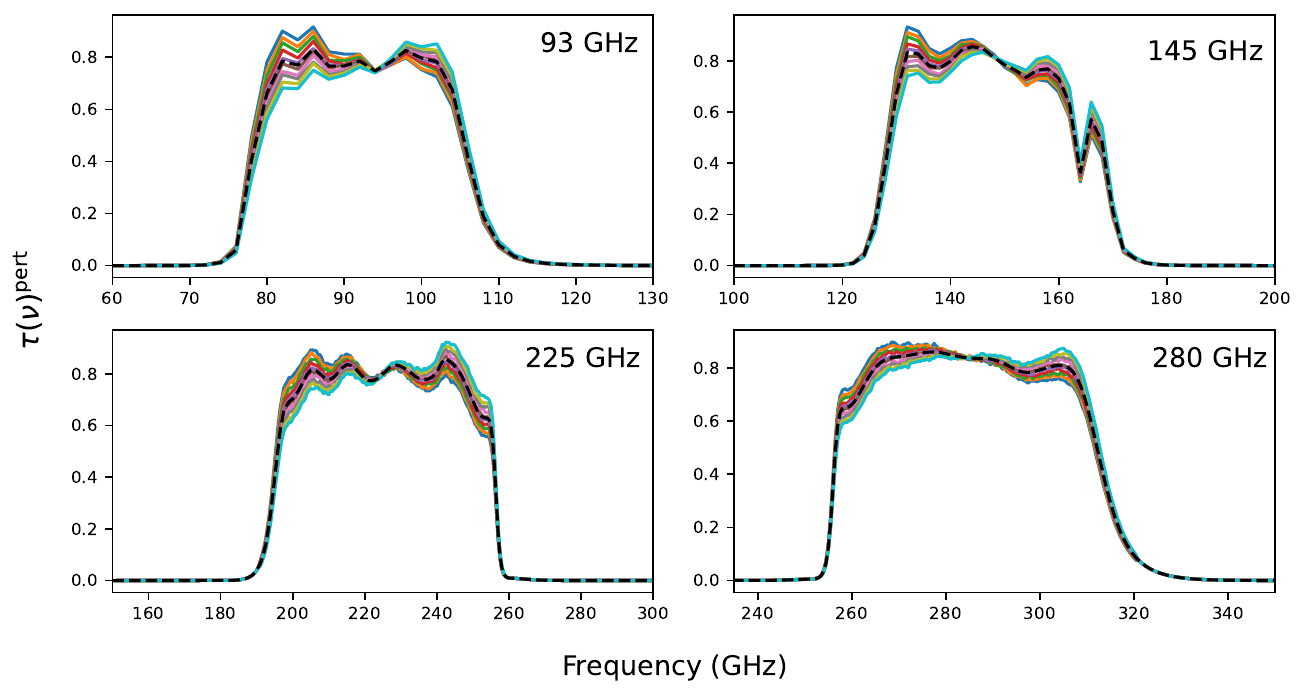}
    \caption{\label{fig:pert_bpasses} Ten perturbed versions of the reference SO SAT bandpasses from \cite{Abitbol_2021} for the MF and UHF frequency bands. The nominal bandpasses (shown with black dashed lines) have been perturbed by a slope ranging from -1 to 1.}
\end{figure}

Figure \ref{fig:pert_bpasses} shows the newly constructed perturbed bandpasses (curves of various colors) for all bands considered in this work, plotted against the reference bandpasses (black dashed lines). The changes in $\beta_{\tau}$ become clearly evident. The amplitude of the bandpass variations allowed by our parametrization is comparable with those originating from realistic atmospheric variations and the frequency-dependent scaling from the sky components presented in \cite{Ward_bpass_variations}. Specifically, by observing Figure 4 of the same work, we deduce that the range of slopes covered in our case is significantly larger. The ratio between each perturbed bandpass integrated over the full frequency range and the corresponding value for the reference bandpass does not exceed $1.5\%$ for the MF bands and $0.6\%$ for the UHF bands. These upper boundaries represent approximately half of the allowed percent gain uncertainty determined in \cite{Abitbol_2021}, ensuring that the bias on r does not exceed the desired $\Delta r = 10^{-3}$ from the bandpass perturbations alone. This will not necessarily be the case when employing both the above bandpasses and the chromatic beams used in the previous sections.

%As shown in Figure 4 of \cite{Ward_bpass_variations}, the amplitude of the bandpass variations allowed by our parametrization is comparable with those originating from realistic atmospheric variations and the frequency-dependent scaling from the sky components, and the range of slopes covered here is significantly larger. The ratio between each perturbed bandpass integrated over the full frequency range and the corresponding value for the reference bandpass does not exceed $1.5\%$ for the MF bands and $0.6\%$ for the UHF bands. These upper boundaries represent approximately half of the allowed percent gain uncertainty determined in \cite{Abitbol_2021}, ensuring that the bias on $r$ does not exceed the desired $\Delta r = 10^{-3}$ limit. Figure \ref{fig:pert_bpasses} shows the newly constructed perturbed bandpasses (curves of various colors) for all bands considered in this work, plotted against the reference bandpasses (black dashed lines). 

We construct ten new versions for each of the ten sets of band-averaged \texttt{beamconv} maps used in the analysis of Sections \ref{Gauss_sims} and \ref{non_Gauss_sims}, one for each of the new perturbed bandpasses. We assume the same input sky realizations as before and quantify the combined impact of beam chromaticity and bandpass fluctuations on the forecast parameters for both the Gaussian and non-Gaussian foreground scenarios. We do so similarly to the previous sections by taking the bias of each parameter with respect to its expected mean and uncertainty. This time, however, we use the mean and uncertainty of the fitted parameters in the chromatic beams and nominal bandpass cases (namely case `C' in Tables \ref{tab:forecast_params_chrom}, \ref{tab:forecast_params_chrom_nong}) for each foreground model. This is because we want to specifically quantify the added bias from the slopped bandpasses with respect to the already established bias due to chromatic beams. 

The bias of each parameter averaged over all sky realizations scales linearly and smoothly as a function of $\beta_{\tau}$ around its expected value in the `C' scenario for both Gaussian and non-Gaussian foregrounds. We will thus quote these biases only at the extreme values of the bandpass spectral index. Note that the reason for considering as many $\beta_{\tau}$ values was to ensure that the pipeline was not finding spurious degeneracies between the frequency scaling from the sky, the beam and the bandpasses for any of the assumed bandpass slopes. At $\beta_{\tau} = \mp 1$ and assuming Gaussian foregrounds, we compute biases of $\pm 0.05\sigma$, $\mp 0.03\sigma$, $\pm 0.07\sigma$, $\pm 0.24\sigma$, $\mp 0.42\sigma$, $\mp 0.83 \sigma$, $\pm 0.27\sigma$, $\mp 0.01 \sigma$, and $\mp 0.01\sigma$ for $A_{\rm lens},\, r,\, \, \beta_{\rm d},\, \varepsilon_{\rm ds}, \, \alpha_{\rm d},\, A_{\rm d}, \, \beta_{\rm s}, \, \alpha_{\rm s},$ and $A_{\rm s}$, respectively. Note once again that we now assume $\sigma=\sigma(\mathrm{C})$. These are symmetric biases around the $\beta_{\tau}=0$ nominal case. We observe comparable results for the non-Gaussian case with average biases of $\pm 0.02\sigma$, $\pm 0.1\sigma$, $\mp 0.04\sigma$, $\mp 0.02$, $\mp 0.48$, $\mp 0.67$, $\mp 0.02$, $\mp 0.01$, and $\pm 0.04$ for the same parameters. 

The tensor-to-scalar ratio, $r$, appears to be insensitive to the simulated bandpass perturbations; even for the non-Gaussian case, the bias is at the $0.1\sigma$ level. Once again, most of the impact is absorbed in the dust spatial parameters independently of the assumed foreground model. In the Gaussian foreground scenario, we also observe smaller biases in the spectral indices of dust and synchrotron ($\beta_{d/s}$). These are consistent with the results of Sections \ref{beam_systematics}, \ref{Gauss_sims}, where we demonstrated that the pipeline aims to account for the convolution of non-Gaussian beam modes with Gaussian sky realizations through shifts in the spectral parameters of the foreground components. In conclusion, the results indicate that even in the fairly exaggerated cases of $\beta_{\tau}=\mp 1$, the added bias from bandpass variations does not exceed the expected uncertainty of any of the fitted parameters estimated assuming chromatic beams for the SATs (and $r$ in particular is remarkably robust).  Note again that the synchrotron parameters should be interpreted carefully, as they are predominantly informed by the LF bands, the maps of which are only convolved with achromatic beams. 

\section{Conclusion and Discussion}
\label{conclusion}
In this paper, we present the method and associated pipeline for quantifying the beam chromaticity impact on
%In this paper, we have quantified the impact of beam chromaticity and non-idealities (cross-polarization, asymmetry) 
the large-scale $B$-mode power spectra of the SO SATs, through the constraints on the tensor-to-scalar ratio, $r$, and foreground parameters derived from them. The method entails:
\begin{itemize}
  \item Simulating five monochromatic PO beams for each of the MF and UHF frequency bands, at frequencies evenly spanning the band's full frequency range.
  \item Performing the time-domain convolution of simulated sky templates with these beams while assuming a SAT-like scan strategy, using {\tt beamconv}.
  \item Estimating the power spectra of these beam-convolved maps after masking and minimizing, correcting for $E$-to-$B$ leakage, and employing $B$-mode purification techniques with \namaster.
  \item Using these spectra to determine the best-fit values for the tensor-to-scalar ratio, lensing amplitude, and foreground parameters, employing \bbpower.
\end{itemize}
We perform the analysis assuming both Gaussian (Section \ref{Gauss_sims}) and non-Gaussian foreground models (Section \ref{non_Gauss_sims}) and also study the interplay between beam chromaticity and bandpass variations (Section \ref{bandpass_uncertainty}).

The beam simulations used in this analysis are generated assuming a three-lens refracting telescope, representing the fundamental optical configuration of the SO SATs (see Section \ref{beam_fdependence}). We find that most non-idealities present in the achromatic beam simulations, like asymmetry or cross-polarization, are not expected to impact significantly the best-fit values and accuracy of the tensor-to-scalar ratio and foreground parameters (as shown in Section \ref{beam_systematics}). Note, however, that these beam non-idealities will be further modeled from real planet observations, and we will then assess again their impact, though we do not expect any major effect for this analysis. The ten sky realizations provided to \texttt{beamconv} contain CMB, Galactic dust, and synchrotron at the same frequencies as the beam simulations and the convolution happens assuming a SAT-like scan strategy. 

When considering Gaussian foregrounds, we find that beam chromaticity leads to a negligible bias on the tensor-to-scalar ratio ($0.02\sigma$). In turn, most of the effect is absorbed by the parameters characterizing the spatial structure of Galactic dust, recovering a bias on the $B$-mode power spectrum amplitude, $A_d$, of $0.77\sigma$ (see Table \ref{tab:forecast_params_chrom}). We find similar results when repeating the analysis on more realistic, non-Gaussian foregrounds (\texttt{PySM} model `d0s0'). The largest impact is again on the spatial dust parameters, with both $A_d$ and $\alpha_d$ biased at the $\sim0.5\sigma$ level (see Table \ref{tab:forecast_params_chrom_nong}). Nevertheless, the constraints on $r$ remain comparably unaffected. This is not unexpected considering that beam chromaticity impacts mostly the higher multipoles (see Figure \ref{fig:f_sed_blms}) while most of the constraining power for $r$ is at large scales ($\ell \sim 80$). Furthermore, the dust-synchrotron correlation increases when beam chromaticity is accounted for in the simulations, and this effect is independent of the employed foreground model. %\nadiacomment{Do we really observe a noticeably larger effect on the dust-synchrotron correlation parameter? The bias of the dust parameters refers to the difference between achromatic and chromatic beams which for $e_ds$ is much smaller than the bias for the same parameter in the Gaussian case.}
%corresponding to a modified black-body distribution and a power-law for the SEDs of Galactic dust and synchrotron, respectively.
%The input values shown for the Gaussian case roughly match the corresponding values for `d0s0' when the mask described in \cite{wolz2023simons} is employed. 
%This is the nominal SAT mask and is marginally wider than the mask used in the present analysis. However, despite this difference being small, we omit the input values of the spatial foreground parameters $A_{d}$, $\alpha_{d}$, $A_{s}$, and $\alpha_{s}$ when presenting the results that refer to non-Gaussian foregrounds in Table \ref{tab:forecast_params_chrom_nong}.

%, From the table, we deduce that the greatest bias is for $A_{d}$, also in this case, and is estimated as $0.53\sigma$. While this decrease in the bias might seem strange considering we are now employing a more complicated dust model, it can be explained when noting the subsequent increase in the bias of $\alpha_{d}$. This is now equal to $0.47\sigma$ as compared to its $0.27\sigma$ Gaussian-case value. The bias does not increase in the non-Gaussian foreground scenario; instead, it spreads almost evenly between the two dust spatial parameters. Furthermore, the dust-synchrotron correlation increases by an order of magnitude when assuming more complex foregrounds and is greater when the maps are convolved with chromatic beams. This contrasts with the Gaussian case, where the algorithm actually benefited from the increased number of frequency points, resulting in cleaner separation between synchrotron and dust.

In addition to quantifying the net effect of beam chromaticity, we also consider variations in the bandpass shape and evaluate how these couple to the beam frequency dependence. We do so by applying a slope perturbation to the fiducial SO SAT bandpasses (the impact of uncertainties in the effective frequencies was studied in \cite{Abitbol_2021}), characterized by a spectral index $\beta_{\tau}$. We compute the bias of the best-fit parameters for a given sky realization when $\beta_{\tau}=\mp 1$ and then average over all realizations. The biases are expressed with respect to the expected uncertainty in the chromatic beams, nominal bandpass scenarios. For the Gaussian case, the largest biases are of the order of $\mp 0.83\sigma$ and $\mp 0.42\sigma$ for $A_{d}$ and $\alpha_d$, respectively, while the spectral indices of dust and synchrotron carry smaller but non-negligible biases from the convolution with asymmetric beams. For the non-Gaussian case, the bandpass variations are fully absorbed in $A_{d}$ and $\alpha_d$ with corresponding biases of $\mp 0.67\sigma$ and $\mp 0.48\sigma$.
%ealizations and estimate the added uncertainty for a single parameter as the standard deviation over the best-fit values of this parameter for the same sky realization and beam simulations but different bandpasses. We then estimate the average of this added uncertainty for every parameter over the full set of sky realizations. The greatest uncertainty once again refers to the amplitude of the dust $B$-mode spectra and is calculated as $0.4\sigma$ and $0.38\sigma$ when the sky templates include Gaussian and non-Gaussian foregrounds, respectively. The scaling factor for the dust spectra also shows non-negligible uncertainty when combining bandpass fluctuations with beam chromaticity, with corresponding values of $0.24\sigma$ and $0.29\sigma$ in the two foreground scenarios. An interesting remark is how the tensor-to-scalar ratio uncertainty increases by an order of magnitude when assuming more complex foreground models. Specifically, the Gaussian-case value of $0.01\sigma$ becomes $0.1\sigma$ when inducing this additional frequency scaling and employing `d0s0'. As for the nominal bandpasses, however, the added uncertainty remains well under $1\sigma$ for all parameters.

While this analysis indicates that in-band beam chromaticity is unlikely to cause significant bias on the tensor-to-scalar ratio and foreground parameters, there are still future directions to explore within the same scope. A key example is the development of more advanced beam simulations to establish a clear link between the complexity of the beam and the resulting bias in the foreground parameters. %\dam{Is this in the same vein as my previous comment? In that case maybe link it to the beamconv precision issues.}. 
%The latter is done by simulating beam maps for pixels placed at the edge of the focal plane. %\dam{This sentence seems super orphan: if these are interesting studies, I would either include them in the main text somewhere, or summarise their main results more thoroughly here or elsewhere.}.
%It would also be instructive to quantify the sidelobe amplitude at which chromatic beam sidelobes may cause non-negligible biases in the analysis, particularly in the context of future, more sensitive experiments, such as CMB S4 \citep{CMBS4_book}. 
Another important component in this context is the SAT HWP, which is not yet included in the simulation setup. Any frequency-dependent systematics associated with the HWP (like the frequency-dependent angle offset that multi-layer HWPs often carry \cite{Matsumura:09, Hill_2016, beamconv_hwp_2021}) can contribute to the total, in-band instrument chromaticity if not properly modeled in advance or as part of the component separation pipeline. The latter will be studied in \citep{TKS_in_prep} for a SAT-like instrument. The performance of the HWP is also vital for mitigating $T$-to-$P$ leakage and consequently any resulting bias from the interplay of this leakage with beam chromaticity. We do, however, expect only a small second-order effect from this interplay. Moreover, one could repeat the analysis employing other component separation approaches, particularly those acting at the map level (e.g. `Pipeline B' and `Pipeline C' in \cite{wolz2023simons}) which have been shown to recover constraints on $r$ that are consistent with those found using \bbpower. Therefore, although we would expect similar results to hold in those cases (e.g. the impact of isotropic beam chromaticity would be absorbed by the foreground components), this would require a more detailed investigation. Finally, the analysis will be verified using SO SATs beam chromaticity data we are obtaining, since December 2024, by drone-borne calibrators with tunable emission frequencies \citep{Nati_2017, Dunner_2021, Coppi:2022qjs, Coppi:2025arxiv}. 

%Ultimately, although the parameter biases quantified here are small, it is the detailed instrument modeling and its integration into the component separation analysis that allows us to safeguard the ambitious scientific goals set for the SO SATs. 

\acknowledgments
 This work was supported in part by a grant from the Simons Foundation (Award $\#$457687, B.K.). ND and JEG acknowledge support from the Swedish Research Council (Reg. no. 2019-03959). ND, GC and FN acknowledge funding from the European Union (ERC, POLOCALC, 101096035). Views and opinions expressed are, however, those of the authors only and do not necessarily reflect those of the EU or the ERC. Neither the EU nor the granting authority can be held responsible for them. KW is supported by the STFC, grant ST/X006344/1, and by a Gianturco Junior Research Fellowship of Linacre College, Oxford. SA is partially funded by a grant from the Simons Foundation (Award $\#$457687, B.K.). DA acknowledges support from UKSA under grant ST/Y005902/1, and from the Beecroft Trust. AEA is supported by the National Science Foundation (Award No.~2153201, UEI GM1XX56LEP58). JEG acknowledges support from the Swedish National Space Agency (SNSA/Rymdstyrelsen) and the Icelandic Research Fund (Grant number: 2410656-051). Funded in part by the European Union (ERC, CMBeam, 101040169). CB acknowledges partial support by the Italian Space Agency LiteBIRD Project (ASI Grants No. 2020-9-HH.0 and 2016-24-H.1-2018), and the Italian Space Agency Euclid Project, as well as the InDark and LiteBIRD Initiative of the National IInstitute for Nuclear Phyiscs, and the RadioForegroundsPlus Project HORIZON-CL4-2023-SPACE-01, GA 101135036. JE acknowledges funding from the SCIPOL project\footnote{\url{scipol.in2p3.fr}} funded by the European Research Council (ERC) under the European Union’s Horizon 2020 research and innovation program (PI: Josquin Errard, Grant agreement No. 101044073). MG is funded by the European Union (ERC, RELiCS, project number 101116027) and by the PRIN (Progetti di ricerca di Rilevante Interesse Nazionale) number 2022WJ9J33. RG would like to acknowledge support from the University of Southern California. CHC acknowledges ANID FONDECYT Postdoc Fellowship 3220255 and BASAL CATA FB210003. SCH is supported by P. J. E. Peebles Fellowship at the Perimeter Institute for Theoretical Physics. Research at Perimeter Institute is supported by the Government of Canada through the Department of Innovation, Science and Economic Development Canada and by the Province of Ontario through the Ministry of Research, Innovation and Science.

\bibliography{references.bib}{}

@ARTICLE{2015PhRvL.114j1301B,
       author = {{BICEP2/Keck Collaboration} and {Planck Collaboration} and {Ade}, P.~A.~R. and {Aghanim}, N. and {Ahmed}, Z. and {Aikin}, R.~W. and {Alexander}, K.~D. and {Arnaud}, M. and {Aumont}, J. and {Baccigalupi}, C. and {Banday}, A.~J. and {Barkats}, D. and {Barreiro}, R.~B. and {Bartlett}, J.~G. and {Bartolo}, N. and {Battaner}, E. and {Benabed}, K. and {Beno{\^\i}t}, A. and {Benoit-L{\'e}vy}, A. and {Benton}, S.~J. and {Bernard}, J. -P. and {Bersanelli}, M. and {Bielewicz}, P. and {Bischoff}, C.~A. and {Bock}, J.~J. and {Bonaldi}, A. and {Bonavera}, L. and {Bond}, J.~R. and {Borrill}, J. and {Bouchet}, F.~R. and {Boulanger}, F. and {Brevik}, J.~A. and {Bucher}, M. and {Buder}, I. and {Bullock}, E. and {Burigana}, C. and {Butler}, R.~C. and {Buza}, V. and {Calabrese}, E. and {Cardoso}, J. -F. and {Catalano}, A. and {Challinor}, A. and {Chary}, R. -R. and {Chiang}, H.~C. and {Christensen}, P.~R. and {Colombo}, L.~P.~L. and {Combet}, C. and {Connors}, J. and {Couchot}, F. and {Coulais}, A. and {Crill}, B.~P. and {Curto}, A. and {Cuttaia}, F. and {Danese}, L. and {Davies}, R.~D. and {Davis}, R.~J. and {de Bernardis}, P. and {De Rosa}, A. and {de Zotti}, G. and {Delabrouille}, J. and {Delouis}, J. -M. and {D{\'e}sert}, F. -X. and {Dickinson}, C. and {Diego}, J.~M. and {Dole}, H. and {Donzelli}, S. and {Dor{\'e}}, O. and {Douspis}, M. and {Dowell}, C.~D. and {Duband}, L. and {Ducout}, A. and {Dunkley}, J. and {Dupac}, X. and {Dvorkin}, C. and {Efstathiou}, G. and {Elsner}, F. and {En{\ss}lin}, T.~A. and {Eriksen}, H.~K. and {Falgarone}, E. and {Filippini}, J.~P. and {Finelli}, F. and {Fliescher}, S. and {Forni}, O. and {Frailis}, M. and {Fraisse}, A.~A. and {Franceschi}, E. and {Frejsel}, A. and {Galeotta}, S. and {Galli}, S. and {Ganga}, K. and {Ghosh}, T. and {Giard}, M. and {Gjerl{\o}w}, E. and {Golwala}, S.~R. and {Gonz{\'a}lez-Nuevo}, J. and {G{\'o}rski}, K.~M. and {Gratton}, S. and {Gregorio}, A. and {Gruppuso}, A. and {Gudmundsson}, J.~E. and {Halpern}, M. and {Hansen}, F.~K. and {Hanson}, D. and {Harrison}, D.~L. and {Hasselfield}, M. and {Helou}, G. and {Henrot-Versill{\'e}}, S. and {Herranz}, D. and {Hildebrandt}, S.~R. and {Hilton}, G.~C. and {Hivon}, E. and {Hobson}, M. and {Holmes}, W.~A. and {Hovest}, W. and {Hristov}, V.~V. and {Huffenberger}, K.~M. and {Hui}, H. and {Hurier}, G. and {Irwin}, K.~D. and {Jaffe}, A.~H. and {Jaffe}, T.~R. and {Jewell}, J. and {Jones}, W.~C. and {Juvela}, M. and {Karakci}, A. and {Karkare}, K.~S. and {Kaufman}, J.~P. and {Keating}, B.~G. and {Kefeli}, S. and {Keih{\"a}nen}, E. and {Kernasovskiy}, S.~A. and {Keskitalo}, R. and {Kisner}, T.~S. and {Kneissl}, R. and {Knoche}, J. and {Knox}, L. and {Kovac}, J.~M. and {Krachmalnicoff}, N. and {Kunz}, M. and {Kuo}, C.~L. and {Kurki-Suonio}, H. and {Lagache}, G. and {L{\"a}hteenm{\"a}ki}, A. and {Lamarre}, J. -M. and {Lasenby}, A. and {Lattanzi}, M. and {Lawrence}, C.~R. and {Leitch}, E.~M. and {Leonardi}, R. and {Levrier}, F. and {Lewis}, A. and {Liguori}, M. and {Lilje}, P.~B. and {Linden-V{\o}rnle}, M. and {L{\'o}pez-Caniego}, M. and {Lubin}, P.~M. and {Lueker}, M. and {Mac{\'\i}as-P{\'e}rez}, J.~F. and {Maffei}, B. and {Maino}, D. and {Mandolesi}, N. and {Mangilli}, A. and {Maris}, M. and {Martin}, P.~G. and {Mart{\'\i}nez-Gonz{\'a}lez}, E. and {Masi}, S. and {Mason}, P. and {Matarrese}, S. and {Megerian}, K.~G. and {Meinhold}, P.~R. and {Melchiorri}, A. and {Mendes}, L. and {Mennella}, A. and {Migliaccio}, M. and {Mitra}, S. and {Miville-Desch{\^e}nes}, M. -A. and {Moneti}, A. and {Montier}, L. and {Morgante}, G. and {Mortlock}, D. and {Moss}, A. and {Munshi}, D. and {Murphy}, J.~A. and {Naselsky}, P. and {Nati}, F. and {Natoli}, P. and {Netterfield}, C.~B. and {Nguyen}, H.~T. and {N{\o}rgaard-Nielsen}, H.~U. and {Noviello}, F. and {Novikov}, D. and {Novikov}, I. and {O'Brient}, R. and {Ogburn}, R.~W. and {Orlando}, A. and {Pagano}, L. and {Pajot}, F. and {Paladini}, R. and {Paoletti}, D. and {Partridge}, B. and {Pasian}, F. and {Patanchon}, G. and {Pearson}, T.~J. and {Perdereau}, O. and {Perotto}, L. and {Pettorino}, V. and {Piacentini}, F. and {Piat}, M. and {Pietrobon}, D. and {Plaszczynski}, S. and {Pointecouteau}, E. and {Polenta}, G. and {Ponthieu}, N. and {Pratt}, G.~W. and {Prunet}, S. and {Pryke}, C. and {Puget}, J. -L. and {Rachen}, J.~P. and {Reach}, W.~T. and {Rebolo}, R. and {Reinecke}, M. and {Remazeilles}, M. and {Renault}, C. and {Renzi}, A. and {Richter}, S. and {Ristorcelli}, I. and {Rocha}, G. and {Rossetti}, M. and {Roudier}, G. and {Rowan-Robinson}, M. and {Rubi{\~n}o-Mart{\'\i}n}, J.~A. and {Rusholme}, B. and {Sandri}, M. and {Santos}, D. and {Savelainen}, M. and {Savini}, G. and {Schwarz}, R. and {Scott}, D. and {Seiffert}, M.~D. and {Sheehy}, C.~D. and {Spencer}, L.~D. and {Staniszewski}, Z.~K. and {Stolyarov}, V. and {Sudiwala}, R. and {Sunyaev}, R. and {Sutton}, D. and {Suur-Uski}, A. -S. and {Sygnet}, J. -F. and {Tauber}, J.~A. and {Teply}, G.~P. and {Terenzi}, L. and {Thompson}, K.~L. and {Toffolatti}, L. and {Tolan}, J.~E. and {Tomasi}, M. and {Tristram}, M. and {Tucci}, M. and {Turner}, A.~D. and {Valenziano}, L. and {Valiviita}, J. and {van Tent}, B. and {Vibert}, L. and {Vielva}, P. and {Vieregg}, A.~G. and {Villa}, F. and {Wade}, L.~A. and {Wandelt}, B.~D. and {Watson}, R. and {Weber}, A.~C. and {Wehus}, I.~K. and {White}, M. and {White}, S.~D.~M. and {Willmert}, J. and {Wong}, C.~L. and {Yoon}, K.~W. and {Yvon}, D. and {Zacchei}, A. and {Zonca}, A. and {Bicep2/Keck} and {Planck Collaborations}},
        title = "{Joint Analysis of BICEP2/Keck Array and Planck Data}",
      journal = {\prl},
         year = 2015,
        month = mar,
       volume = {114},
       number = {10},
          eid = {101301},
        pages = {101301},
          doi = {10.1103/PhysRevLett.114.101301},
archivePrefix = {arXiv},
       eprint = {1502.00612},
 primaryClass = {astro-ph.CO},
       adsurl = {https://ui.adsabs.harvard.edu/abs/2015PhRvL.114j1301B}
}

@ARTICLE{2006PhRvD..74h3002S,
       author = {{Smith}, Kendrick M.},
        title = "{Pseudo-C$_{{\ensuremath{\ell}}}$ estimators which do not mix E and B modes}",
      journal = {\prd},
         year = 2006,
        month = oct,
       volume = {74},
       number = {8},
          eid = {083002},
        pages = {083002},
          doi = {10.1103/PhysRevD.74.083002},
archivePrefix = {arXiv},
       eprint = {astro-ph/0511629},
 primaryClass = {astro-ph},
       adsurl = {https://ui.adsabs.harvard.edu/abs/2006PhRvD..74h3002S}
}

@article{camb,
      author         = "Lewis, Antony and Challinor, Anthony and Lasenby,
                        Anthony",
      title          = "{Efficient computation of CMB anisotropies in closed FRW
                        models}",
      journal        = "\apj",
      volume         = "538",
      year           = "2000",
      pages          = "473-476",
      doi            = "10.1086/309179",
      eprint         = "astro-ph/9911177",
      archivePrefix  = "arXiv",
      primaryClass   = "astro-ph",
      SLACcitation   = "%%CITATION = ASTRO-PH/9911177;%%"
}

@ARTICLE{Dachlythra_2024,
       author = {{Dachlythra}, Nadia and {Duivenvoorden}, Adriaan J. and {Gudmundsson}, Jon E. and {Hasselfield}, Matthew and {Coppi}, Gabriele and {Adler}, Alexandre E. and {Alonso}, David and {Azzoni}, Susanna and {Chesmore}, Grace E. and {Fabbian}, Giulio and {Ganga}, Ken and {Gerras}, Remington G. and {Jaffe}, Andrew H. and {Johnson}, Bradley R. and {Keating}, Brian and {Keskitalo}, Reijo and {Kisner}, Theodore S. and {Krachmalnicoff}, Nicoletta and {Lungu}, Marius and {Matsuda}, Frederick and {Naess}, Sigurd and {Page}, Lyman and {Puddu}, Roberto and {Puglisi}, Giuseppe and {Simon}, Sara M. and {Teply}, Grant and {Tsan}, Tran and {Wollack}, Edward J. and {Wolz}, Kevin and {Xu}, Zhilei},
        title = "{The Simons Observatory: Beam Characterization for the Small Aperture Telescopes}",
      journal = {\apj},
         year = 2024,
        month = jan,
       volume = {961},
       number = {1},
          eid = {138},
        pages = {138},
          doi = {10.3847/1538-4357/ad0969},
archivePrefix = {arXiv},
       eprint = {2304.08995},
 primaryClass = {astro-ph.IM},
       adsurl = {https://ui.adsabs.harvard.edu/abs/2024ApJ...961..138D}
}

@article{Abitbol_2021,
   title={The Simons Observatory: gain, bandpass and polarization-angle calibration requirements for B-mode searches},
   volume={2021},
   ISSN={1475-7516},
   url={http://dx.doi.org/10.1088/1475-7516/2021/05/032},
   DOI={10.1088/1475-7516/2021/05/032},
   number={05},
   journal={Journal of Cosmology and Astroparticle Physics},
   publisher={IOP Publishing},
   author={Abitbol, Maximilian H. and Alonso, David and Simon, Sara M. and Lashner, Jack and Crowley, Kevin T. and Ali, Aamir M. and Azzoni, Susanna and Baccigalupi, Carlo and Barron, Darcy and Brown, Michael L. and et al.},
   year={2021},
   month={May},
   pages={032}
}

@article{planck_diff_sep_comp_2020,
   title={Planck2018 results},
   volume={641},
   ISSN={1432-0746},
   url={http://dx.doi.org/10.1051/0004-6361/201833881},
   DOI={10.1051/0004-6361/201833881},
   journal={Astronomy \& Astrophysics},
   publisher={EDP Sciences},
   author={{The Planck Collaboration} and Akrami, Y. and Ashdown, M. and Aumont, J. and Baccigalupi, C. and Ballardini, M. and Banday, A. J. and Barreiro, R. B. and Bartolo, N. and Basak, S. and et al.},
   year={2020},
   month={Sep},
   pages={A4}
}

@ARTICLE{Planck_2018-6,
       author = {{Planck Collaboration} and {Aghanim}, N. and {Akrami}, Y. and {Ashdown}, M. and {Aumont}, J. and {Baccigalupi}, C. and {Ballardini}, M. and {Banday}, A.~J. and {Barreiro}, R.~B. and {Bartolo}, N. and {Basak}, S. and {Battye}, R. and {Benabed}, K. and {Bernard}, J. -P. and {Bersanelli}, M. and {Bielewicz}, P. and {Bock}, J.~J. and {Bond}, J.~R. and {Borrill}, J. and {Bouchet}, F.~R. and {Boulanger}, F. and {Bucher}, M. and {Burigana}, C. and {Butler}, R.~C. and {Calabrese}, E. and {Cardoso}, J. -F. and {Carron}, J. and {Challinor}, A. and {Chiang}, H.~C. and {Chluba}, J. and {Colombo}, L.~P.~L. and {Combet}, C. and {Contreras}, D. and {Crill}, B.~P. and {Cuttaia}, F. and {de Bernardis}, P. and {de Zotti}, G. and {Delabrouille}, J. and {Delouis}, J. -M. and {Di Valentino}, E. and {Diego}, J.~M. and {Dor{\'e}}, O. and {Douspis}, M. and {Ducout}, A. and {Dupac}, X. and {Dusini}, S. and {Efstathiou}, G. and {Elsner}, F. and {En{\ss}lin}, T.~A. and {Eriksen}, H.~K. and {Fantaye}, Y. and {Farhang}, M. and {Fergusson}, J. and {Fernandez-Cobos}, R. and {Finelli}, F. and {Forastieri}, F. and {Frailis}, M. and {Fraisse}, A.~A. and {Franceschi}, E. and {Frolov}, A. and {Galeotta}, S. and {Galli}, S. and {Ganga}, K. and {G{\'e}nova-Santos}, R.~T. and {Gerbino}, M. and {Ghosh}, T. and {Gonz{\'a}lez-Nuevo}, J. and {G{\'o}rski}, K.~M. and {Gratton}, S. and {Gruppuso}, A. and {Gudmundsson}, J.~E. and {Hamann}, J. and {Handley}, W. and {Hansen}, F.~K. and {Herranz}, D. and {Hildebrandt}, S.~R. and {Hivon}, E. and {Huang}, Z. and {Jaffe}, A.~H. and {Jones}, W.~C. and {Karakci}, A. and {Keih{\"a}nen}, E. and {Keskitalo}, R. and {Kiiveri}, K. and {Kim}, J. and {Kisner}, T.~S. and {Knox}, L. and {Krachmalnicoff}, N. and {Kunz}, M. and {Kurki-Suonio}, H. and {Lagache}, G. and {Lamarre}, J. -M. and {Lasenby}, A. and {Lattanzi}, M. and {Lawrence}, C.~R. and {Le Jeune}, M. and {Lemos}, P. and {Lesgourgues}, J. and {Levrier}, F. and {Lewis}, A. and {Liguori}, M. and {Lilje}, P.~B. and {Lilley}, M. and {Lindholm}, V. and {L{\'o}pez-Caniego}, M. and {Lubin}, P.~M. and {Ma}, Y. -Z. and {Mac{\'\i}as-P{\'e}rez}, J.~F. and {Maggio}, G. and {Maino}, D. and {Mandolesi}, N. and {Mangilli}, A. and {Marcos-Caballero}, A. and {Maris}, M. and {Martin}, P.~G. and {Martinelli}, M. and {Mart{\'\i}nez-Gonz{\'a}lez}, E. and {Matarrese}, S. and {Mauri}, N. and {McEwen}, J.~D. and {Meinhold}, P.~R. and {Melchiorri}, A. and {Mennella}, A. and {Migliaccio}, M. and {Millea}, M. and {Mitra}, S. and {Miville-Desch{\^e}nes}, M. -A. and {Molinari}, D. and {Montier}, L. and {Morgante}, G. and {Moss}, A. and {Natoli}, P. and {N{\o}rgaard-Nielsen}, H.~U. and {Pagano}, L. and {Paoletti}, D. and {Partridge}, B. and {Patanchon}, G. and {Peiris}, H.~V. and {Perrotta}, F. and {Pettorino}, V. and {Piacentini}, F. and {Polastri}, L. and {Polenta}, G. and {Puget}, J. -L. and {Rachen}, J.~P. and {Reinecke}, M. and {Remazeilles}, M. and {Renzi}, A. and {Rocha}, G. and {Rosset}, C. and {Roudier}, G. and {Rubi{\~n}o-Mart{\'\i}n}, J.~A. and {Ruiz-Granados}, B. and {Salvati}, L. and {Sandri}, M. and {Savelainen}, M. and {Scott}, D. and {Shellard}, E.~P.~S. and {Sirignano}, C. and {Sirri}, G. and {Spencer}, L.~D. and {Sunyaev}, R. and {Suur-Uski}, A. -S. and {Tauber}, J.~A. and {Tavagnacco}, D. and {Tenti}, M. and {Toffolatti}, L. and {Tomasi}, M. and {Trombetti}, T. and {Valenziano}, L. and {Valiviita}, J. and {Van Tent}, B. and {Vibert}, L. and {Vielva}, P. and {Villa}, F. and {Vittorio}, N. and {Wandelt}, B.~D. and {Wehus}, I.~K. and {White}, M. and {White}, S.~D.~M. and {Zacchei}, A. and {Zonca}, A.},
        title = "{Planck 2018 results. VI. Cosmological parameters}",
      journal = {\aap},
         year = 2020,
        month = sep,
       volume = {641},
          eid = {A6},
        pages = {A6},
          doi = {10.1051/0004-6361/201833910},
archivePrefix = {arXiv},
       eprint = {1807.06209},
 primaryClass = {astro-ph.CO},
       adsurl = {https://ui.adsabs.harvard.edu/abs/2020A&A...641A...6P}
}

@ARTICLE{wolz2023simons,
       author = {{Wolz}, Kevin and {Azzoni}, Susanna and {Herv{\'\i}as-Caimapo}, Carlos and {Errard}, Josquin and {Krachmalnicoff}, Nicoletta and {Alonso}, David and {Baccigalupi}, Carlo and {Baleato Lizancos}, Ant{\'o}n and {Brown}, Michael L. and {Calabrese}, Erminia and {Chluba}, Jens and {Dunkley}, Jo and {Fabbian}, Giulio and {Galitzki}, Nicholas and {Jost}, Baptiste and {Morshed}, Magdy and {Nati}, Federico},
        title = "{The Simons Observatory: Pipeline comparison and validation for large-scale B-modes}",
      journal = {\aap},
         year = 2024,
        month = jun,
       volume = {686},
          eid = {A16},
        pages = {A16},
          doi = {10.1051/0004-6361/202346105},
archivePrefix = {arXiv},
       eprint = {2302.04276},
 primaryClass = {astro-ph.CO},
       adsurl = {https://ui.adsabs.harvard.edu/abs/2024A&A...686A..16W}
}

@ARTICLE{Ade2019,
       author = {{The Simons Observatory} and {Ade}, Peter and {Aguirre}, James and {Ahmed}, Zeeshan and {Aiola}, Simone and {Ali}, Aamir and {Alonso}, David and {Alvarez}, Marcelo A. and {Arnold}, Kam and {Ashton}, Peter and {Austermann}, Jason and {Awan}, Humna and {Baccigalupi}, Carlo and {Baildon}, Taylor and {Barron}, Darcy and {Battaglia}, Nick and {Battye}, Richard and {Baxter}, Eric and {Bazarko}, Andrew and {Beall}, James A. and {Bean}, Rachel and {Beck}, Dominic and {Beckman}, Shawn and {Beringue}, Benjamin and {Bianchini}, Federico and {Boada}, Steven and {Boettger}, David and {Bond}, J. Richard and {Borrill}, Julian and {Brown}, Michael L. and {Bruno}, Sarah Marie and {Bryan}, Sean and {Calabrese}, Erminia and {Calafut}, Victoria and {Calisse}, Paolo and {Carron}, Julien and {Challinor}, Anthony and {Chesmore}, Grace and {Chinone}, Yuji and {Chluba}, Jens and {Cho}, Hsiao-Mei Sherry and {Choi}, Steve and {Coppi}, Gabriele and {Cothard}, Nicholas F. and {Coughlin}, Kevin and {Crichton}, Devin and {Crowley}, Kevin D. and {Crowley}, Kevin T. and {Cukierman}, Ari and {D'Ewart}, John M. and {D{\"u}nner}, Rolando and {de Haan}, Tijmen and {Devlin}, Mark and {Dicker}, Simon and {Didier}, Joy and {Dobbs}, Matt and {Dober}, Bradley and {Duell}, Cody J. and {Duff}, Shannon and {Duivenvoorden}, Adri and {Dunkley}, Jo and {Dusatko}, John and {Errard}, Josquin and {Fabbian}, Giulio and {Feeney}, Stephen and {Ferraro}, Simone and {Flux{\`a}}, Pedro and {Freese}, Katherine and {Frisch}, Josef C. and {Frolov}, Andrei and {Fuller}, George and {Fuzia}, Brittany and {Galitzki}, Nicholas and {Gallardo}, Patricio A. and {Tomas Galvez Ghersi}, Jose and {Gao}, Jiansong and {Gawiser}, Eric and {Gerbino}, Martina and {Gluscevic}, Vera and {Goeckner-Wald}, Neil and {Golec}, Joseph and {Gordon}, Sam and {Gralla}, Megan and {Green}, Daniel and {Grigorian}, Arpi and {Groh}, John and {Groppi}, Chris and {Guan}, Yilun and {Gudmundsson}, Jon E. and {Han}, Dongwon and {Hargrave}, Peter and {Hasegawa}, Masaya and {Hasselfield}, Matthew and {Hattori}, Makoto and {Haynes}, Victor and {Hazumi}, Masashi and {He}, Yizhou and {Healy}, Erin and {Henderson}, Shawn W. and {Hervias-Caimapo}, Carlos and {Hill}, Charles A. and {Hill}, J. Colin and {Hilton}, Gene and {Hilton}, Matt and {Hincks}, Adam D. and {Hinshaw}, Gary and {Hlo{\v{z}}ek}, Ren{\'e}e and {Ho}, Shirley and {Ho}, Shuay-Pwu Patty and {Howe}, Logan and {Huang}, Zhiqi and {Hubmayr}, Johannes and {Huffenberger}, Kevin and {Hughes}, John P. and {Ijjas}, Anna and {Ikape}, Margaret and {Irwin}, Kent and {Jaffe}, Andrew H. and {Jain}, Bhuvnesh and {Jeong}, Oliver and {Kaneko}, Daisuke and {Karpel}, Ethan D. and {Katayama}, Nobuhiko and {Keating}, Brian and {Kernasovskiy}, Sarah S. and {Keskitalo}, Reijo and {Kisner}, Theodore and {Kiuchi}, Kenji and {Klein}, Jeff and {Knowles}, Kenda and {Koopman}, Brian and {Kosowsky}, Arthur and {Krachmalnicoff}, Nicoletta and {Kuenstner}, Stephen E. and {Kuo}, Chao-Lin and {Kusaka}, Akito and {Lashner}, Jacob and {Lee}, Adrian and {Lee}, Eunseong and {Leon}, David and {Leung}, Jason S. -Y. and {Lewis}, Antony and {Li}, Yaqiong and {Li}, Zack and {Limon}, Michele and {Linder}, Eric and {Lopez-Caraballo}, Carlos and {Louis}, Thibaut and {Lowry}, Lindsay and {Lungu}, Marius and {Madhavacheril}, Mathew and {Mak}, Daisy and {Maldonado}, Felipe and {Mani}, Hamdi and {Mates}, Ben and {Matsuda}, Frederick and {Maurin}, Lo{\"\i}c and {Mauskopf}, Phil and {May}, Andrew and {McCallum}, Nialh and {McKenney}, Chris and {McMahon}, Jeff and {Meerburg}, P. Daniel and {Meyers}, Joel and {Miller}, Amber and {Mirmelstein}, Mark and {Moodley}, Kavilan and {Munchmeyer}, Moritz and {Munson}, Charles and {Naess}, Sigurd and {Nati}, Federico and {Navaroli}, Martin and {Newburgh}, Laura and {Nguyen}, Ho Nam and {Niemack}, Michael and {Nishino}, Haruki and {Orlowski-Scherer}, John and {Page}, Lyman and {Partridge}, Bruce and {Peloton}, Julien and {Perrotta}, Francesca and {Piccirillo}, Lucio and {Pisano}, Giampaolo and {Poletti}, Davide and {Puddu}, Roberto and {Puglisi}, Giuseppe and {Raum}, Chris and {Reichardt}, Christian L. and {Remazeilles}, Mathieu and {Rephaeli}, Yoel and {Riechers}, Dominik and {Rojas}, Felipe and {Roy}, Anirban and {Sadeh}, Sharon and {Sakurai}, Yuki and {Salatino}, Maria and {Sathyanarayana Rao}, Mayuri and {Schaan}, Emmanuel and {Schmittfull}, Marcel and {Sehgal}, Neelima and {Seibert}, Joseph and {Seljak}, Uros and {Sherwin}, Blake and {Shimon}, Meir and {Sierra}, Carlos and {Sievers}, Jonathan and {Sikhosana}, Precious and {Silva-Feaver}, Maximiliano and {Simon}, Sara M. and {Sinclair}, Adrian and {Siritanasak}, Praween and {Smith}, Kendrick and {Smith}, Stephen R. and {Spergel}, David and {Staggs}, Suzanne T. and {Stein}, George and {Stevens}, Jason R. and {Stompor}, Radek and {Suzuki}, Aritoki and {Tajima}, Osamu and {Takakura}, Satoru and {Teply}, Grant and {Thomas}, Daniel B. and {Thorne}, Ben and {Thornton}, Robert and {Trac}, Hy and {Tsai}, Calvin and {Tucker}, Carole and {Ullom}, Joel and {Vagnozzi}, Sunny and {van Engelen}, Alexander and {Van Lanen}, Jeff and {Van Winkle}, Daniel D. and {Vavagiakis}, Eve M. and {Verg{\`e}s}, Clara and {Vissers}, Michael and {Wagoner}, Kasey and {Walker}, Samantha and {Ward}, Jon and {Westbrook}, Ben and {Whitehorn}, Nathan and {Williams}, Jason and {Williams}, Joel and {Wollack}, Edward J. and {Xu}, Zhilei and {Yu}, Byeonghee and {Yu}, Cyndia and {Zago}, Fernando and {Zhang}, Hezi and {Zhu}, Ningfeng and {Simons Observatory Collaboration}},
        title = "{The Simons Observatory: science goals and forecasts}",
      journal = {\jcap},
         year = 2019,
        month = feb,
       volume = {2019},
       number = {2},
          eid = {056},
        pages = {056},
          doi = {10.1088/1475-7516/2019/02/056},
archivePrefix = {arXiv},
       eprint = {1808.07445},
 primaryClass = {astro-ph.CO},
       adsurl = {https://ui.adsabs.harvard.edu/abs/2019JCAP...02..056A}
}

@article{beamconv_2018,
   title={Full-sky beam convolution for cosmic microwave background applications},
   volume={486},
   ISSN={1365-2966},
   url={http://dx.doi.org/10.1093/mnras/stz1143},
   DOI={10.1093/mnras/stz1143},
   number={4},
   journal={Monthly Notices of the Royal Astronomical Society},
   publisher={Oxford University Press (OUP)},
   author={Duivenvoorden, Adriaan J and Gudmundsson, Jon E and Rahlin, Alexandra S},
   year={2019},
   month={May},
   pages={5448–5467}
}

@article{Azzoni_2021,
   title={A minimal power-spectrum-based moment expansion for CMB B-mode searches},
   volume={2021},
   ISSN={1475-7516},
   url={http://dx.doi.org/10.1088/1475-7516/2021/05/047},
   DOI={10.1088/1475-7516/2021/05/047},
   number={05},
   journal={Journal of Cosmology and Astroparticle Physics},
   publisher={IOP Publishing},
   author={Azzoni, S. and Abitbol, M.H. and Alonso, D. and Gough, A. and Katayama, N. and Matsumura, T.},
   year={2021},
   month=may, pages={047} }

@ARTICLE{Azzoni_2023,
       author = {{Azzoni}, S. and {Alonso}, D. and {Abitbol}, M.~H. and {Errard}, J. and {Krachmalnicoff}, N.},
        title = "{A hybrid map-C$_{{\ensuremath{\ell}}}$ component separation method for primordial CMB B-mode searches}",
      journal = {\jcap},
         year = 2023,
        month = mar,
       volume = {2023},
       number = {3},
          eid = {035},
        pages = {035},
          doi = {10.1088/1475-7516/2023/03/035},
archivePrefix = {arXiv},
       eprint = {2210.14838},
 primaryClass = {astro-ph.CO},
       adsurl = {https://ui.adsabs.harvard.edu/abs/2023JCAP...03..035A}
}

@article{Lungu_2022,
   title={The Atacama Cosmology Telescope: measurement and analysis of 1D beams for DR4},
   volume={2022},
   ISSN={1475-7516},
   url={http://dx.doi.org/10.1088/1475-7516/2022/05/044},
   DOI={10.1088/1475-7516/2022/05/044},
   number={05},
   journal={Journal of Cosmology and Astroparticle Physics},
   publisher={IOP Publishing},
   author={Lungu, Marius and Storer, Emilie R. and Hasselfield, Matthew and Duivenvoorden, Adriaan J. and Calabrese, Erminia and Chesmore, Grace E. and Choi, Steve K. and Dunkley, Jo and Dünner, Rolando and Gallardo, Patricio A. and Golec, Joseph E. and Guan, Yilun and Hill, J. Colin and Hincks, Adam D. and Hubmayr, Johannes and Madhavacheril, Mathew S. and Mallaby-Kay, Maya and McMahon, Jeff and Moodley, Kavilan and Naess, Sigurd and Nati, Federico and Niemack, Michael D. and Page, Lyman A. and Partridge, Bruce and Puddu, Roberto and Schillaci, Alessandro and Sifón, Cristóbal and Staggs, Suzanne and Sunder, Dhaneshwar D. and Wollack, Edward J. and Xu, Zhilei},
   year={2022},
   month=may, pages={044} }

@article{Alonso_2019,
   title={A unified pseudo-C$\ell$ framework},
   volume={484},
   ISSN={1365-2966},
   url={http://dx.doi.org/10.1093/mnras/stz093},
   DOI={10.1093/mnras/stz093},
   number={3},
   journal={Monthly Notices of the Royal Astronomical Society},
   publisher={Oxford University Press (OUP)},
   author={Alonso, David and Sanchez, Javier and Slosar, Anže},
   year={2019},
   month=jan, pages={4127–4151} }

@ARTICLE{AnneXFaster,
       author = {{Gambrel}, A.~E. and {Rahlin}, A.~S. and {Song}, X. and {Contaldi}, C.~R. and {Ade}, P.~A.~R. and {Amiri}, M. and {Benton}, S.~J. and {Bergman}, A.~S. and {Bihary}, R. and {Bock}, J.~J. and {Bond}, J.~R. and {Bonetti}, J.~A. and {Bryan}, S.~A. and {Chiang}, H.~C. and {Duivenvoorden}, A.~J. and {Eriksen}, H.~K. and {Farhang}, M. and {Filippini}, J.~P. and {Fraisse}, A.~A. and {Freese}, K. and {Galloway}, M. and {Gandilo}, N.~N. and {Gualtieri}, R. and {Gudmundsson}, J.~E. and {Halpern}, M. and {Hartley}, J. and {Hasselfield}, M. and {Hilton}, G. and {Holmes}, W. and {Hristov}, V.~V. and {Huang}, Z. and {Irwin}, K.~D. and {Jones}, W.~C. and {Karakci}, A. and {Kuo}, C.~L. and {Kermish}, Z.~D. and {Leung}, J.~S. -Y. and {Li}, S. and {Mak}, D.~S.~Y. and {Mason}, P.~V. and {Megerian}, K. and {Moncelsi}, L. and {Morford}, T.~A. and {Nagy}, J.~M. and {Netterfield}, C.~B. and {Nolta}, M. and {O’Brient}, R. and {Osherson}, B. and {Padilla}, I.~L. and {Racine}, B. and {Reintsema}, C. and {Ruhl}, J.~E. and {Ruud}, T.~M. and {Shariff}, J.~A. and {Shaw}, E.~C. and {Shiu}, C. and {Soler}, J.~D. and {Trangsrud}, A. and {Tucker}, C. and {Tucker}, R.~S. and {Turner}, A.~D. and {List}, J.~F. van der and {Weber}, A.~C. and {Wehus}, I.~K. and {Wen}, S. and {Wiebe}, D.~V. and {Young}, E.~Y.},
        title = "{The XFaster Power Spectrum and Likelihood Estimator for the Analysis of Cosmic Microwave Background Maps}",
      journal = {ApJ},
         year = 2021,
        month = dec,
       volume = {922},
       number = {2},
          eid = {132},
        pages = {132},
          doi = {10.3847/1538-4357/ac230b},
archivePrefix = {arXiv},
       eprint = {2104.01172},
 primaryClass = {astro-ph.CO},
       adsurl = {https://ui.adsabs.harvard.edu/abs/2021ApJ...922..132G}
}

@ARTICLE{Aumont_smica,
       author = {{Aumont}, J. and {Mac{\'\i}as-P{\'e}rez}, J.~F.},
        title = "{Blind component separation for polarized observations of the cosmic microwave background}",
      journal = {\mnras},
         year = 2007,
        month = apr,
       volume = {376},
       number = {2},
        pages = {739-758},
          doi = {10.1111/j.1365-2966.2007.11470.x},
archivePrefix = {arXiv},
       eprint = {astro-ph/0603044},
 primaryClass = {astro-ph},
       adsurl = {https://ui.adsabs.harvard.edu/abs/2007MNRAS.376..739A}
}

@ARTICLE{Commander3,
       author = {{Galloway}, M. and {Andersen}, K.~J. and {Aurlien}, R. and {Banerji}, R. and {Bersanelli}, M. and {Bertocco}, S. and {Brilenkov}, M. and {Carbone}, M. and {Colombo}, L.~P.~L. and {Eriksen}, H.~K. and {Eskilt}, J.~R. and {Foss}, M.~K. and {Franceschet}, C. and {Fuskeland}, U. and {Galeotta}, S. and {Gerakakis}, S. and {Gjerl{\o}w}, E. and {Hensley}, B. and {Herman}, D. and {Iacobellis}, M. and {Ieronymaki}, M. and {Ihle}, H.~T. and {Jewell}, J.~B. and {Karakci}, A. and {Keih{\"a}nen}, E. and {Keskitalo}, R. and {Maggio}, G. and {Maino}, D. and {Maris}, M. and {Mennella}, A. and {Paradiso}, S. and {Partridge}, B. and {Reinecke}, M. and {San}, M. and {Suur-Uski}, A. -S. and {Svalheim}, T.~L. and {Tavagnacco}, D. and {Thommesen}, H. and {Watts}, D.~J. and {Wehus}, I.~K. and {Zacchei}, A.},
        title = "{BEYONDPLANCK. III. Commander3}",
      journal = {\aap},
         year = 2023,
        month = jul,
       volume = {675},
          eid = {A3},
        pages = {A3},
          doi = {10.1051/0004-6361/202243137},
archivePrefix = {arXiv},
       eprint = {2201.03509},
 primaryClass = {astro-ph.CO},
       adsurl = {https://ui.adsabs.harvard.edu/abs/2023A&A...675A...3G}
}

@ARTICLE{Crittenden_1993,
       author = {{Crittenden}, Robert and {Davis}, Richard L. and {Steinhardt}, Paul J.},
        title = "{Polarization of the Microwave Background Due to Primordial Gravitational Waves}",
      journal = {\apjl},
         year = {1993},
        month = {nov},
       volume = {417},
        pages = {L13},
          doi = {10.1086/187082},
archivePrefix = {arXiv},
       eprint = {astro-ph/9306027},
 primaryClass = {astro-ph},
       adsurl = {https://ui.adsabs.harvard.edu/abs/1993ApJ...417L..13C}
}

@ARTICLE{Delabrouille_2009,
       author = {{Delabrouille}, J. and {Cardoso}, J. -F. and {Le Jeune}, M. and {Betoule}, M. and {Fay}, G. and {Guilloux}, F.},
        title = "{A full sky, low foreground, high resolution CMB map from WMAP}",
      journal = {\aap},
         year = 2009,
        month = jan,
       volume = {493},
       number = {3},
        pages = {835-857},
          doi = {10.1051/0004-6361:200810514},
archivePrefix = {arXiv},
       eprint = {0807.0773},
 primaryClass = {astro-ph},
       adsurl = {https://ui.adsabs.harvard.edu/abs/2009A&A...493..835D}
}

@ARTICLE{Didier_2019,
       author = {{Didier}, Joy and {Miller}, Amber D. and {Araujo}, Derek and {Aubin}, Fran{\c{c}}ois and {Geach}, Christopher and {Johnson}, Bradley and {Korotkov}, Andrei and {Raach}, Kate and {Westbrook}, Benjamin and {Young}, Karl and {Aboobaker}, Asad M. and {Ade}, Peter and {Baccigalupi}, Carlo and {Bao}, Chaoyun and {Chapman}, Daniel and {Dobbs}, Matt and {Grainger}, Will and {Hanany}, Shaul and {Helson}, Kyle and {Hillbrand}, Seth and {Hubmayr}, Johannes and {Jaffe}, Andrew and {Jones}, Terry J. and {Klein}, Jeff and {Lee}, Adrian and {Limon}, Michele and {MacDermid}, Kevin and {Milligan}, Michael and {Pascale}, Enzo and {Reichborn-Kjennerud}, Britt and {Sagiv}, Ilan and {Tucker}, Carole and {Tucker}, Gregory S. and {Zilic}, Kyle},
        title = "{Intensity-coupled Polarization in Instruments with a Continuously Rotating Half-wave Plate}",
      journal = {\apj},
         year = 2019,
        month = may,
       volume = {876},
       number = {1},
          eid = {54},
        pages = {54},
          doi = {10.3847/1538-4357/ab0f36},
       adsurl = {https://ui.adsabs.harvard.edu/abs/2019ApJ...876...54D}
}

@ARTICLE{emcee_code,
       author = {{Foreman-Mackey}, Daniel and {Hogg}, David W. and {Lang}, Dustin and {Goodman}, Jonathan},
        title = "{emcee: The MCMC Hammer}",
      journal = {PASP},
         year = 2013,
        month = mar,
       volume = {125},
       number = {925},
        pages = {306},
          doi = {10.1086/670067},
archivePrefix = {arXiv},
       eprint = {1202.3665},
 primaryClass = {astro-ph.IM},
       adsurl = {https://ui.adsabs.harvard.edu/abs/2013PASP..125..306F}
}

@article{Ward_bpass_variations,
   title={The Effects of Bandpass Variations on Foreground Removal Forecasts for Future CMB Experiments},
   volume={861},
   ISSN={1538-4357},
   url={http://dx.doi.org/10.3847/1538-4357/aac71f},
   DOI={10.3847/1538-4357/aac71f},
   number={2},
   journal={The Astrophysical Journal},
   publisher={American Astronomical Society},
   author={Ward, J. T. and Alonso, D. and Errard, J. and Devlin, M. J. and Hasselfield, M.},
   year={2018},
   month=jul, pages={82} }

@article{Minami_cb,
   title={New Extraction of the Cosmic Birefringence from the Planck 2018 Polarization Data},
   volume={125},
   ISSN={1079-7114},
   url={http://dx.doi.org/10.1103/PhysRevLett.125.221301},
   DOI={10.1103/physrevlett.125.221301},
   number={22},
   journal={Physical Review Letters},
   publisher={American Physical Society (APS)},
   author={Minami, Yuto and Komatsu, Eiichiro},
   year={2020},
   month=nov }

@ARTICLE{Fixsen_2009,
       author = {{Fixsen}, D.~J.},
        title = "{The Temperature of the Cosmic Microwave Background}",
      journal = {\apj},
         year = 2009,
        month = dec,
       volume = {707},
       number = {2},
        pages = {916-920},
          doi = {10.1088/0004-637X/707/2/916},
archivePrefix = {arXiv},
       eprint = {0911.1955},
 primaryClass = {astro-ph.CO},
       adsurl = {https://ui.adsabs.harvard.edu/abs/2009ApJ...707..916F}
}

@article{Hensley_2022,
   title={The Simons Observatory: Galactic Science Goals and Forecasts},
   volume={929},
   ISSN={1538-4357},
   url={http://dx.doi.org/10.3847/1538-4357/ac5e36},
   DOI={10.3847/1538-4357/ac5e36},
   number={2},
   journal={The Astrophysical Journal},
   publisher={American Astronomical Society},
   author={Hensley, Brandon S. and Clark, Susan E. and Fanfani, Valentina and Krachmalnicoff, Nicoletta and Fabbian, Giulio and Poletti, Davide and Puglisi, Giuseppe and Coppi, Gabriele and Nibauer, Jacob and Gerasimov, Roman and Galitzki, Nicholas and Choi, Steve K. and Ashton, Peter C. and Baccigalupi, Carlo and Baxter, Eric and Burkhart, Blakesley and Calabrese, Erminia and Chluba, Jens and Errard, Josquin and Frolov, Andrei V. and Hervías-Caimapo, Carlos and Huffenberger, Kevin M. and Johnson, Bradley R. and Jost, Baptiste and Keating, Brian and McCarrick, Heather and Nati, Federico and Sathyanarayana Rao, Mayuri and van Engelen, Alexander and Walker, Samantha and Wolz, Kevin and Xu, Zhilei and Zhu, Ningfeng and Zonca, Andrea},
   year={2022},
   month=apr, pages={166} }

@ARTICLE{Hu_1997,
       author = {{Hu}, Wayne and {White}, Martin},
        title = "{A CMB polarization primer}",
      journal = {New Astronomy},
         year = 1997,
        month = oct,
       volume = {2},
       number = {4},
        pages = {323-344},
          doi = {10.1016/S1384-1076(97)00022-5},
archivePrefix = {arXiv},
       eprint = {astro-ph/9706147},
 primaryClass = {astro-ph},
       adsurl = {https://ui.adsabs.harvard.edu/abs/1997NewA....2..323H}
}

@ARTICLE{Leloup_2024,
       author = {{Leloup}, C. and {Patanchon}, G. and {Errard}, J. and {Franceschet}, C. and {Gudmundsson}, J.~E. and {Henrot-Versill{\'e}}, S. and {Imada}, H. and {Ishino}, H. and {Matsumura}, T. and {Puglisi}, G. and {Wang}, W. and {Adler}, A. and {Aumont}, J. and {Aurlien}, R. and {Baccigalupi}, C. and {Ballardini}, M. and {Banday}, A.~J. and {Barreiro}, R.~B. and {Bartolo}, N. and {Basyrov}, A. and {Bersanelli}, M. and {Blinov}, D. and {Bortolami}, M. and {Brinckmann}, T. and {Campeti}, P. and {Carones}, A. and {Carralot}, F. and {Casas}, F.~J. and {Cheung}, K. and {Clermont}, L. and {Columbro}, F. and {Conenna}, G. and {Coppolecchia}, A. and {Cuttaia}, F. and {Dachlythra}, N. and {D'Alessandro}, G. and {de Bernardis}, P. and {de Haan}, T. and {De Petris}, M. and {Della Torre}, S. and {Diego-Palazuelos}, P. and {Eriksen}, H.~K. and {Finelli}, F. and {Fuskeland}, U. and {Galloni}, G. and {Galloway}, M. and {Georges}, M. and {Gerbino}, M. and {Gervasi}, M. and {G{\'e}nova-Santos}, R.~T. and {Ghigna}, T. and {Giardiello}, S. and {Gimeno-Amo}, C. and {Gjerl{\o}w}, E. and {Gruppuso}, A. and {Hazumi}, M. and {Hergt}, L.~T. and {Herranz}, D. and {Hivon}, E. and {Hoang}, T.~D. and {Jost}, B. and {Kohri}, K. and {Krachmalnicoff}, N. and {Lee}, A.~T. and {Lembo}, M. and {Levrier}, F. and {Lonappan}, A.~I. and {L{\'o}pez-Caniego}, M. and {Macias-Perez}, J. and {Mart{\'\i}nez-Gonz{\'a}lez}, E. and {Masi}, S. and {Matarrese}, S. and {Micheli}, S. and {Monelli}, M. and {Montier}, L. and {Morgante}, G. and {Mot}, B. and {Mousset}, L. and {Namikawa}, T. and {Natoli}, P. and {Novelli}, A. and {Noviello}, F. and {Obata}, I. and {Odagiri}, K. and {Pagano}, L. and {Paiella}, A. and {Paoletti}, D. and {Pascual-Cisneros}, G. and {Pavlidou}, V. and {Piacentini}, F. and {Piccirilli}, G. and {Pisano}, G. and {Polenta}, G. and {Raffuzzi}, N. and {Remazeilles}, M. and {Ritacco}, A. and {Rizzieri}, A. and {Ruiz-Granda}, M. and {Sakurai}, Y. and {Shiraishi}, M. and {Stever}, S.~L. and {Takase}, Y. and {Tassis}, K. and {Terenzi}, L. and {Thompson}, K.~L. and {Tristram}, M. and {Vacher}, L. and {Vielva}, P. and {Wehus}, I.~K. and {Weymann-Despres}, G. and {Zannoni}, M. and {Zhou}, Y. and {LiteBIRD Collaboration}},
        title = "{Impact of beam far side-lobe knowledge in the presence of foregrounds for LiteBIRD}",
      journal = {\jcap},
         year = 2024,
        month = jun,
       volume = {2024},
       number = {6},
          eid = {011},
        pages = {011},
          doi = {10.1088/1475-7516/2024/06/011},
archivePrefix = {arXiv},
       eprint = {2312.09001},
 primaryClass = {astro-ph.CO},
       adsurl = {https://ui.adsabs.harvard.edu/abs/2024JCAP...06..011L}
}

@ARTICLE{Naess_2023,
       author = {{Naess}, Sigurd and {Louis}, Thibaut},
        title = "{Large-scale power loss in ground-based CMB mapmaking}",
      journal = {The Open Journal of Astrophysics},
         year = 2023,
        month = jul,
       volume = {6},
          eid = {21},
        pages = {21},
          doi = {10.21105/astro.2210.02243},
archivePrefix = {arXiv},
       eprint = {2210.02243},
 primaryClass = {astro-ph.IM},
       adsurl = {https://ui.adsabs.harvard.edu/abs/2023OJAp....6E..21N}
}

@ARTICLE{Stompor_2009,
       author = {{Stompor}, Radek and {Leach}, Samuel and {Stivoli}, Federico and {Baccigalupi}, Carlo},
        title = "{Maximum likelihood algorithm for parametric component separation in cosmic microwave background experiments}",
      journal = {\mnras},
         year = 2009,
        month = jan,
       volume = {392},
       number = {1},
        pages = {216-232},
          doi = {10.1111/j.1365-2966.2008.14023.x},
archivePrefix = {arXiv},
       eprint = {0804.2645},
 primaryClass = {astro-ph},
       adsurl = {https://ui.adsabs.harvard.edu/abs/2009MNRAS.392..216S}
}

@ARTICLE{SO_overview_2019,
       author = {{Ade}, Peter and {Aguirre}, James and {Ahmed}, Zeeshan and {Aiola}, Simone and {Ali}, Aamir and {Alonso}, David and {Alvarez}, Marcelo A. and {Arnold}, Kam and {Ashton}, Peter and {Austermann}, Jason and {Awan}, Humna and {Baccigalupi}, Carlo and {Baildon}, Taylor and {Barron}, Darcy and {Battaglia}, Nick and {Battye}, Richard and {Baxter}, Eric and {Bazarko}, Andrew and {Beall}, James A. and {Bean}, Rachel and {Beck}, Dominic and {Beckman}, Shawn and {Beringue}, Benjamin and {Bianchini}, Federico and {Boada}, Steven and {Boettger}, David and {Bond}, J. Richard and {Borrill}, Julian and {Brown}, Michael L. and {Bruno}, Sarah Marie and {Bryan}, Sean and {Calabrese}, Erminia and {Calafut}, Victoria and {Calisse}, Paolo and {Carron}, Julien and {Challinor}, Anthony and {Chesmore}, Grace and {Chinone}, Yuji and {Chluba}, Jens and {Cho}, Hsiao-Mei Sherry and {Choi}, Steve and {Coppi}, Gabriele and {Cothard}, Nicholas F. and {Coughlin}, Kevin and {Crichton}, Devin and {Crowley}, Kevin D. and {Crowley}, Kevin T. and {Cukierman}, Ari and {D'Ewart}, John M. and {D{\"u}nner}, Rolando and {de Haan}, Tijmen and {Devlin}, Mark and {Dicker}, Simon and {Didier}, Joy and {Dobbs}, Matt and {Dober}, Bradley and {Duell}, Cody J. and {Duff}, Shannon and {Duivenvoorden}, Adri and {Dunkley}, Jo and {Dusatko}, John and {Errard}, Josquin and {Fabbian}, Giulio and {Feeney}, Stephen and {Ferraro}, Simone and {Flux{\`a}}, Pedro and {Freese}, Katherine and {Frisch}, Josef C. and {Frolov}, Andrei and {Fuller}, George and {Fuzia}, Brittany and {Galitzki}, Nicholas and {Gallardo}, Patricio A. and {Tomas Galvez Ghersi}, Jose and {Gao}, Jiansong and {Gawiser}, Eric and {Gerbino}, Martina and {Gluscevic}, Vera and {Goeckner-Wald}, Neil and {Golec}, Joseph and {Gordon}, Sam and {Gralla}, Megan and {Green}, Daniel and {Grigorian}, Arpi and {Groh}, John and {Groppi}, Chris and {Guan}, Yilun and {Gudmundsson}, Jon E. and {Han}, Dongwon and {Hargrave}, Peter and {Hasegawa}, Masaya and {Hasselfield}, Matthew and {Hattori}, Makoto and {Haynes}, Victor and {Hazumi}, Masashi and {He}, Yizhou and {Healy}, Erin and {Henderson}, Shawn W. and {Hervias-Caimapo}, Carlos and {Hill}, Charles A. and {Hill}, J. Colin and {Hilton}, Gene and {Hilton}, Matt and {Hincks}, Adam D. and {Hinshaw}, Gary and {Hlo{\v{z}}ek}, Ren{\'e}e and {Ho}, Shirley and {Ho}, Shuay-Pwu Patty and {Howe}, Logan and {Huang}, Zhiqi and {Hubmayr}, Johannes and {Huffenberger}, Kevin and {Hughes}, John P. and {Ijjas}, Anna and {Ikape}, Margaret and {Irwin}, Kent and {Jaffe}, Andrew H. and {Jain}, Bhuvnesh and {Jeong}, Oliver and {Kaneko}, Daisuke and {Karpel}, Ethan D. and {Katayama}, Nobuhiko and {Keating}, Brian and {Kernasovskiy}, Sarah S. and {Keskitalo}, Reijo and {Kisner}, Theodore and {Kiuchi}, Kenji and {Klein}, Jeff and {Knowles}, Kenda and {Koopman}, Brian and {Kosowsky}, Arthur and {Krachmalnicoff}, Nicoletta and {Kuenstner}, Stephen E. and {Kuo}, Chao-Lin and {Kusaka}, Akito and {Lashner}, Jacob and {Lee}, Adrian and {Lee}, Eunseong and {Leon}, David and {Leung}, Jason S. -Y. and {Lewis}, Antony and {Li}, Yaqiong and {Li}, Zack and {Limon}, Michele and {Linder}, Eric and {Lopez-Caraballo}, Carlos and {Louis}, Thibaut and {Lowry}, Lindsay and {Lungu}, Marius and {Madhavacheril}, Mathew and {Mak}, Daisy and {Maldonado}, Felipe and {Mani}, Hamdi and {Mates}, Ben and {Matsuda}, Frederick and {Maurin}, Lo{\"\i}c and {Mauskopf}, Phil and {May}, Andrew and {McCallum}, Nialh and {McKenney}, Chris and {McMahon}, Jeff and {Meerburg}, P. Daniel and {Meyers}, Joel and {Miller}, Amber and {Mirmelstein}, Mark and {Moodley}, Kavilan and {Munchmeyer}, Moritz and {Munson}, Charles and {Naess}, Sigurd and {Nati}, Federico and {Navaroli}, Martin and {Newburgh}, Laura and {Nguyen}, Ho Nam and {Niemack}, Michael and {Nishino}, Haruki and {Orlowski-Scherer}, John and {Page}, Lyman and {Partridge}, Bruce and {Peloton}, Julien and {Perrotta}, Francesca and {Piccirillo}, Lucio and {Pisano}, Giampaolo and {Poletti}, Davide and {Puddu}, Roberto and {Puglisi}, Giuseppe and {Raum}, Chris and {Reichardt}, Christian L. and {Remazeilles}, Mathieu and {Rephaeli}, Yoel and {Riechers}, Dominik and {Rojas}, Felipe and {Roy}, Anirban and {Sadeh}, Sharon and {Sakurai}, Yuki and {Salatino}, Maria and {Sathyanarayana Rao}, Mayuri and {Schaan}, Emmanuel and {Schmittfull}, Marcel and {Sehgal}, Neelima and {Seibert}, Joseph and {Seljak}, Uros and {Sherwin}, Blake and {Shimon}, Meir and {Sierra}, Carlos and {Sievers}, Jonathan and {Sikhosana}, Precious and {Silva-Feaver}, Maximiliano and {Simon}, Sara M. and {Sinclair}, Adrian and {Siritanasak}, Praween and {Smith}, Kendrick and {Smith}, Stephen R. and {Spergel}, David and {Staggs}, Suzanne T. and {Stein}, George and {Stevens}, Jason R. and {Stompor}, Radek and {Suzuki}, Aritoki and {Tajima}, Osamu and {Takakura}, Satoru and {Teply}, Grant and {Thomas}, Daniel B. and {Thorne}, Ben and {Thornton}, Robert and {Trac}, Hy and {Tsai}, Calvin and {Tucker}, Carole and {Ullom}, Joel and {Vagnozzi}, Sunny and {van Engelen}, Alexander and {Van Lanen}, Jeff and {Van Winkle}, Daniel D. and {Vavagiakis}, Eve M. and {Verg{\`e}s}, Clara and {Vissers}, Michael and {Wagoner}, Kasey and {Walker}, Samantha and {Ward}, Jon and {Westbrook}, Ben and {Whitehorn}, Nathan and {Williams}, Jason and {Williams}, Joel and {Wollack}, Edward J. and {Xu}, Zhilei and {Yu}, Byeonghee and {Yu}, Cyndia and {Zago}, Fernando and {Zhang}, Hezi and {Zhu}, Ningfeng and {Simons Observatory Collaboration}},
        title = "{The Simons Observatory: science goals and forecasts}",
      journal = {\jcap},
         year = 2019,
        month = feb,
       volume = {2019},
       number = {2},
          eid = {056},
        pages = {056},
          doi = {10.1088/1475-7516/2019/02/056},
archivePrefix = {arXiv},
       eprint = {1808.07445},
 primaryClass = {astro-ph.CO},
       adsurl = {https://ui.adsabs.harvard.edu/abs/2019JCAP...02..056A}
}

@article{Galitzki_2024,
doi = {10.3847/1538-4365/ad64c9},
url = {https://dx.doi.org/10.3847/1538-4365/ad64c9},
year = {2024},
month = {sep},
publisher = {The American Astronomical Society},
volume = {274},
number = {2},
pages = {33},
author = {Galitzki, Nicholas and Tsan, Tran and Spisak, Jake and Randall, Michael and Silva-Feaver, Max and Seibert, Joseph and Lashner, Jacob and Adachi, Shunsuke and Adkins, Sean M. and Alford, Thomas and Arnold, Kam and Ashton, Peter C. and Austermann, Jason E. and Baccigalupi, Carlo and Bazarko, Andrew and Beall, James A. and Bhimani, Sanah and Bixler, Bryce and Coppi, Gabriele and Corbett, Lance and Crowley, Kevin D. and Crowley, Kevin T. and Day-Weiss, Samuel and Devlin, Mark J. and Dicker, Simon and DiGia, Brooke and Dow, Peter N. and Duell, Cody J. and Duff, Shannon M. and Gerras, Remington G. and Groh, John C. and Gudmundsson, Jon E. and Harrington, Kathleen and Hasegawa, Masaya and Healy, Erin and Henderson, Shawn W. and Hubmayr, Johannes and Iuliano, Jeffrey and Johnson, Bradley R. and Keating, Brian and Keller, Ben and Kiuchi, Kenji and Kofman, Anna M. and Koopman, Brian J. and Kusaka, Akito and Lee, Adrian T. and Lew, Richard A. and Lin, Lawrence T. and Link, Michael J. and Lucas, Tammy J. and Lungu, Marius and Mangu, Aashrita and McMahon, Jeffrey J and Miller, Amber D. and Moore, Jenna E. and Morshed, Magdy and Nakata, Hironobu and Nati, Federico and Newburgh, Laura B. and Nguyen, David V. and Niemack, Michael D. and Page, Lyman A. and Sakaguri, Kana and Sakurai, Yuki and Sathyanarayana Rao, Mayuri and Saunders, Lauren J. and Shroyer, Jordan E. and Sugiyama, Junna and Tajima, Osamu and Takeuchi, Atsuto and Bua, Refilwe Tanah and Teply, Grant and Terasaki, Tomoki and Ullom, Joel N. and Van Lanen, Jeffrey L. and Vavagiakis, Eve M. and Vissers, Michael R and Walters, Liam and Wang, Yuhan and Xu, Zhilei and Yamada, Kyohei and Zheng, Kaiwen},
title = {The Simons Observatory: Design, Integration, and Testing of the Small Aperture Telescopes},
journal = {The Astrophysical Journal Supplement Series}
}

@ARTICLE{SO_LATr_2021,
       author = {{Zhu}, Ningfeng and {Bhandarkar}, Tanay and {Coppi}, Gabriele and {Kofman}, Anna M. and {Orlowski-Scherer}, John L. and {Xu}, Zhilei and {Adachi}, Shunsuke and {Ade}, Peter and {Aiola}, Simone and {Austermann}, Jason and {Bazarko}, Andrew O. and {Beall}, James A. and {Bhimani}, Sanah and {Bond}, J. Richard and {Chesmore}, Grace E. and {Choi}, Steve K. and {Connors}, Jake and {Cothard}, Nicholas F. and {Devlin}, Mark and {Dicker}, Simon and {Dober}, Bradley and {Duell}, Cody J. and {Duff}, Shannon M. and {D{\"u}nner}, Rolando and {Fabbian}, Giulio and {Galitzki}, Nicholas and {Gallardo}, Patricio A. and {Golec}, Joseph E. and {Haridas}, Saianeesh K. and {Harrington}, Kathleen and {Healy}, Erin and {Ho}, Shuay-Pwu Patty and {Huber}, Zachary B. and {Hubmayr}, Johannes and {Iuliano}, Jeffrey and {Johnson}, Bradley R. and {Keating}, Brian and {Kiuchi}, Kenji and {Koopman}, Brian J. and {Lashner}, Jack and {Lee}, Adrian T. and {Li}, Yaqiong and {Limon}, Michele and {Link}, Michael and {Lucas}, Tammy J. and {McCarrick}, Heather and {Moore}, Jenna and {Nati}, Federico and {Newburgh}, Laura B. and {Niemack}, Michael D. and {Pierpaoli}, Elena and {Randall}, Michael J. and {Sarmiento}, Karen Perez and {Saunders}, Lauren J. and {Seibert}, Joseph and {Sierra}, Carlos and {Sonka}, Rita and {Spisak}, Jacob and {Sutariya}, Shreya and {Tajima}, Osamu and {Teply}, Grant P. and {Thornton}, Robert J. and {Tsan}, Tran and {Tucker}, Carole and {Ullom}, Joel and {Vavagiakis}, Eve M. and {Vissers}, Michael R. and {Walker}, Samantha and {Westbrook}, Benjamin and {Wollack}, Edward J. and {Zannoni}, Mario},
        title = "{The Simons Observatory Large Aperture Telescope Receiver}",
      journal = {\apjs},
         year = 2021,
        month = sep,
       volume = {256},
       number = {1},
          eid = {23},
        pages = {23},
          doi = {10.3847/1538-4365/ac0db7},
archivePrefix = {arXiv},
       eprint = {2103.02747},
 primaryClass = {astro-ph.IM},
       adsurl = {https://ui.adsabs.harvard.edu/abs/2021ApJS..256...23Z}
}

@ARTICLE{Zaldarriaga_1997,
       author = {{Zaldarriaga}, Matias and {Spergel}, David N. and {Seljak}, Uro{\v{s}}},
        title = "{Microwave Background Constraints on Cosmological Parameters}",
      journal = {\apj},
         year = 1997,
        month = oct,
       volume = {488},
       number = {1},
        pages = {1-13},
          doi = {10.1086/304692},
archivePrefix = {arXiv},
       eprint = {astro-ph/9702157},
 primaryClass = {astro-ph},
       adsurl = {https://ui.adsabs.harvard.edu/abs/1997ApJ...488....1Z}
}

@ARTICLE{Planck_2013-9,
       author = {{Planck Collaboration} and {Ade}, P.~A.~R. and {Aghanim}, N. and {Armitage-Caplan}, C. and {Arnaud}, M. and {Ashdown}, M. and {Atrio-Barandela}, F. and {Aumont}, J. and {Baccigalupi}, C. and {Banday}, A.~J. and {Barreiro}, R.~B. and {Battaner}, E. and {Benabed}, K. and {Beno{\^\i}t}, A. and {Benoit-L{\'e}vy}, A. and {Bernard}, J. -P. and {Bersanelli}, M. and {Bielewicz}, P. and {Bobin}, J. and {Bock}, J.~J. and {Bond}, J.~R. and {Borrill}, J. and {Bouchet}, F.~R. and {Boulanger}, F. and {Bridges}, M. and {Bucher}, M. and {Burigana}, C. and {Cardoso}, J. -F. and {Catalano}, A. and {Challinor}, A. and {Chamballu}, A. and {Chary}, R. -R. and {Chen}, X. and {Chiang}, H.~C. and {Chiang}, L. -Y. and {Christensen}, P.~R. and {Church}, S. and {Clements}, D.~L. and {Colombi}, S. and {Colombo}, L.~P.~L. and {Combet}, C. and {Comis}, B. and {Couchot}, F. and {Coulais}, A. and {Crill}, B.~P. and {Curto}, A. and {Cuttaia}, F. and {Danese}, L. and {Davies}, R.~D. and {de Bernardis}, P. and {de Rosa}, A. and {de Zotti}, G. and {Delabrouille}, J. and {Delouis}, J. -M. and {D{\'e}sert}, F. -X. and {Dickinson}, C. and {Diego}, J.~M. and {Dole}, H. and {Donzelli}, S. and {Dor{\'e}}, O. and {Douspis}, M. and {Dupac}, X. and {Efstathiou}, G. and {En{\ss}lin}, T.~A. and {Eriksen}, H.~K. and {Falgarone}, E. and {Finelli}, F. and {Forni}, O. and {Frailis}, M. and {Franceschi}, E. and {Galeotta}, S. and {Ganga}, K. and {Giard}, M. and {Giraud-H{\'e}raud}, Y. and {Gonz{\'a}lez-Nuevo}, J. and {G{\'o}rski}, K.~M. and {Gratton}, S. and {Gregorio}, A. and {Gruppuso}, A. and {Hansen}, F.~K. and {Hanson}, D. and {Harrison}, D. and {Henrot-Versill{\'e}}, S. and {Hern{\'a}ndez-Monteagudo}, C. and {Herranz}, D. and {Hildebrandt}, S.~R. and {Hivon}, E. and {Hobson}, M. and {Holmes}, W.~A. and {Hornstrup}, A. and {Hovest}, W. and {Huffenberger}, K.~M. and {Hurier}, G. and {Jaffe}, A.~H. and {Jaffe}, T.~R. and {Jones}, W.~C. and {Juvela}, M. and {Keih{\"a}nen}, E. and {Keskitalo}, R. and {Kisner}, T.~S. and {Kneissl}, R. and {Knoche}, J. and {Knox}, L. and {Kunz}, M. and {Kurki-Suonio}, H. and {Lagache}, G. and {Lamarre}, J. -M. and {Lasenby}, A. and {Laureijs}, R.~J. and {Lawrence}, C.~R. and {Leahy}, J.~P. and {Leonardi}, R. and {Leroy}, C. and {Lesgourgues}, J. and {Liguori}, M. and {Lilje}, P.~B. and {Linden-V{\o}rnle}, M. and {L{\'o}pez-Caniego}, M. and {Lubin}, P.~M. and {Mac{\'\i}as-P{\'e}rez}, J.~F. and {Maffei}, B. and {Mandolesi}, N. and {Maris}, M. and {Marshall}, D.~J. and {Martin}, P.~G. and {Mart{\'\i}nez-Gonz{\'a}lez}, E. and {Masi}, S. and {Massardi}, M. and {Matarrese}, S. and {Matthai}, F. and {Mazzotta}, P. and {McGehee}, P. and {Melchiorri}, A. and {Mendes}, L. and {Mennella}, A. and {Migliaccio}, M. and {Mitra}, S. and {Miville-Desch{\^e}nes}, M. -A. and {Moneti}, A. and {Montier}, L. and {Morgante}, G. and {Mortlock}, D. and {Munshi}, D. and {Murphy}, J.~A. and {Naselsky}, P. and {Nati}, F. and {Natoli}, P. and {Netterfield}, C.~B. and {N{\o}rgaard-Nielsen}, H.~U. and {North}, C. and {Noviello}, F. and {Novikov}, D. and {Novikov}, I. and {Osborne}, S. and {Oxborrow}, C.~A. and {Paci}, F. and {Pagano}, L. and {Pajot}, F. and {Paoletti}, D. and {Pasian}, F. and {Patanchon}, G. and {Perdereau}, O. and {Perotto}, L. and {Perrotta}, F. and {Piacentini}, F. and {Piat}, M. and {Pierpaoli}, E. and {Pietrobon}, D. and {Plaszczynski}, S. and {Pointecouteau}, E. and {Polenta}, G. and {Ponthieu}, N. and {Popa}, L. and {Poutanen}, T. and {Pratt}, G.~W. and {Pr{\'e}zeau}, G. and {Prunet}, S. and {Puget}, J. -L. and {Rachen}, J.~P. and {Reinecke}, M. and {Remazeilles}, M. and {Renault}, C. and {Ricciardi}, S. and {Riller}, T. and {Ristorcelli}, I. and {Rocha}, G. and {Rosset}, C. and {Roudier}, G. and {Rusholme}, B. and {Santos}, D. and {Savini}, G. and {Scott}, D. and {Shellard}, E.~P.~S. and {Spencer}, L.~D. and {Starck}, J. -L. and {Stolyarov}, V. and {Stompor}, R. and {Sudiwala}, R. and {Sureau}, F. and {Sutton}, D. and {Suur-Uski}, A. -S. and {Sygnet}, J. -F. and {Tauber}, J.~A. and {Tavagnacco}, D. and {Terenzi}, L. and {Tomasi}, M. and {Tristram}, M. and {Tucci}, M. and {Umana}, G. and {Valenziano}, L. and {Valiviita}, J. and {Van Tent}, B. and {Vielva}, P. and {Villa}, F. and {Vittorio}, N. and {Wade}, L.~A. and {Wandelt}, B.~D. and {Yvon}, D. and {Zacchei}, A. and {Zonca}, A.},
        title = "{Planck 2013 results. IX. HFI spectral response}",
      journal = {\aap},
         year = 2014,
        month = nov,
       volume = {571},
          eid = {A9},
        pages = {A9},
          doi = {10.1051/0004-6361/201321531},
archivePrefix = {arXiv},
       eprint = {1303.5070},
 primaryClass = {astro-ph.IM},
       adsurl = {https://ui.adsabs.harvard.edu/abs/2014A&A...571A...9P}
}

@ARTICLE{Planck_dustfore_2015,
       author = {{Planck Collaboration} and {Ade}, P.~A.~R. and {Alves}, M.~I.~R. and {Aniano}, G. and {Armitage-Caplan}, C. and {Arnaud}, M. and {Atrio-Barandela}, F. and {Aumont}, J. and {Baccigalupi}, C. and {Banday}, A.~J. and {Barreiro}, R.~B. and {Battaner}, E. and {Benabed}, K. and {Benoit-L{\'e}vy}, A. and {Bernard}, J. -P. and {Bersanelli}, M. and {Bielewicz}, P. and {Bock}, J.~J. and {Bond}, J.~R. and {Borrill}, J. and {Bouchet}, F.~R. and {Boulanger}, F. and {Burigana}, C. and {Cardoso}, J. -F. and {Catalano}, A. and {Chamballu}, A. and {Chiang}, H.~C. and {Colombo}, L.~P.~L. and {Combet}, C. and {Couchot}, F. and {Coulais}, A. and {Crill}, B.~P. and {Curto}, A. and {Cuttaia}, F. and {Danese}, L. and {Davies}, R.~D. and {Davis}, R.~J. and {de Bernardis}, P. and {de Zotti}, G. and {Delabrouille}, J. and {D{\'e}sert}, F. -X. and {Dickinson}, C. and {Diego}, J.~M. and {Donzelli}, S. and {Dor{\'e}}, O. and {Douspis}, M. and {Dunkley}, J. and {Dupac}, X. and {En{\ss}lin}, T.~A. and {Eriksen}, H.~K. and {Falgarone}, E. and {Finelli}, F. and {Forni}, O. and {Frailis}, M. and {Fraisse}, A.~A. and {Franceschi}, E. and {Galeotta}, S. and {Ganga}, K. and {Ghosh}, T. and {Giard}, M. and {Gonz{\'a}lez-Nuevo}, J. and {G{\'o}rski}, K.~M. and {Gregorio}, A. and {Gruppuso}, A. and {Guillet}, V. and {Hansen}, F.~K. and {Harrison}, D.~L. and {Helou}, G. and {Hern{\'a}ndez-Monteagudo}, C. and {Hildebrandt}, S.~R. and {Hivon}, E. and {Hobson}, M. and {Holmes}, W.~A. and {Hornstrup}, A. and {Jaffe}, A.~H. and {Jaffe}, T.~R. and {Jones}, W.~C. and {Keih{\"a}nen}, E. and {Keskitalo}, R. and {Kisner}, T.~S. and {Kneissl}, R. and {Knoche}, J. and {Kunz}, M. and {Kurki-Suonio}, H. and {Lagache}, G. and {Lamarre}, J. -M. and {Lasenby}, A. and {Lawrence}, C.~R. and {Leahy}, J.~P. and {Leonardi}, R. and {Levrier}, F. and {Liguori}, M. and {Lilje}, P.~B. and {Linden-V{\o}rnle}, M. and {L{\'o}pez-Caniego}, M. and {Lubin}, P.~M. and {Mac{\'\i}as-P{\'e}rez}, J.~F. and {Maffei}, B. and {Magalh{\~a}es}, A.~M. and {Maino}, D. and {Mandolesi}, N. and {Maris}, M. and {Marshall}, D.~J. and {Martin}, P.~G. and {Mart{\'\i}nez-Gonz{\'a}lez}, E. and {Masi}, S. and {Matarrese}, S. and {Mazzotta}, P. and {Melchiorri}, A. and {Mendes}, L. and {Mennella}, A. and {Migliaccio}, M. and {Miville-Desch{\^e}nes}, M. -A. and {Moneti}, A. and {Montier}, L. and {Morgante}, G. and {Mortlock}, D. and {Munshi}, D. and {Murphy}, J.~A. and {Naselsky}, P. and {Nati}, F. and {Natoli}, P. and {Netterfield}, C.~B. and {Noviello}, F. and {Novikov}, D. and {Novikov}, I. and {Oppermann}, N. and {Oxborrow}, C.~A. and {Pagano}, L. and {Pajot}, F. and {Paoletti}, D. and {Pasian}, F. and {Perdereau}, O. and {Perotto}, L. and {Perrotta}, F. and {Piacentini}, F. and {Pietrobon}, D. and {Plaszczynski}, S. and {Pointecouteau}, E. and {Polenta}, G. and {Popa}, L. and {Pratt}, G.~W. and {Rachen}, J.~P. and {Reach}, W.~T. and {Reinecke}, M. and {Remazeilles}, M. and {Renault}, C. and {Ricciardi}, S. and {Riller}, T. and {Ristorcelli}, I. and {Rocha}, G. and {Rosset}, C. and {Roudier}, G. and {Rubi{\~n}o-Mart{\'\i}n}, J.~A. and {Rusholme}, B. and {Salerno}, E. and {Sandri}, M. and {Savini}, G. and {Scott}, D. and {Spencer}, L.~D. and {Stolyarov}, V. and {Stompor}, R. and {Sudiwala}, R. and {Sutton}, D. and {Suur-Uski}, A. -S. and {Sygnet}, J. -F. and {Tauber}, J.~A. and {Terenzi}, L. and {Toffolatti}, L. and {Tomasi}, M. and {Tristram}, M. and {Tucci}, M. and {Valenziano}, L. and {Valiviita}, J. and {Van Tent}, B. and {Vielva}, P. and {Villa}, F. and {Wandelt}, B.~D. and {Zacchei}, A. and {Zonca}, A.},
        title = "{Planck intermediate results. XXII. Frequency dependence of thermal emission from Galactic dust in intensity and polarization}",
      journal = {\aap},
         year = 2015,
        month = apr,
       volume = {576},
          eid = {A107},
        pages = {A107},
          doi = {10.1051/0004-6361/201424088},
archivePrefix = {arXiv},
       eprint = {1405.0874},
 primaryClass = {astro-ph.GA},
       adsurl = {https://ui.adsabs.harvard.edu/abs/2015A&A...576A.107P}
}

@ARTICLE{Planck_2018_11,
       author = {{Planck Collaboration} and {Akrami}, Y. and {Ashdown}, M. and {Aumont}, J. and {Baccigalupi}, C. and {Ballardini}, M. and {Banday}, A.~J. and {Barreiro}, R.~B. and {Bartolo}, N. and {Basak}, S. and {Benabed}, K. and {Bernard}, J. -P. and {Bersanelli}, M. and {Bielewicz}, P. and {Bond}, J.~R. and {Borrill}, J. and {Bouchet}, F.~R. and {Boulanger}, F. and {Bracco}, A. and {Bucher}, M. and {Burigana}, C. and {Calabrese}, E. and {Cardoso}, J. -F. and {Carron}, J. and {Chiang}, H.~C. and {Combet}, C. and {Crill}, B.~P. and {de Bernardis}, P. and {de Zotti}, G. and {Delabrouille}, J. and {Delouis}, J. -M. and {Di Valentino}, E. and {Dickinson}, C. and {Diego}, J.~M. and {Ducout}, A. and {Dupac}, X. and {Efstathiou}, G. and {Elsner}, F. and {En{\ss}lin}, T.~A. and {Falgarone}, E. and {Fantaye}, Y. and {Ferri{\`e}re}, K. and {Finelli}, F. and {Forastieri}, F. and {Frailis}, M. and {Fraisse}, A.~A. and {Franceschi}, E. and {Frolov}, A. and {Galeotta}, S. and {Galli}, S. and {Ganga}, K. and {G{\'e}nova-Santos}, R.~T. and {Ghosh}, T. and {Gonz{\'a}lez-Nuevo}, J. and {G{\'o}rski}, K.~M. and {Gruppuso}, A. and {Gudmundsson}, J.~E. and {Guillet}, V. and {Handley}, W. and {Hansen}, F.~K. and {Herranz}, D. and {Huang}, Z. and {Jaffe}, A.~H. and {Jones}, W.~C. and {Keih{\"a}nen}, E. and {Keskitalo}, R. and {Kiiveri}, K. and {Kim}, J. and {Krachmalnicoff}, N. and {Kunz}, M. and {Kurki-Suonio}, H. and {Lamarre}, J. -M. and {Lasenby}, A. and {Le Jeune}, M. and {Levrier}, F. and {Liguori}, M. and {Lilje}, P.~B. and {Lindholm}, V. and {L{\'o}pez-Caniego}, M. and {Lubin}, P.~M. and {Ma}, Y. -Z. and {Mac{\'\i}as-P{\'e}rez}, J.~F. and {Maggio}, G. and {Maino}, D. and {Mandolesi}, N. and {Mangilli}, A. and {Martin}, P.~G. and {Mart{\'\i}nez-Gonz{\'a}lez}, E. and {Matarrese}, S. and {McEwen}, J.~D. and {Meinhold}, P.~R. and {Melchiorri}, A. and {Migliaccio}, M. and {Miville-Desch{\^e}nes}, M. -A. and {Molinari}, D. and {Moneti}, A. and {Montier}, L. and {Morgante}, G. and {Natoli}, P. and {Pagano}, L. and {Paoletti}, D. and {Pettorino}, V. and {Piacentini}, F. and {Polenta}, G. and {Puget}, J. -L. and {Rachen}, J.~P. and {Reinecke}, M. and {Remazeilles}, M. and {Renzi}, A. and {Rocha}, G. and {Rosset}, C. and {Roudier}, G. and {Rubi{\~n}o-Mart{\'\i}n}, J.~A. and {Ruiz-Granados}, B. and {Salvati}, L. and {Sandri}, M. and {Savelainen}, M. and {Scott}, D. and {Soler}, J.~D. and {Spencer}, L.~D. and {Tauber}, J.~A. and {Tavagnacco}, D. and {Toffolatti}, L. and {Tomasi}, M. and {Trombetti}, T. and {Valiviita}, J. and {Vansyngel}, F. and {Van Tent}, B. and {Vielva}, P. and {Villa}, F. and {Vittorio}, N. and {Wehus}, I.~K. and {Zacchei}, A. and {Zonca}, A.},
        title = "{Planck 2018 results. XI. Polarized dust foregrounds}",
      journal = {\aap},
         year = 2020,
        month = sep,
       volume = {641},
          eid = {A11},
        pages = {A11},
          doi = {10.1051/0004-6361/201832618},
archivePrefix = {arXiv},
       eprint = {1801.04945},
 primaryClass = {astro-ph.GA},
       adsurl = {https://ui.adsabs.harvard.edu/abs/2020A&A...641A..11P}
}

@INPROCEEDINGS{Shitvov_2022,
       author = {{Shitvov}, Alexey and {Savini}, Giorgio and {Hargrave}, Peter C. and {Ade}, Peter A.~R. and {Tucker}, Carole E. and {Sudiwala}, Rashmi V. and {Zhang}, Jin and {Gudmundsson}, Jon E. and {Winter}, Berend and {Pisano}, Giampaolo and {van der Vorst}, Maarten},
        title = "{Broadband coated lens solutions for FIR-mm-wave instruments}",
    booktitle = {Millimeter, Submillimeter, and Far-Infrared Detectors and Instrumentation for Astronomy XI},
         year = 2022,
       editor = {{Zmuidzinas}, Jonas and {Gao}, Jian-Rong},
       series = {Society of Photo-Optical Instrumentation Engineers (SPIE) Conference Series},
       volume = {12190},
        month = aug,
          eid = {121900D},
        pages = {121900D},
          doi = {10.1117/12.2629968},
       adsurl = {https://ui.adsabs.harvard.edu/abs/2022SPIE12190E..0DS}
}

@ARTICLE{WMAP_3_fore,
       author = {{Kogut}, A. and {Dunkley}, J. and {Bennett}, C.~L. and {Dor{\'e}}, O. and {Gold}, B. and {Halpern}, M. and {Hinshaw}, G. and {Jarosik}, N. and {Komatsu}, E. and {Nolta}, M.~R. and {Odegard}, N. and {Page}, L. and {Spergel}, D.~N. and {Tucker}, G.~S. and {Weiland}, J.~L. and {Wollack}, E. and {Wright}, E.~L.},
        title = "{Three-Year Wilkinson Microwave Anisotropy Probe (WMAP) Observations: Foreground Polarization}",
      journal = {\apj},
         year = 2007,
        month = aug,
       volume = {665},
       number = {1},
        pages = {355-362},
          doi = {10.1086/519754},
archivePrefix = {arXiv},
       eprint = {0704.3991},
 primaryClass = {astro-ph},
       adsurl = {https://ui.adsabs.harvard.edu/abs/2007ApJ...665..355K}
}

@article{beamconv_hwp_2021,
   title={Probing frequency-dependent half-wave plate systematics for CMB experiments with full-sky beam convolution simulations},
   volume={502},
   ISSN={1365-2966},
   url={http://dx.doi.org/10.1093/mnras/stab317},
   DOI={10.1093/mnras/stab317},
   number={3},
   journal={Monthly Notices of the Royal Astronomical Society},
   publisher={Oxford University Press (OUP)},
   author={Duivenvoorden, Adriaan J and Adler, Alexandre E and Billi, Matteo and Dachlythra, Nadia and Gudmundsson, Jon E},
   year={2021},
   month={Feb},
   pages={4526–4539}
}

@article{Miller_beam_asymmetry,
   title={CMB polarization systematics due to beam asymmetry: Impact on cosmological birefringence},
   volume={79},
   ISSN={1550-2368},
   url={http://dx.doi.org/10.1103/PhysRevD.79.103002},
   DOI={10.1103/physrevd.79.103002},
   number={10},
   journal={Physical Review D},
   publisher={American Physical Society (APS)},
   author={Miller, N. J. and Shimon, M. and Keating, B. G.},
   year={2009},
   month=may }

@article{Shimon_2008,
   title={CMB polarization systematics due to beam asymmetry: Impact on inflationary science},
   volume={77},
   ISSN={1550-2368},
   url={http://dx.doi.org/10.1103/PhysRevD.77.083003},
   DOI={10.1103/physrevd.77.083003},
   number={8},
   journal={Physical Review D},
   publisher={American Physical Society (APS)},
   author={Shimon, Meir and Keating, Brian and Ponthieu, Nicolas and Hivon, Eric},
   year={2008},
   month=apr }

@article{bicep2_systematics,
   title={Bicep2. III. INSTRUMENTAL SYSTEMATICS},
   volume={814},
   ISSN={1538-4357},
   url={http://dx.doi.org/10.1088/0004-637X/814/2/110},
   DOI={10.1088/0004-637x/814/2/110},
   number={2},
   journal={The Astrophysical Journal},
   publisher={American Astronomical Society},
   author="{The BICEP Collaboration III}",
   year={2015},
   month=nov, pages={110} }

@ARTICLE{Giardello_2024,
       author = {{Giardiello}, S. and {Duivenvoorden}, A.~J. and {Calabrese}, E. and {Galloni}, G. and {Hasselfield}, M. and {Hill}, J.~C. and {La Posta}, A. and {Louis}, T. and {Madhavacheril}, M. and {Pagano}, L.},
        title = "{Modeling beam chromaticity for high-resolution CMB analyses}",
      journal = {\prd},
         year = 2025,
        month = feb,
       volume = {111},
       number = {4},
          eid = {043502},
        pages = {043502},
          doi = {10.1103/PhysRevD.111.043502},
archivePrefix = {arXiv},
       eprint = {2411.10124},
 primaryClass = {astro-ph.CO},
       adsurl = {https://ui.adsabs.harvard.edu/abs/2025PhRvD.111d3502G}}

@INPROCEEDINGS{Hill_2016,
   title={Design and development of an ambient-temperature continuously-rotating achromatic half-wave plate for CMB polarization modulation on the POLARBEAR-2 experiment},
   author={Hill, Charles A. and Beckman, Shawn and Chinone, Yuji and Goeckner-Wald, Neil and Hazumi, Masashi and Keating, Brian and Kusaka, Akito and Lee, Adrian T. and Matsuda, Frederick and Plambeck, Richard and Suzuki, Aritoki and Takakura, Satoru},
   series = {Society of Photo-Optical Instrumentation Engineers (SPIE) Conference Series},
   volume={9914},
   booktitle={Millimeter, Submillimeter, and Far-Infrared Detectors and Instrumentation for Astronomy VIII},
   ISSN={0277-786X},
   url={http://dx.doi.org/10.1117/12.2232280},
   DOI={10.1117/12.2232280},
   publisher={SPIE},
   editor={Holland, Wayne S. and Zmuidzinas, Jonas},
   year={2016},
   month=jul, pages={99142U} }

@article{Nati_2017,
	doi = {10.1142/s2251171717400086},  
	url = {https://doi.org/10.1142%2Fs2251171717400086},
  	year = 2017,
	month = {may},  
	publisher = {World Scientific Pub Co Pte Ltd}, 
	volume = {06}, 
	number = {02}, 
	author = {Federico Nati and Mark J. Devlin and Martina Gerbino and Bradley R. Johnson and Brian Keating and Luca Pagano and Grant Teply},
  
	title = {{POLOCALC}: A Novel Method to Measure the Absolute Polarization Orientation of the Cosmic Microwave Background},
  
	journal = {Journal of Astronomical Instrumentation}
}

@INPROCEEDINGS{Dunner_2021,
  author={Dünner, Rolando and Fluxá, Juan and Best, Sergio and Carrero, Felipe and Boettger, David},
  booktitle={2021 15th European Conference on Antennas and Propagation (EuCAP)}, 
  title={Drone-based polarization calibration source for mm-wave telescopes}, 
  year={2021},
  volume={},
  number={},
  pages={1-5},
  doi={10.23919/EuCAP51087.2021.9411058}}

@inproceedings{Coppi:2022qjs,
author = {Gabriele Coppi and Giulia Conenna and Sofia Savorgnano and Felipe Carrero and Rolando D{\"u}nner Planella and Nicholas Galitzki and Federico Nati and Mario Zannoni},
title = {{PROTOCALC: an artificial calibrator source for CMB telescopes}},
volume = {12190},
booktitle = {Millimeter, Submillimeter, and Far-Infrared Detectors and Instrumentation for Astronomy XI},
editor = {Jonas Zmuidzinas and Jian-Rong Gao},
organization = {International Society for Optics and Photonics},
publisher = {SPIE},
series = {Society of Photo-Optical Instrumentation Engineers (SPIE) Conference Series},
pages = {1219015},
year = {2022},
doi = {10.1117/12.2628312},
URL = {https://doi.org/10.1117/12.2628312}
}

@ARTICLE{Coppi:2025arxiv,
       author = {{Coppi}, Gabriele and {Dachlythra}, Nadia and {Nati}, Federico and {D{\"u}nner-Planella}, Rolando and {Adler}, Alexandre E. and {Errard}, Josquin and {Galitzki}, Nicholas and {Li}, Yunyang and {Petroff}, Matthew A. and {Simon}, Sara M. and {Tsang King Sang}, Ema and {Villarrubia Aguilar}, Amalia and {Wollack}, Edward J. and {Zannoni}, Mario},
        title = "{PROTOCALC, A W-band polarized calibrator for CMB Telescopes: application to Simons Observatory and CLASS}",
      journal = {arXiv e-prints},
         year = 2025,
        month = feb,
          eid = {arXiv:2502.14473},
        pages = {arXiv:2502.14473},
          doi = {10.48550/arXiv.2502.14473},
archivePrefix = {arXiv},
       eprint = {2502.14473},
 primaryClass = {astro-ph.IM},
       adsurl = {https://ui.adsabs.harvard.edu/abs/2025arXiv250214473C}
}

@article{Matsumura:09,
author = {Tomotake Matsumura and Shaul Hanany and Peter Ade and Bradley R. Johnson and Terry J. Jones and Prashanth Jonnalagadda and Giorgio Savini},
journal = {Appl. Opt.},
number = {19},
pages = {3614--3625},
publisher = {Optica Publishing Group},
title = {Performance of three- and five-stack achromatic half-wave plates at millimeter wavelengths},
volume = {48},
month = {Jul},
year = {2009},
url = {https://opg.optica.org/ao/abstract.cfm?URI=ao-48-19-3614},
doi = {10.1364/AO.48.003614},
}

@article{Burleigh:16,
author = {M. R. Burleigh and C. R. Richey and S. A. Rinehart and M. A. Quijada and E. J. Wollack},
journal = {Appl. Opt.},
number = {29},
pages = {8201--8206},
publisher = {Optica Publishing Group},
title = {Spectrometer baseline control via spatial filtering},
volume = {55},
month = {Oct},
year = {2016},
url = {https://opg.optica.org/ao/abstract.cfm?URI=ao-55-29-8201},
doi = {10.1364/AO.55.008201},
}

@ARTICLE{Takakura_2017,
       author = {{Takakura}, Satoru and {Aguilar}, Mario and {Akiba}, Yoshiki and {Arnold}, Kam and {Baccigalupi}, Carlo and {Barron}, Darcy and {Beckman}, Shawn and {Boettger}, David and {Borrill}, Julian and {Chapman}, Scott and {Chinone}, Yuji and {Cukierman}, Ari and {Ducout}, Anne and {Elleflot}, Tucker and {Errard}, Josquin and {Fabbian}, Giulio and {Fujino}, Takuro and {Galitzki}, Nicholas and {Goeckner-Wald}, Neil and {Halverson}, Nils W. and {Hasegawa}, Masaya and {Hattori}, Kaori and {Hazumi}, Masashi and {Hill}, Charles and {Howe}, Logan and {Inoue}, Yuki and {Jaffe}, Andrew H. and {Jeong}, Oliver and {Kaneko}, Daisuke and {Katayama}, Nobuhiko and {Keating}, Brian and {Keskitalo}, Reijo and {Kisner}, Theodore and {Krachmalnicoff}, Nicoletta and {Kusaka}, Akito and {Lee}, Adrian T. and {Leon}, David and {Lowry}, Lindsay and {Matsuda}, Frederick and {Matsumura}, Tomotake and {Navaroli}, Martin and {Nishino}, Haruki and {Paar}, Hans and {Peloton}, Julien and {Poletti}, Davide and {Puglisi}, Giuseppe and {Reichardt}, Christian L. and {Ross}, Colin and {Siritanasak}, Praween and {Suzuki}, Aritoki and {Tajima}, Osamu and {Takatori}, Sayuri and {Teply}, Grant},
        title = "{Performance of a continuously rotating half-wave plate on the POLARBEAR telescope}",
      journal = {\jcap},
         year = 2017,
        month = may,
       volume = {2017},
       number = {5},
          eid = {008},
        pages = {008},
          doi = {10.1088/1475-7516/2017/05/008},
archivePrefix = {arXiv},
       eprint = {1702.07111},
 primaryClass = {astro-ph.IM},
       adsurl = {https://ui.adsabs.harvard.edu/abs/2017JCAP...05..008T}
}

@ARTICLE{hfi_beams_2015,
       author = {{The Planck Collaboration VII} and {Adam}, R. and {Ade}, P.~A.~R. and {Aghanim}, N. and {Arnaud}, M. and {Ashdown}, M. and {Aumont}, J. and {Baccigalupi}, C. and {Banday}, A.~J. and {Barreiro}, R.~B. and {Bartolo}, N. and {Battaner}, E. and {Benabed}, K. and {Beno{\^\i}t}, A. and {Benoit-L{\'e}vy}, A. and {Bernard}, J. -P. and {Bersanelli}, M. and {Bertincourt}, B. and {Bielewicz}, P. and {Bock}, J.~J. and {Bonavera}, L. and {Bond}, J.~R. and {Borrill}, J. and {Bouchet}, F.~R. and {Boulanger}, F. and {Bucher}, M. and {Burigana}, C. and {Calabrese}, E. and {Cardoso}, J. -F. and {Catalano}, A. and {Challinor}, A. and {Chamballu}, A. and {Chary}, R. -R. and {Chiang}, H.~C. and {Christensen}, P.~R. and {Clements}, D.~L. and {Colombi}, S. and {Colombo}, L.~P.~L. and {Combet}, C. and {Couchot}, F. and {Coulais}, A. and {Crill}, B.~P. and {Curto}, A. and {Cuttaia}, F. and {Danese}, L. and {Davies}, R.~D. and {Davis}, R.~J. and {de Bernardis}, P. and {de Rosa}, A. and {de Zotti}, G. and {Delabrouille}, J. and {Delouis}, J. -M. and {D{\'e}sert}, F. -X. and {Diego}, J.~M. and {Dole}, H. and {Donzelli}, S. and {Dor{\'e}}, O. and {Douspis}, M. and {Ducout}, A. and {Dupac}, X. and {Efstathiou}, G. and {Elsner}, F. and {En{\ss}lin}, T.~A. and {Eriksen}, H.~K. and {Falgarone}, E. and {Fergusson}, J. and {Finelli}, F. and {Forni}, O. and {Frailis}, M. and {Fraisse}, A.~A. and {Franceschi}, E. and {Frejsel}, A. and {Galeotta}, S. and {Galli}, S. and {Ganga}, K. and {Ghosh}, T. and {Giard}, M. and {Giraud-H{\'e}raud}, Y. and {Gjerl{\o}w}, E. and {Gonz{\'a}lez-Nuevo}, J. and {G{\'o}rski}, K.~M. and {Gratton}, S. and {Gruppuso}, A. and {Gudmundsson}, J.~E. and {Hansen}, F.~K. and {Hanson}, D. and {Harrison}, D.~L. and {Henrot-Versill{\'e}}, S. and {Herranz}, D. and {Hildebrandt}, S.~R. and {Hivon}, E. and {Hobson}, M. and {Holmes}, W.~A. and {Hornstrup}, A. and {Hovest}, W. and {Huffenberger}, K.~M. and {Hurier}, G. and {Jaffe}, A.~H. and {Jaffe}, T.~R. and {Jones}, W.~C. and {Juvela}, M. and {Keih{\"a}nen}, E. and {Keskitalo}, R. and {Kisner}, T.~S. and {Kneissl}, R. and {Knoche}, J. and {Kunz}, M. and {Kurki-Suonio}, H. and {Lagache}, G. and {Lamarre}, J. -M. and {Lasenby}, A. and {Lattanzi}, M. and {Lawrence}, C.~R. and {Le Jeune}, M. and {Leahy}, J.~P. and {Lellouch}, E. and {Leonardi}, R. and {Lesgourgues}, J. and {Levrier}, F. and {Liguori}, M. and {Lilje}, P.~B. and {Linden-V{\o}rnle}, M. and {L{\'o}pez-Caniego}, M. and {Lubin}, P.~M. and {Mac{\'\i}as-P{\'e}rez}, J.~F. and {Maggio}, G. and {Maino}, D. and {Mandolesi}, N. and {Mangilli}, A. and {Maris}, M. and {Martin}, P.~G. and {Mart{\'\i}nez-Gonz{\'a}lez}, E. and {Masi}, S. and {Matarrese}, S. and {McGehee}, P. and {Melchiorri}, A. and {Mendes}, L. and {Mennella}, A. and {Migliaccio}, M. and {Mitra}, S. and {Miville-Desch{\^e}nes}, M. -A. and {Moneti}, A. and {Montier}, L. and {Moreno}, R. and {Morgante}, G. and {Mortlock}, D. and {Moss}, A. and {Mottet}, S. and {Munshi}, D. and {Murphy}, J.~A. and {Naselsky}, P. and {Nati}, F. and {Natoli}, P. and {Netterfield}, C.~B. and {N{\o}rgaard-Nielsen}, H.~U. and {Noviello}, F. and {Novikov}, D. and {Novikov}, I. and {Oxborrow}, C.~A. and {Paci}, F. and {Pagano}, L. and {Pajot}, F. and {Paoletti}, D. and {Pasian}, F. and {Patanchon}, G. and {Pearson}, T.~J. and {Perdereau}, O. and {Perotto}, L. and {Perrotta}, F. and {Pettorino}, V. and {Piacentini}, F. and {Piat}, M. and {Pierpaoli}, E. and {Pietrobon}, D. and {Plaszczynski}, S. and {Pointecouteau}, E. and {Polenta}, G. and {Pratt}, G.~W. and {Pr{\'e}zeau}, G. and {Prunet}, S. and {Puget}, J. -L. and {Rachen}, J.~P. and {Reinecke}, M. and {Remazeilles}, M. and {Renault}, C. and {Renzi}, A. and {Ristorcelli}, I. and {Rocha}, G. and {Rosset}, C. and {Rossetti}, M. and {Roudier}, G. and {Rowan-Robinson}, M. and {Rusholme}, B. and {Sandri}, M. and {Santos}, D. and {Sauv{\'e}}, A. and {Savelainen}, M. and {Savini}, G. and {Scott}, D. and {Seiffert}, M.~D.},
        title = "{Planck 2015 results. VII. High Frequency Instrument data processing: Time-ordered information and beams}",
      journal = {\aap},
         year = 2016,
        month = sep,
       volume = {594},
          eid = {A7},
        pages = {A7},
          doi = {10.1051/0004-6361/201525844},
archivePrefix = {arXiv},
       eprint = {1502.01586},
 primaryClass = {astro-ph.IM},
       adsurl = {https://ui.adsabs.harvard.edu/abs/2016A&A...594A...7P}
}

@article{Zhu_2021,
doi = {10.3847/1538-4365/ac0db7},
url = {https://dx.doi.org/10.3847/1538-4365/ac0db7},
year = {2021},
month = {sep},
publisher = {The American Astronomical Society},
volume = {256},
number = {1},
pages = {23},
author = {Zhu, Ningfeng and Bhandarkar, Tanay and Coppi, Gabriele and Kofman, Anna M. and Orlowski-Scherer, John L. and Xu, Zhilei and Adachi, Shunsuke and Ade, Peter and Aiola, Simone and Austermann, Jason and Bazarko, Andrew O. and Beall, James A. and Bhimani, Sanah and Bond, J. Richard and Chesmore, Grace E. and Choi, Steve K. and Connors, Jake and Cothard, Nicholas F. and Devlin, Mark and Dicker, Simon and Dober, Bradley and Duell, Cody J. and Duff, Shannon M. and Dünner, Rolando and Fabbian, Giulio and Galitzki, Nicholas and Gallardo, Patricio A. and Golec, Joseph E. and Haridas, Saianeesh K. and Harrington, Kathleen and Healy, Erin and Ho, Shuay-Pwu Patty and Huber, Zachary B. and Hubmayr, Johannes and Iuliano, Jeffrey and Johnson, Bradley R. and Keating, Brian and Kiuchi, Kenji and Koopman, Brian J. and Lashner, Jack and Lee, Adrian T. and Li, Yaqiong and Limon, Michele and Link, Michael and Lucas, Tammy J and McCarrick, Heather and Moore, Jenna and Nati, Federico and Newburgh, Laura B. and Niemack, Michael D. and Pierpaoli, Elena and Randall, Michael J. and Sarmiento, Karen Perez and Saunders, Lauren J. and Seibert, Joseph and Sierra, Carlos and Sonka, Rita and Spisak, Jacob and Sutariya, Shreya and Tajima, Osamu and Teply, Grant P. and Thornton, Robert J. and Tsan, Tran and Tucker, Carole and Ullom, Joel and Vavagiakis, Eve M. and Vissers, Michael R. and Walker, Samantha and Westbrook, Benjamin and Wollack, Edward J. and Zannoni, Mario},
title = {The Simons Observatory Large Aperture Telescope Receiver},
journal = {The Astrophysical Journal Supplement Series}
}

@article{TKS_in_prep,
  author = {{Tsang King Sang}, Ema and {et al.,}},
  title  = "{Half-wave-plate non idealities propagated to component separated CMB B-modes}",
  year   = {2025},
  journal = {in prep.}
}

@article{Gorski_2005,
doi = {10.1086/427976},
url = {https://dx.doi.org/10.1086/427976},
year = {2005},
month = {apr},
publisher = {},
volume = {622},
number = {2},
pages = {759},
author = {Górski, K. M. and Hivon, E. and Banday, A. J. and Wandelt, B. D. and Hansen, F. K. and Reinecke, M. and Bartelmann, M.},
title = {HEALPix: A Framework for High-Resolution Discretization and Fast Analysis of Data Distributed on the Sphere},
journal = {The Astrophysical Journal}
}

@article{Zonca2019,
  doi = {10.21105/joss.01298},
  url = {https://doi.org/10.21105/joss.01298},
  year = {2019},
  month = mar,
  publisher = {The Open Journal},
  volume = {4},
  number = {35},
  pages = {1298},
  author = {Andrea Zonca and Leo Singer and Daniel Lenz and Martin Reinecke and Cyrille Rosset and Eric Hivon and Krzysztof Gorski},
  title = {healpy: equal area pixelization and spherical harmonics transforms for data on the sphere in Python},
  journal = {Journal of Open Source Software}
}
\bibliographystyle{mnras}
\appendix

\section{Bias from the simulated \texttt{beamconv} approach}
\label{cl_based}

To assess the bias induced by integrating \texttt{beamconv} into the pipeline, we run two different kinds of simulations. Both sets refer only to the center frequencies of the SO bands (there is no chromaticity effect included) and employ Gaussian beams. In the first case, the beam convolution is performed using Gaussian smoothing, while in the second, the Gaussian beams are convolved with the maps in \texttt{beamconv} assuming the scan strategy described in Section \ref{beamconv_sims}. In an additional set of simulations, we provide to \texttt{beamconv} Gaussian beams only for the LF bands while the MF and UHF beams now correspond to the symmetric part of the simulated PO beams which are used throughout the paper. This last set aims at capturing the significance of the deviation of the symmetric PO beams from their Gaussian approximation. The maps generated in all ways described above are masked using the same custom mask discussed in Section \ref{beamconv_sims}, and the resulting beam-deconvolved spectra are provided to \bbpower{} to obtain the best-fit estimates for the nine model parameters. For consistency with the tests performed in Section \ref{beam_systematics}, we employ Gaussian foregrounds and present the maximum likelihood posteriors of the tensor-to-scalar ratio and other parameters, averaged over ten different sky realizations in Figure \ref{fig:posteriors_beamconv_bias}. 

\begin{figure}
    \includegraphics[width=\textwidth]{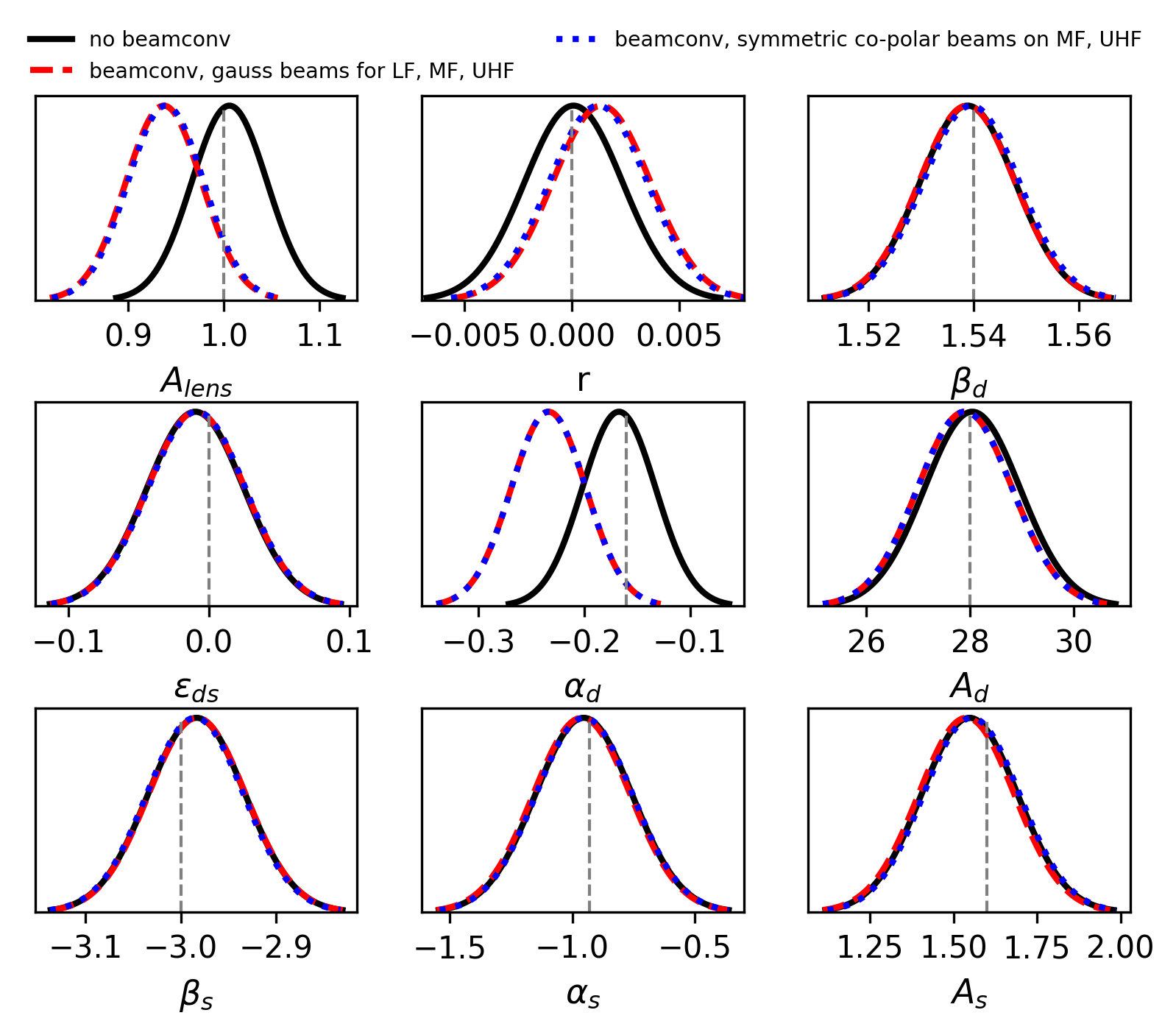} \caption{\label{fig:posteriors_beamconv_bias}The parameter posterior distributions estimated from simulations where the beam-convolution was performed (i) as Gaussian smoothing to all frequency bands (black), (ii) using \texttt{beamconv} simulations with Gaussian beams for all bands (red), and (iii) using \texttt{beamconv} simulations with symmetric co-polar PO beams for the MF, UHF bands and Gaussian smoothing for the LF bands (blue).}
    \end{figure}

From the figure, we observe biases when including \texttt{beamconv} in the pipeline (red and blue curves) with respect to posteriors of the Gaussian smoothing case (black curves) which are closely centered on the input parameters (vertical gray dashed lines). The largest discrepancies refer to the lensing amplitude, $A_{\mathrm{lens}}$, and scaling factor of the dust $B$-mode spectra, $\alpha_{d}$, and to a lesser extent to the tensor-to-scalar ratio, $r$. %These are estimated as 1.7$\sigma$, $1.9\sigma$, and $0.4\sigma$,  for $A_{\mathrm{lens}}$, $\alpha_{d}$, and $r$, respectively, where $\sigma$ here refers to the uncertainty of the most trivial case where we apply a Gaussian smoothing to all frequency bands. 
The main reason for these biases lies in the sparsity of the focal plane of the simulated optical configuration. In order to reduce the required simulation time, we choose to only use two hundred detector pairs across the \SI{35}{\degree} simulated focal plane as opposed to the six thousand pairs employed for the actual SAT focal plane. This inevitably leads to suppression of the smaller scales that carry most of the constraining power for the lensing amplitude and the dust spatial parameters. Part of this bias is also leaking to the tensor-to-scalar ratio as the scale suppression starts around $\ell \sim$ 130 and hence at scales still somewhat relevant to primordial $B$-modes (expected to peak at $\ell \sim 80$). The replacement of the Gaussian beams applied to the MF, UHF bands with the symmetric co-polar ones (blue curves) does not seem to shift the parameter posteriors. Note that the biases induced with \texttt{beamconv} are not a threat to the success of this analysis. Since we will be exclusively working within the context of simulations performed with this software, this shift in the estimated parameters will be consistently included as a systematic in both chromatic and achromatic beam scenarios. Consequently, it will also be consistently eliminated when computing the difference in the best-fit parameters between the two cases. 

\section{Cross-polarization impact on $EB$ correlation}
\label{eb_corr}

As discussed in Section \ref{beam_systematics}, the component separation analysis results do not seem to vary significantly when introducing the cross-polar component of the simulated beam. Nevertheless, it is important to assess the cross-polarization impact on the resulting $EB$ spectra. A rotation of the $EB$ spectra due to cross-polarization could interfere with cosmic birefringence measurements or affect the polarization angle calibration  \citep{beamconv_2018}. In contrast to beam asymmetry, the effect of beam cross-polarization can not be mitigated in the case where a HWP is included in the instrumental setup \citep{beamconv_hwp_2021}. Figure \ref{fig:eb_spectra} shows the auto- and cross-channel $EB$ spectra estimated from six center-frequency \texttt{beamconv} maps of a single sky realization with \namaster. In the case of MF and UHF bands, the maps have been convolved either with strictly co-polar beams (blue) or with beams carrying both the co- and cross-polar components. The LF beams are still approximated with Gaussian distributions. We run simulations using both the cross-polar beams generated with \texttt{TICRA TOOLS} (orange) and a modified version, where the cross-polarization amplitude is intentionally exaggerated by an order of magnitude (green) for comparison.

The sky maps, whose spectra are shown in this figure, consist of CMB and Gaussian foregrounds. They have been masked and corrected for mode-mixing. While it would be more appropriate to use simulations of a CMB-only sky model for this assessment, the spectra convolved with beams that include the unscaled cross-polarization component appear identical to those convolved with only co-polar beams. Non-negligible $EB$ values for these two cases are observed only due to foregrounds and cut-sky effects. The contribution of foregrounds to the $EB$ spectra of maps convolved with either co-polar beams or beams that include the nominal cross-polar component becomes apparent when examining their auto-spectra. Notably, the spectra from the lowest (\SI{27}{\giga\hertz}) and highest frequency channels (\SI{280}{\giga\hertz}) exhibit the largest amplitudes. The MF channels (\SI{93}{} and \SI{145}{\giga\hertz}), which refer to the least foreground-contaminated frequency regime, exhibit $EB$ spectra amplitudes on the order of $10^{-2} \mu K_{\mathrm{CMB}}$ and do not appear to deviate significantly from the spectra shown in \cite{Minami_cb}. To quantify any systematic $E$-to-$B$ leakage from \texttt{beamconv} simulations, the best-fit birefringence angle will be estimated in the near future from maps that do not contain any foreground emission. 

When we scale up the amplitude of the cross-polar beam by an order of magnitude we observe the $EB$ correlation notably increasing, in particular for the highest frequencies. After the scaling, the cross-polar beam becomes roughly just an order of magnitude fainter than the co-polar beam, resulting in some $E$-to-$B$ leakage in the power spectra, primarily impacting the higher multipoles. The degree-scale spectra, where $B$-modes are expected to peak, do not appear to be significantly threatened by this (large) increase in the cross-polarization amplitude. 

\begin{figure}[h!]
    \centering
    \includegraphics[width=\textwidth]{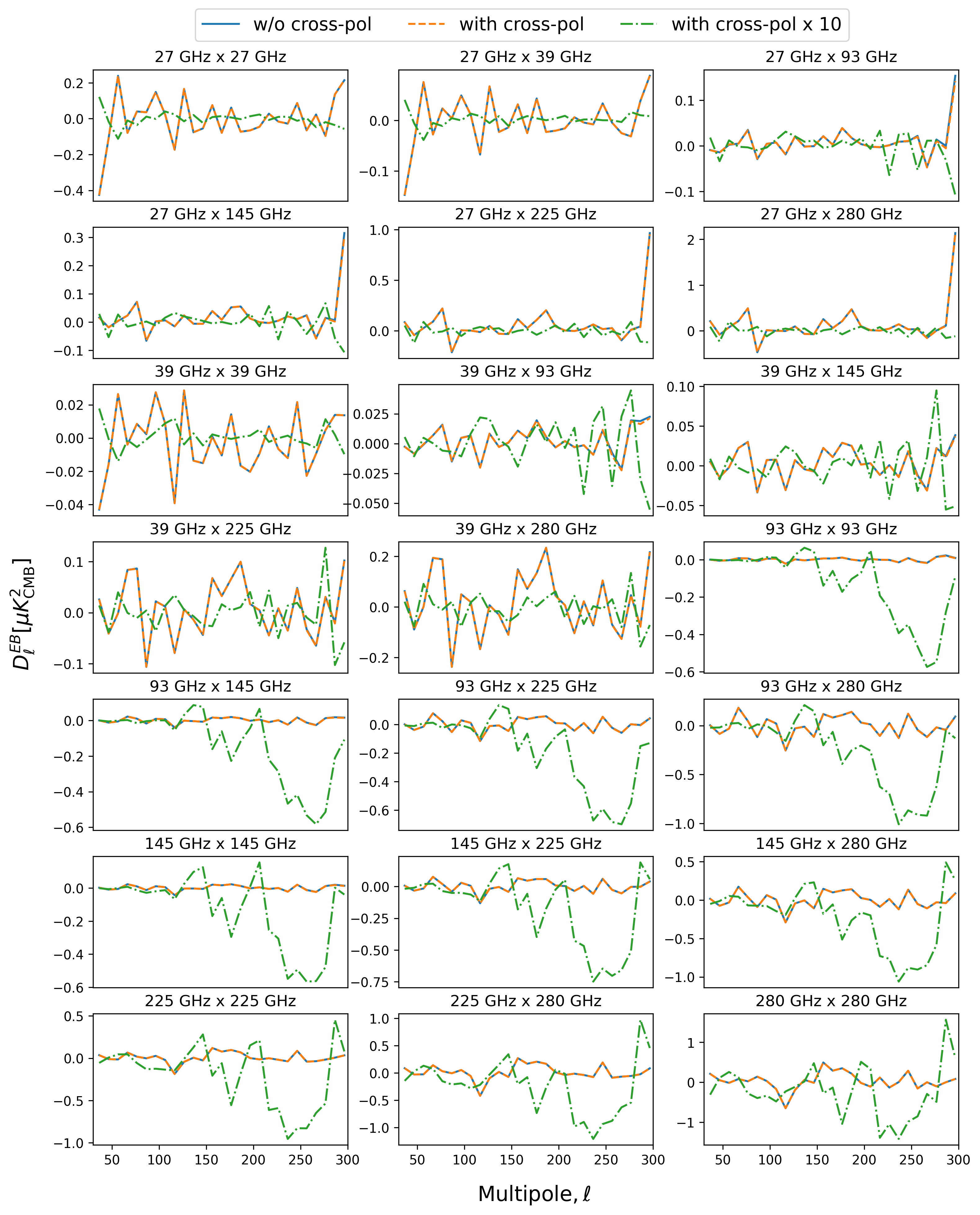}
    \caption{\label{fig:eb_spectra} The $EB$ auto- and cross-channel spectra of a single set of six \texttt{beamconv} maps. The MF and UHF maps have been convolved with beams that either have no cross-polar components (blue), incorporate the original simulated cross-polar components (orange), or include cross-polar components with an amplitude ten times greater than the original simulation (green). The LF beams are assumed to be Gaussian, and the illustrated spectra refer only to the center frequencies of the SO bands. The blue and orange spectra appear identical due to the small cross-polarization in the simulated beams (2-3 orders of magnitude smaller than the co-polar component).}
\end{figure}

%\section{Extended systematic effects}
\section{Dependence on the detector location}
\label{det_loc}

The beam model depends on the position of the detector on the telescope's focal plane. A detector positioned at the edge of the SAT focal plane produces a significantly more elliptical beam pattern compared to a detector placed on the boresight, due to the asymmetrical illumination of the telescope aperture. Specifically, SAT beam simulations indicate a $1-2\%$ variation in beam size and nearly a $50\%$ increase in beam ellipticity between the center and edge pixel cases \citep{Dachlythra_2024}. Figure \ref{fig:center_vs_edge} shows the logarithmic beam profiles of two detectors located at the boresight and edge of the simulated SAT focal plane for the center frequencies of the MF and UHF frequency bands, which have been uniformly averaged over the beam azimuthal angle, $\phi$. Upon examining the various sub-panels of this figure, we deduce that the edge-pixel beam profiles appear more diffuse compared to the sharper main peaks we see for the center pixels. Note that the figure illustrates the power closer to the center of the beam where the main contribution comes from. For the time-domain simulations, however, we use the wide edge-pixel beams (extending to \SI{12}{\degree}) to achieve a direct comparison with the analysis described in Section \ref{non_Gauss_sims} where the input beam models were simulated for boresight pixels only. 

Applying the same component separation method on the $B$-mode spectra of non-Gaussian foreground maps, which have been convolved with edge-pixel beam models, yields parameter values that closely match those presented in Table \ref{tab:forecast_params_chrom_nong}. The most significantly affected parameter is $\beta_{d}$, for which the beam chromaticity bias is now estimated as 0.13$\sigma$, compared to just 0.01$\sigma$ in the center-pixel case. This minor difference, however, may not fully capture the realistic scenario, as the pixels at the edges of the SAT focal plane could also experience additional sidelobe power due to reflections from filters and other surface elements of the optical setup, which are not accounted for in the simulations. 

\section{Chromaticity bias for different beam spectral response}
\label{increased_chrommaticity}

Let us now assume that we have underestimated the intrinsic beam frequency scaling. In the simplest parametrization, we only account for the main beam which we describe solely by its best-fit FWHM. Specifically, we fit a first-order polynomial to the beam sizes of the five monochromatic frequencies employed for the MF and UHF frequency bands. We then increase the fitted slope by $10\%$ and by an exaggerated $25\%$ with respect to its original value while keeping the FWHM value of the center frequency of each band fixed. In this way, we increase the relative beam size difference inside the bands with respect to the corresponding band centers. We run \texttt{beamconv} simulations for the two newly constructed cases employing Gaussian beams and assuming Gaussian foregrounds. Table \ref{tab:increased_beam_spectral_slope} presents the best-fit parameters in the cases where the beam spectral slope increases by $10\%$ and $25\%$, averaged over ten sky realizations. The corresponding parameters for the nominal assumed spectral dependence of the beam are also shown as a reference, along with their input values. 

\begin{table}[h!]
\begin{center}
\begin{tabular}{ |c|c|c|c|c| } 
  \hline
 Parameter &  Input & Chromatic & $10\% $ Chromatic & $25\%$ Chromatic  \\
  \hline
$A_{\mathrm{lens}}$ & 1 & $0.93 \pm 0.04$ & $0.93 \pm 0.04$ & $0.77 \pm 0.07$ \\
$r$ $(10^{-3})$ & 0 & $1.369 \pm 2.305$ & $2.081 \pm 2.352$ & $1.120 \pm 2.0$ \\
$\beta_{d}$ & 1.54 & $1.538 \pm 0.008$ & $1.537 \pm 0.008$ & $1.25 \pm 0.06$ \\
$\varepsilon_{ds}$ $(10^{-3})$ & 0 & $-0.35 \pm 32.33$ & $-9.677 \pm 42.31$ & $-5.314 \pm 29.49$  \\
$\alpha_{d}$ & -0.16 & $-0.239 \pm 0.031$ & $-0.244 \pm 0.031$ & $-0.198 \pm 0.034$ \\
$A_{d}$ & 28 & $27.1 \pm 0.7$ & $27.3 \pm 0.8$ & $22.1 \pm 1.68$ \\
$\beta_{s}$ & -3.0 & $-2.979 \pm 0.051$ & $-2.866 \pm 0.061$ & $-2.388 \pm 0.154$ \\
$\alpha_{s}$ & -0.93 & $-0.920 \pm 0.194$ & $-1.540 \pm 0.246$ & $-0.892 \pm 0.220$ \\
$A_{s}$ & 1.6 & $1.56 \pm 0.14$ & $0.79 \pm 1.30$ & $1.08 \pm 0.53$ \\
\hline
\end{tabular}
\end{center}
\caption{\label{tab:increased_beam_spectral_slope}The average best-fit parameters over ten sky realizations for the cases where we increased the spectral slope of the MF/UHF beams  by $10\%$ and $25\%$. The fitted parameters for the nominal case are also shown together with their input values.}
\end{table}

For the more reasonable case of the $10\%$ increase in chromaticity, we observe the largest biases in the synchrotron parameters. This is due to the component separation algorithm assigning the additional frequency scaling to errors in the assumed spectral dependence of synchrotron which more closely resembles the scaled spectral dependence of the beam than does that of dust. On the other hand, for the exaggerated case of $25\%$, it becomes evident that we can no longer recover the lensing amplitude and foreground parameters. The scaled beam spectral dependence is now very different from the ones of both dust and synchrotron and, thus, the code can not properly integrate it in the fitting model. The spectral indices of dust and synchrotron absorb most of the chromaticity bias. The tensor-to-scalar ratio increases in the first case (corresponding to a bias of $\sim 0.2\sigma$ compared to achromatic beams), yet remains largely unaffected in the second. Note that, even though we assumed Gaussian beams, we did the beam-convolution in the time-domain for consistency in the used method and associated biases throughout the paper.
 
\begin{figure}
    \centering
    \includegraphics[width=0.8\linewidth]{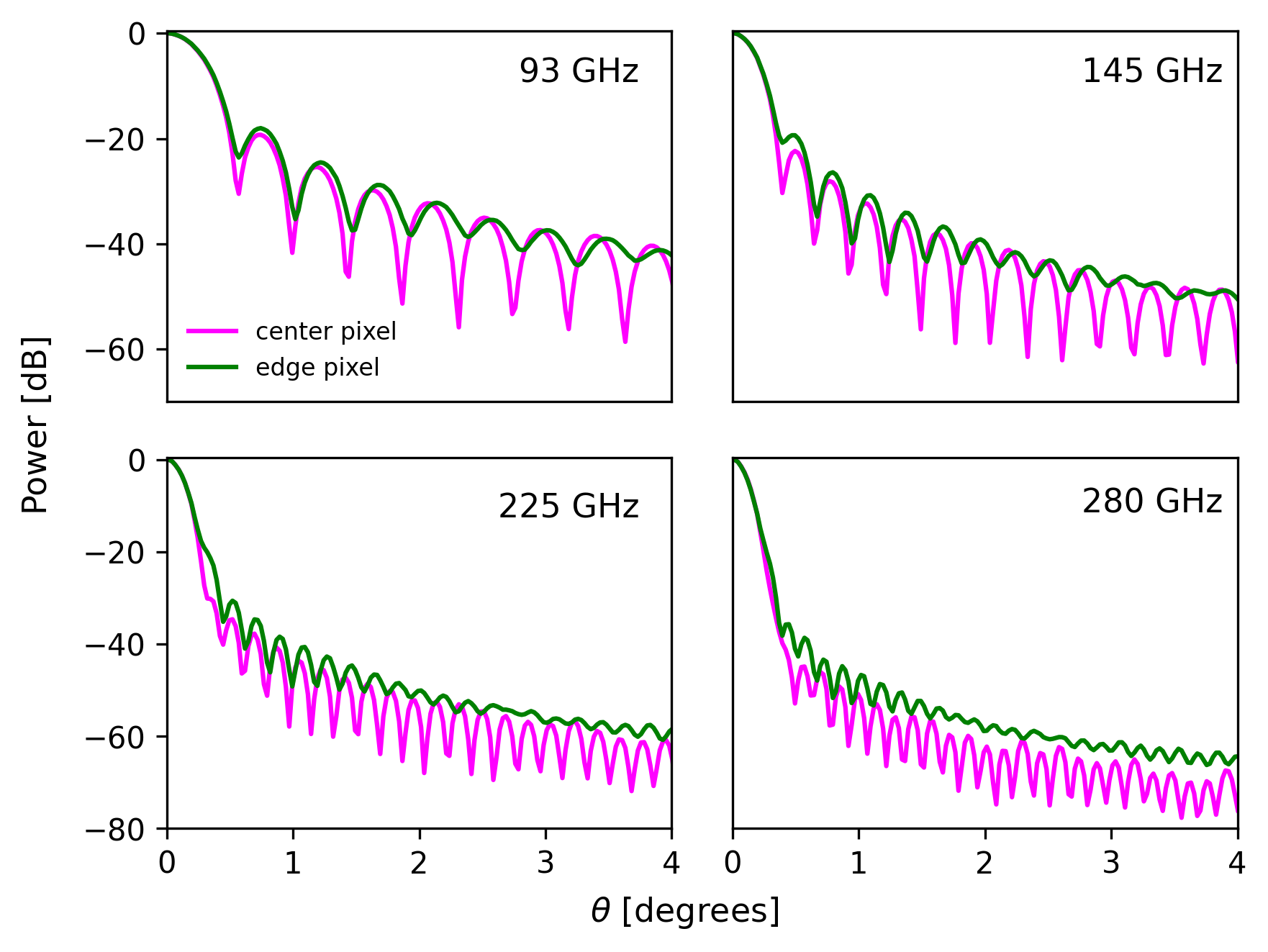}
    \caption{\label{fig:center_vs_edge}Azimuthally averaged radial profiles of far-field beams simulated assuming a detector at the center (purple curve) and a detector at the edge of the SAT focal plane (green curve) for the center frequencies of the four bands shown in logarithmic scale.}
\end{figure}

\end{document}